\documentclass[twocolumn]{aastex631}

\received{July 29, 2024}
\revised{October 24, 2024}
\accepted{November 7, 2024}

\shorttitle{DPConCFil for identifying and analysing filament}
\shortauthors{Jiang et al.}

\graphicspath{{./}{figures/}}
\usepackage{mathrsfs}
\usepackage{amsmath}
\usepackage{appendix}
\usepackage{array}
\usepackage{url}
\usepackage{soul} 
\usepackage{color, xcolor} 
\usepackage{makecell}
\usepackage{booktabs}
\usepackage{threeparttable}

\usepackage{graphicx}
\usepackage{enumitem}

\shortauthors{Jiang et al.}

\begin{document}

\title{Investigations of MWISP Filaments. I. Filament Identification and Analysis Algorithms, and Source Catalog}

\author[0000-0002-3549-5029]{Yu Jiang}
\affiliation{Purple Mountain Observatory and Key Laboratory of Radio Astronomy, Chinese Academy of Sciences, 10 Yuanhua Road, Nanjing 210034, People’s Republic of China }
\affiliation{School of Astronomy and Space Science, University of Science and Technology of China, 96 Jinzhai Road, Hefei 230026, People’s Republic of China}
\affiliation{Center for Astronomy and Space Sciences, China Three Gorges University, 8 University Road, Yichang 443002, People’s Republic of China}

\author[0000-0003-3151-8964]{Xuepeng Chen}
\affiliation{Purple Mountain Observatory and Key Laboratory of Radio Astronomy, Chinese Academy of Sciences, 10 Yuanhua Road, Nanjing 210034, People’s Republic of China }
\affiliation{School of Astronomy and Space Science, University of Science and Technology of China, 96 Jinzhai Road, Hefei 230026, People’s Republic of China}

\author{Sheng Zheng}
\affiliation{Center for Astronomy and Space Sciences, China Three Gorges University, 8 University Road, Yichang 443002,  People’s Republic of China}
\affiliation{College of Science, China Three Gorges University, 8 University Road, Yichang 443002, People’s Republic of China}

\author{Zhibo Jiang}
\affiliation{Purple Mountain Observatory and Key Laboratory of Radio Astronomy, Chinese Academy of Sciences, 10 Yuanhua Road, Nanjing 210034, People’s Republic of China }

\author[0000-0003-0849-0692]{Zhiwei Chen}
\affiliation{Purple Mountain Observatory and Key Laboratory of Radio Astronomy, Chinese Academy of Sciences, 10 Yuanhua Road, Nanjing 210034, People’s Republic of China }

\author{Yao Huang}
\affiliation{Center for Astronomy and Space Sciences, China Three Gorges University, 8 University Road, Yichang 443002, People’s Republic of China}
\affiliation{College of Science, China Three Gorges University, 8 University Road, Yichang 443002, People’s Republic of China}

\author[0000-0002-0197-470X]{Yang Su}
\affiliation{Purple Mountain Observatory and Key Laboratory of Radio Astronomy, Chinese Academy of Sciences, 10 Yuanhua Road, Nanjing 210034, People’s Republic of China }

\author[0000-0002-0197-470X]{Li Sun}
\affiliation{Purple Mountain Observatory and Key Laboratory of Radio Astronomy, Chinese Academy of Sciences, 10 Yuanhua Road, Nanjing 210034, People’s Republic of China }

\author[0009-0000-3311-0159]{Jiancheng Feng}
\affiliation{Purple Mountain Observatory and Key Laboratory of Radio Astronomy, Chinese Academy of Sciences, 10 Yuanhua Road, Nanjing 210034, People’s Republic of China }
\affiliation{School of Astronomy and Space Science, University of Science and Technology of China, 96 Jinzhai Road, Hefei 230026, People’s Republic of China}

\author[0000-0003-1714-0600]{Haoran Feng}
\affiliation{Purple Mountain Observatory and Key Laboratory of Radio Astronomy, Chinese Academy of Sciences, 10 Yuanhua Road, Nanjing 210034, People’s Republic of China }
\affiliation{School of Astronomy and Space Science, University of Science and Technology of China, 96 Jinzhai Road, Hefei 230026, People’s Republic of China}

\author[0000-0001-7768-7320]{Ji Yang}
\affiliation{Purple Mountain Observatory and Key Laboratory of Radio Astronomy, Chinese Academy of Sciences, 10 Yuanhua Road, Nanjing 210034, People’s Republic of China }

%
%

%
%
%
%
%
\correspondingauthor{Xuepeng Chen}
\email{xpchen@pmo.ac.cn}

\correspondingauthor{Yu Jiang}
\email{yujiang@pmo.ac.cn}






\begin{abstract}
Filaments play a crucial role in providing the necessary environmental conditions for star formation, actively participating in the process. To facilitate the identification and analysis of filaments, we introduce DPConCFil (Directional and Positional Consistency between Clumps and Filaments), a suite of algorithms comprising one identification method and two analysis methods. The first method, the consistency-based identification approach, uses directional and positional consistency among neighboring clumps and local filament axes to identify filaments in the PPV datacube. The second method employs a graph-based skeletonization technique to extract the filament intensity skeletons. The third method, a graph-based substructuring approach, allows the decomposition of complex filaments into simpler sub-filaments. We demonstrate the effectiveness of DPConCFil by applying the identification method to the clumps detected in the Milky Way Imaging Scroll Painting (MWISP) survey dataset by \href{https://github.com/JiangYuTS/FacetClumps}{FacetClumps}, successfully identifying a batch of filaments across various scales within $10^{\circ} \leq l \leq 20^{\circ}$, $-5.25^{\circ} \leq b \leq 5.25^{\circ}$ and -200 km s$^{-1}$ $\leq v \leq$ 200 km s$^{-1}$. Subsequently, we apply the analysis methods to the identified filaments, presenting a catalog with basic parameters and conducting statistics of their galactic distribution and properties. DPConCFil is openly available on \href{https://github.com/JiangYuTS/DPConCFil}{GitHub}, accompanied by a \href{https://github.com/JiangYuTS/DPConCFil/blob/main/Manuals/DPConCFil_Manual.ipynb}{manual}.

\end{abstract} 

\keywords{radio lines: ISM - ISM: molecules, structure - stars: formation - method: data analysis - techniques: image processing - catalogs }


\section{Introduction}
Molecular cloud filaments are elongated structures that stretch across interstellar space and are composed of gas and dust. Investigating filaments is of paramount importance in unraveling the evolution of interstellar matter, the birth of stars, and the circulation of materials in the galaxies. Research on filaments has become a central topic in the study of the interstellar medium (ISM) and star formation, focusing primarily on identifying filaments and analyzing their intrinsic properties \citep[e.g.][]{Review_1}. 

Filament formation in the ISM is driven by a complex interplay of anisotropic processes, including large-scale gas motions, turbulence-induced shocks, magnetic fields, feedback, and gravitational effects \citep[e.g.][]{Filament_Formation_10,Filament_Formation_9,Arzoumanian_2013,Profile_Region_1,Filament_Formation_7,Arzoumanian_2019,Filament_Formation_8,Filament_Formation_6,Filament_Formation_5}. The formation of filaments is a dynamic process, gravitational contraction plays a key role, especially in molecular clouds, while thermal instability and shock compression contribute in different ISM phases \citep[e.g.][]{Filament_Formation_2,Filament_Formation_1,Filament_Formation_3,Filament_Formation_4}. The widespread presence of filamentary structures across different wavelengths and their continuous evolution make pinpointing a single dominant mechanism challenging \citep[see][for reviews]{Review_4,Review_1,Review_2}.

Owing to the presence of filamentary structures, material within molecular clouds can gather into filaments, giving rise to high-density regions \citep[e.g.][]{Filament_Core_0, Filament_Core_4, Filament_Core_5, Filament_Core_6, Filament_Core_7}. The formation of prestellar dense cores is primarily driven by cloud fragmentation among high-density regions of filaments \citep[e.g.][]{Filament_Core_1, Filament_Core_2, Filament_Core_3, Core_Formation_Filaments_1,Filament_Core_9,Review_2,Filament_Core_10}. Molecular clumps are localized high-density regions within molecular clouds. Under the influence of gravity, gas and dust gradually condense into denser regions within these clumps, evolving into multiple cores over time \citep[e.g.][]{Review_3}. These clumps are considered the fundamental units for star formation, as they supply the essential material and conditions for the birth of protostars \citep[e.g.][]{FIVe}. Clumps serve as a critical intermediary in the process of both low-mass and high-mass star formation from filaments. Therefore, utilizing the properties of clumps as a reference to further identify the filamentary structure is a possible scheme. 

The Milky Way Imaging Scroll Painting survey \citep[MWISP,][]{CO_Survey_MWISP} is an extensive CO survey of the Galactic plane conducted by the Purple Mountain Observatory (PMO). This survey offers advantages such as wide sky coverage, multiple spectral lines, and high sensitivity. FacetClumps is an innovative algorithm proposed by \cite{FacetClumps} for extracting and analyzing clump in molecular clouds. Numerous experiments have confirmed that FacetClumps exhibits superior performance in detecting clumps. The Minimum Spanning Tree (MST) algorithm \citep{MST} first utilizes the positional information of the Bolocam Galactic Plane Survey \citep{BGPS_Clumps} clumps to identify coherent filaments in the position-position-velocity (PPV) space. This algorithm has also been successfully applied to other clump catalogs \citep{MST_Apply_1,MST_Apply_2,MST_Apply_3}. The availability of new observational data and clump detection techniques has greatly encouraged us to conduct a more extensive investigation of filaments. 

This paper firstly introduces DPConCFil, a collection of new algorithms designed for the identification and analysis of filaments. DPConCFil identifies filaments exhibiting spatial and velocity connectivity in the PPV space by leveraging the consistency between the direction and position of clumps and those of the local filament axis \citep[e.g.,][]{Filament_Inertia_Matrix,Core_Formation_Filaments_5,Core_Formation_Filaments_2,Core_Formation_Filaments_6}. Clump properties, such as directions, positions, and regional masks, are extracted from MWISP using FacetClumps. The filament region is inherited from the associated clumps, and DPConCFil then applies a graph-theoretical technique to determine the intensity skeleton of the filaments within this region. Utilizing graph theory and a custom recursive function, DPConCFil is capable of identifying substructures within complex filaments. Finally, DPConCFil is applied to the $^{13}$CO emission data cube from MWISP to generate a catalog of filaments. 

The paper is organized as follows: Section \ref{Data} describes the data and the algorithm used for clump extraction. Section \ref{DPConCFil} combines text and schematic diagrams to illustrate the processes and details of DPConCFil. In Section \ref{ApplicationAnalysis}, we present the application of DPConCFil to MWISP and perform a concise statistical analysis of the catalog. Finally, a summary of our work is presented in Section \ref{Summary}. 

\section{Data} \label{Data}
\subsection{The MWISP Survey and $^{13}$CO Emission}\label{Data_CO}
MWISP\footnote{\href{http://english.dlh.pmo.cas.cn/ic/in/}{http://english.dlh.pmo.cas.cn/ic/in/}} is an unbiased Galactic plane CO survey in the northern sky, conducted with the PMO 13.7 meter millimeter-wavelength telescope at Delingha, China, targeting simultaneous observations of the $^{12}$CO, $^{13}$CO, and C$^{18}$O ($J = 1-0$)  emission lines. The mapping area of MWISP spans a range of Galactic longitudes from $9^{\circ}.75$ to $230^{\circ}.25$ and Galactic latitudes from $-5^{\circ}.25$ to $5^{\circ}.25$ (Phase I). The half-power beamwidth (HPBW) for $^{12}$CO emission in MWISP is \textasciitilde52$\arcsec$, with a grid spacing of 30$\arcsec$. The spectral resolution achieved for $^{12}$CO is approximately 0.16 km s$^{-1}$, with an average RMS noise about 0.5 K per channel. Similarly, the HPBW of $^{13}$CO and C$^{18}$O emission in MWISP is \textasciitilde55$\arcsec$, with a grid spacing of 30$\arcsec$. The spectral resolutions obtained for $^{13}$CO and C$^{18}$O are approximately 0.17 km s$^{-1}$, with an average RMS noise about 0.3 K per channel. 

\begin{figure}
\centering
\centerline{\includegraphics[width=4.2in]{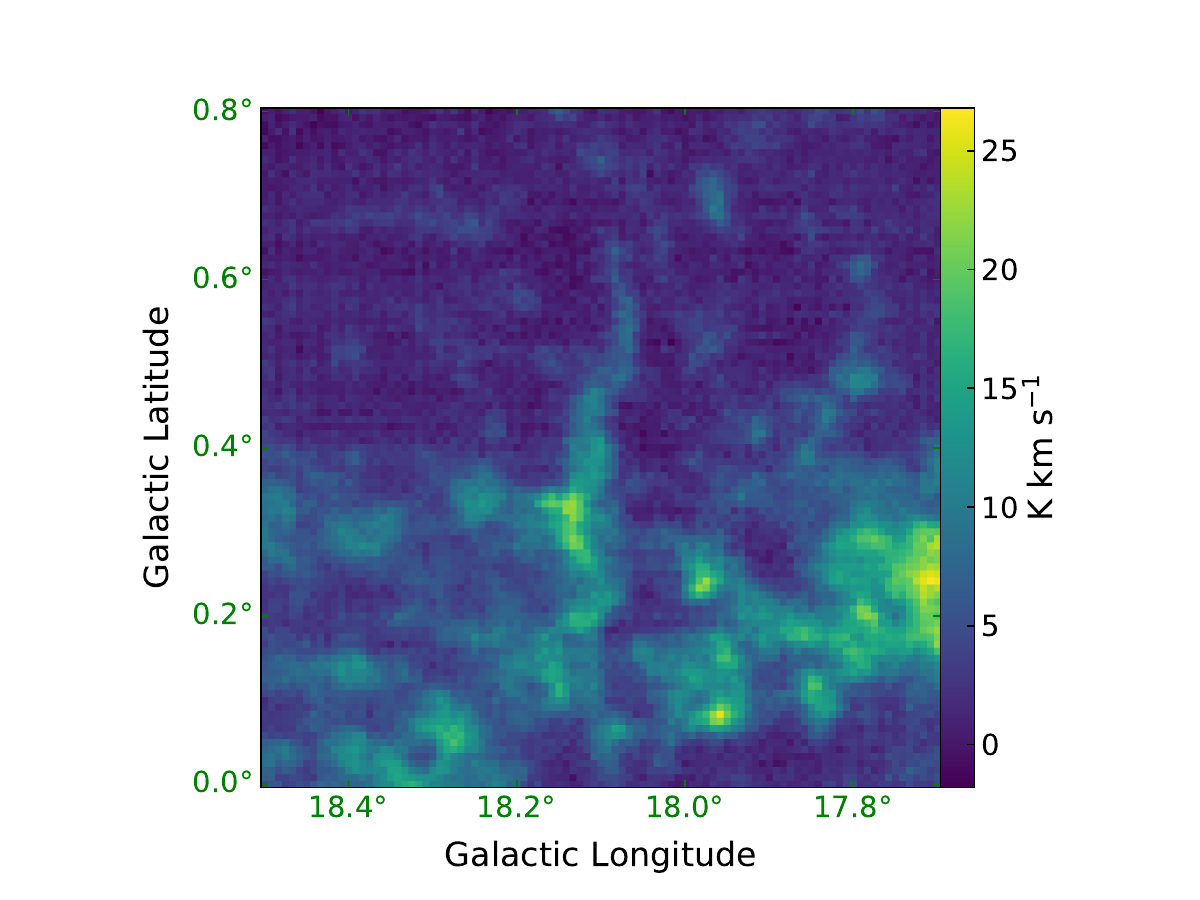}}
\caption{An example data to illustrate DPConCFil. The data is the $^{13}$CO emission of MWISP within $17.7^{\circ} \leq l \leq 18.5^{\circ}$, $0^{\circ} \leq b \leq 0.8^{\circ}$ and 5 km s$^{-1}$ $\leq v \leq$ 30 km s$^{-1}$.  }
\label{Example_Data}
\end{figure}

The $^{12}$CO emission is diffuse, while C$^{18}$O emission traces regions of higher density. Consequently, this work concentrates on identifying filaments using $^{13}$CO emission, considering both density and completeness. To illustrate our filament identification and analysis algorithms DPConCFil, we select the $^{13}$CO emission from MWISP within $17.7^{\circ} \leq l \leq 18.5^{\circ}$, $0^{\circ} \leq b \leq 0.8^{\circ}$ and 5 km s$^{-1}$ $\leq v \leq$ 30 km s$^{-1}$ as an example data, as visualized in Figure \ref{Example_Data}. Moreover, we choose the $^{13}$CO emission from MWISP within $10^{\circ} \leq l \leq 20^{\circ}$, $-5.25^{\circ} \leq b \leq 5.25^{\circ}$ and -200 km s$^{-1}$ $\leq v \leq$ 200 km s$^{-1}$ as an application data. This particular region is known to be one of the most complex regions in MWISP. We will present the identification results of this region in Section \ref{ApplicationAnalysis}. 

\subsection{The Clump Detection Algorithm and Clumps Information}\label{Data_Clump}
The FacetClumps algorithm integrates the facet model and the extremum determination theorem of multivariate functions to automatically locate the center of clumps within the preprocessed signal regions. Subsequently, it employs a connectivity-based minimum distance clustering method to merge local regions segmented by local gradients, thereby identifying the region corresponding to each clump. FacetClumps has proven to be highly effective in extracting clumps from MWISP data \citep{FacetClumps}. It demonstrates superior performance compared to other clump detection algorithms, particularly in terms of location and region segmentation accuracy, which are crucial for subsequent filament identification and analysis. Furthermore, through a comparison among clump detection algorithms of Dendrogram \citep{Dendrogram}, FellWalker \citep{FellWalker}, ConBased \citep{ConBased}, and FacetClumps, we have discovered that the clumps obtained from FacetClumps significantly enhance the performance of DPConCFil in filament identification. Hence, we have opted to incorporate FacetClumps for clump detection. 

\begin{figure}
\centering
\centerline{\includegraphics[width=3in]{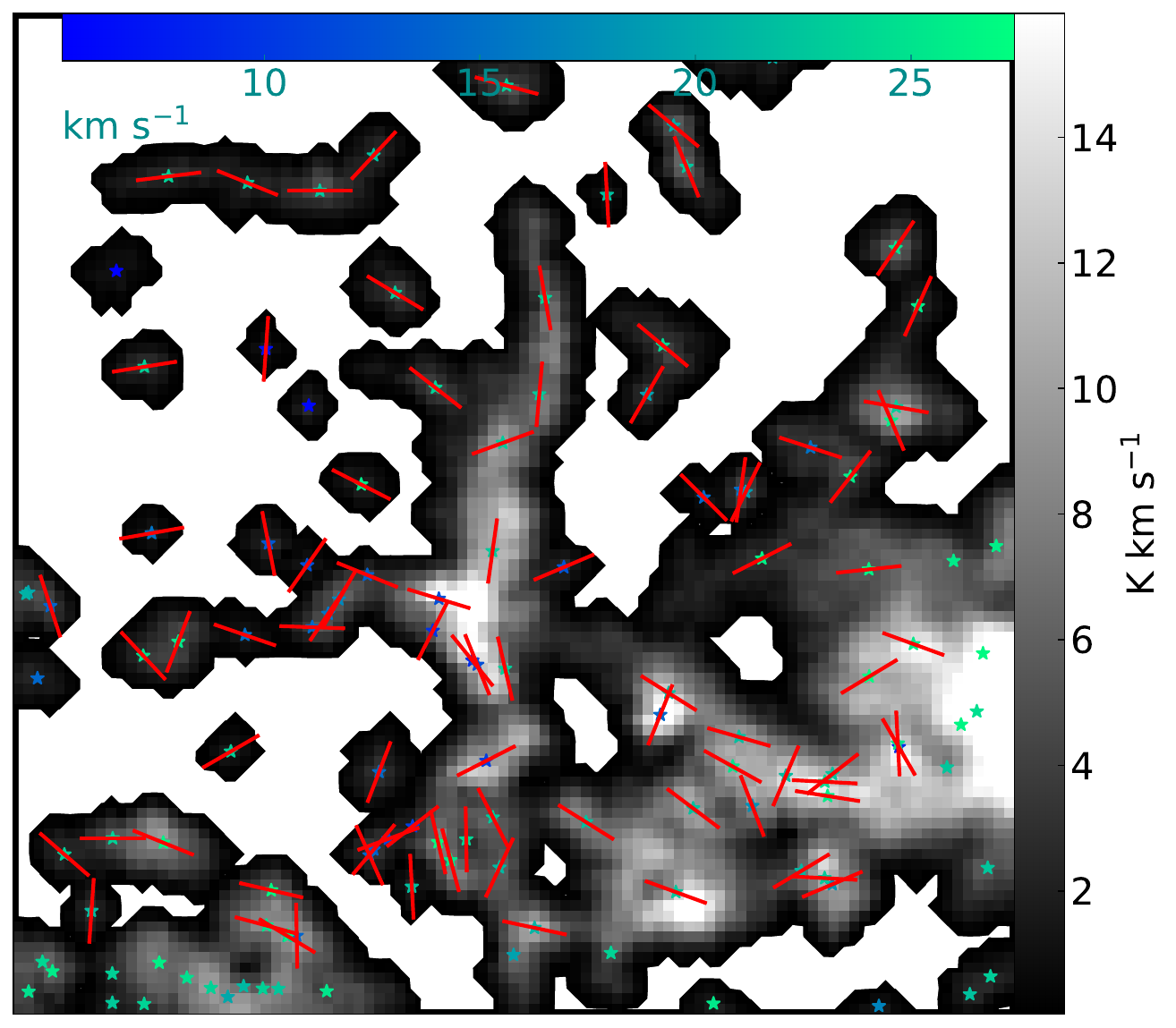}}
\caption{The direction and position of clumps. The total number of clumps is 126, with 88 of them not touching the edge. The asterisks denote the spatial positions of the clumps, and the different colors of the asterisks denote different velocity positions. The red lines illustrate the direction of the principal axis of the clumps at the untouched edges. }
\label{Clumps_Infor}
\end{figure}

Regarding example data, we have detected a total of 126 clumps, out of which 88 do not touch the edge. As for the application data, a larger number of 11,812 clumps have been detected. The direction of the clumps in FacetClumps is obtained by diagonalizing the moment of inertia matrix \citep{Angle}. To enhance the accuracy of both the direction and position of the clumps, a single Gaussian fitting is applied to each clump using its integrated intensity map, thereby improving the performance of DPConCFil. Figure \ref{Clumps_Infor} shows the spatial and velocity positions of the clumps of example data (denoted by different colored asterisks) as well as the directions of their principal axes (represented by red lines) for those clumps that do not touch the edges. The parameters of FacetClumps employed in this study can be found in Table \ref{FacetClumps parameters} of Appendix \ref{Parameters_FacetClumps}. 

\section{The Filament Identification and Analysis Algorithms}\label{DPConCFil}
DPConCFil comprises a new filament identification method, a new filament skeleton analysis method, and a new filament substructure analysis method. The first sub-method is the consistency-based identification method, which first utilizes directional consistency among neighboring clumps to identify local filament structures and axes. It then connects more clumps on filaments based on positional consistency between neighboring clumps and local filament axes. Finally, it merges the records containing the same clumps to obtain the clump IDs for all filaments, thus obtaining the regions of the filaments. Neighboring clumps refer to two clumps whose regions meet the morphological criterion of connectivity \citep{FacetClumps}. The second sub-method is the graph-based skeletonization method. The weights of the graph and tree is developed by considering the spatial distance and integrated intensity between neighboring points. The third sub-method is the graph-based sub-structuring method. In this case, the weights of the graph and tree is developed by comprehensively considering the distance of spatial direction and velocity channel, and mean intensity between neighboring clumps. Then, a recursive function is devised to continuously solve for the longest shortest paths of the updated tree, thereby decomposing arbitrarily complex filaments into sub-filaments. 

The sub-methods can be utilized as stand-alone applications. We have shared the code on Github\footnote{\href{https://github.com/JiangYuTS/DPConCFil}{https://github.com/JiangYuTS/DPConCFil}} under a permissive MIT license, made it publicly accessible as a Python
package called DPConCFil\footnote{\href{https://pypi.org/project/DPConCFil}{https://pypi.org/project/DPConCFil}}, and deposited the latest version to Zenodo \citep{DPConCFilZenodo}. For a more comprehensive manual and illustrative examples on utilizing DPConCFil, detailed instructions are readily available on the Github repository. We will continue to develop DPConCFil and warmly welcome community contributions for its optimization. 

\subsection{The Consistency-based Identification Method}\label{Identification}
Unlike dealing with data of the position-position (PP) space and the position-position-position (PPP) space, identifying filaments in molecular spectral lines is a complex problem in the PPV space. It requires consideration of the different physical meanings of spatial directions and velocity channels. The consistency between clumps and filaments is a widely observed phenomenon, and we use this property to identify filaments in the PPV space.

\subsubsection{The Identification Criteria of Filament}\label{Definition}
The definition methods for filaments can be mainly categorized into two types: visual-based definition and algorithm-based\footnote{For detailed descriptions of the algorithms, please refer to the review in \cite{Review_1}, Appendix \ref{Comparison}, and the papers.} definition \citep[e.g.][]{DisPerse,Getfilaments,FIVe,HessianMatrix_1,HessianMatrix_2,FilFinder,MST,CRISPy,FilDReaMS}. Additionally, the characteristics of filaments can vary depending on the observational tracers employed \citep{Review_1}. Generally, a filament is an elongated, continuous, and overdense structure \citep{MapPPVPPP,CRISPy,Velocity_Coheren_1}. 

When identifying elongated features is conducted on an integrated map of PPV data, such as the GetFilaments \citep{Getfilaments}, Hessian matrix \citep{HessianMatrix_1,HessianMatrix_2}, FilFinder \citep{FilFinder}, and FilDReaMS \citep{FilDReaMS} algorithms, the integration effect has the potential to obscure regions where elongated features are only visible in a limited number of velocity channels. Additionally, structures exhibiting discernible distinctions in PPV space may overlap, ultimately presenting an elongated structure overall \citep{Confusion_PPV}. 

When identifying elongated features is directly conducted on PPV data, such as the DisPerSE \citep{DisPerse} algorithm, one encounters challenges in reconciling the physical length properties between spatial direction and velocity channel, and there is currently no unified method for balancing the physical properties at these two scales \citep{Confusion_PPV_1,MapPPVPPP}. In particular, with high velocity resolution data, there is a potential decrease in reliability during the identification process. This occurs because the stretched morphology along the velocity channel may gain prominence, consequently diminishing the spatial features. However, it is the elongated features in the spatial direction that holds greater significance. 

Although the filaments identified in the PPV space are not closely associated with the filament identified in the PPP space, their projected morphologies in the PP space are largely similar \citep{MapPPVPPP}. It has been observed that dense clumps can indeed form along the filament axis \citep[e.g.][]{Core_Formation_Filaments_1,Filament_Core_2,Core_Formation_Filaments_5,Core_Formation_Filaments_2,Core_Formation_Filaments_3,Core_Formation_Filaments_4,Core_Formation_Filaments_6}. Moreover, MHD and hydrodynamical simulations have demonstrated a tendency for alignment between the filament axis and the direction of stretching \citep{Filament_Inertia_Matrix}.  The connecting lines between neighboring clump positions can represent the local axis of filaments, while the direction of the clumps can indicate the direction of local stretching. 

Taking into account the aforementioned factors, in this work, the identification criteria or the definition for a filament are as follows: 

\begin{enumerate}
\itemsep=2pt
\item In the PP space, a filament is a structure characterized by elongation. 
\item In the PPV space, a filament is a structure characterized by both spatial continuity and velocity continuity. 
\item A filament consists of at least two neighboring clumps, wherein the directions of these two neighboring clumps exhibit consistency with the direction of the local axis of filaments. If clumps within a filament do not exhibit directional consistency, their positions should exhibit consistency with the local axis established by the neighboring clumps that display directional consistency. 
\end{enumerate}

\begin{figure}
\centering
\centerline{\includegraphics[width=4in]{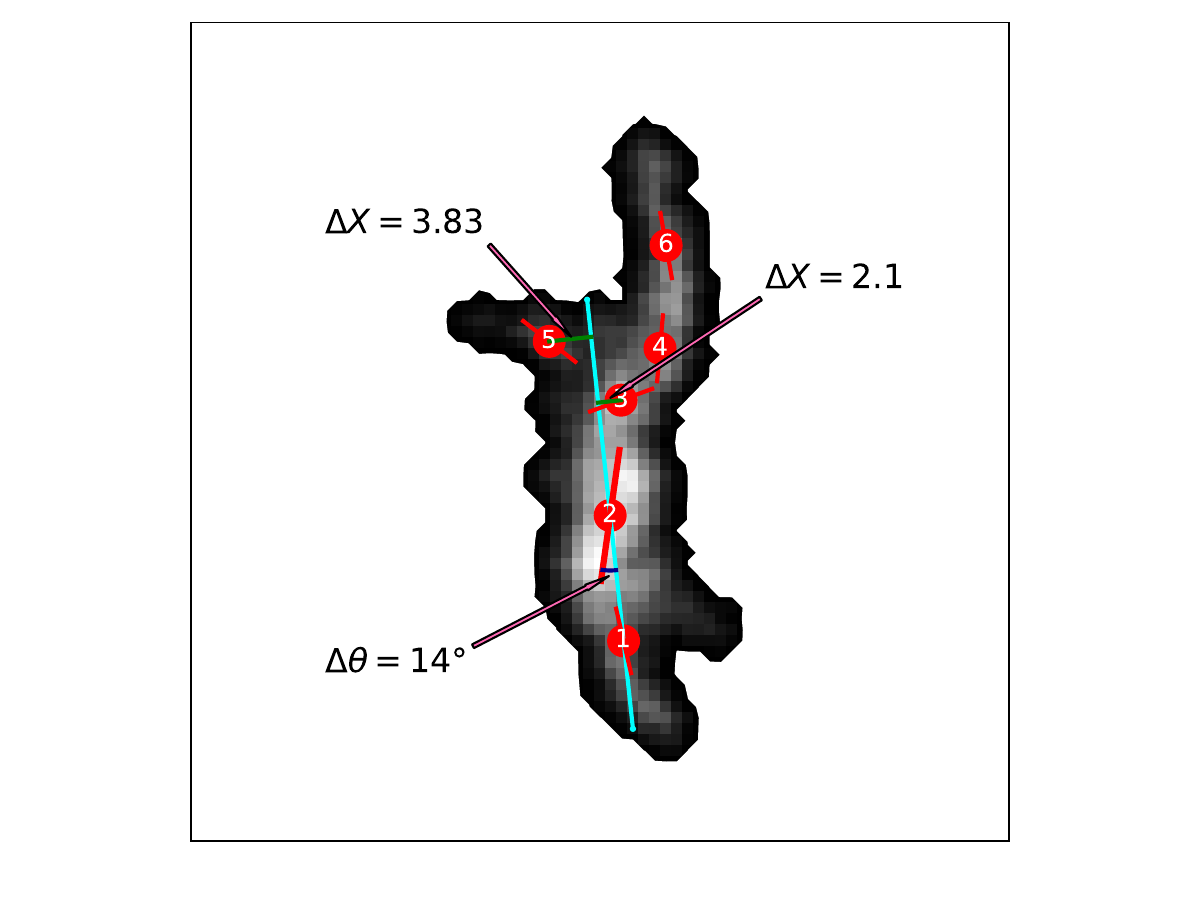}}
\caption{An example diagram illustrating the identification process. The background features a velocity-integrated intensity map of a filament, as identified in Figure \ref{Example_Data}. Red circles indicate the positions of the clumps, numbered accordingly, while red lines represent the direction of each clump's principal axis. Clumps 1 and 2 are neighboring clumps, with the cyan line connecting their positions. The blue curve shows the angle between the direction of clump 2 and the cyan line. Notably, clumps 1 and 2 demonstrate directional consistency with the local filament axis represented by the cyan line. Clumps 2 and 3, as well as clumps 3 and 5, are also neighboring pairs. The green lines are perpendicular to the cyan line for clumps 3 and 5. It can be observed that clumps 3 and 5 have positional consistency with the local filament axis.}
\label{Filament_Detect_Description}
\end{figure}

\begin{figure*}
\centering
\vspace{0cm}
\begin{minipage}[t]{0.3\textwidth}
    \centering
    \centerline{\includegraphics[width=3.2in]{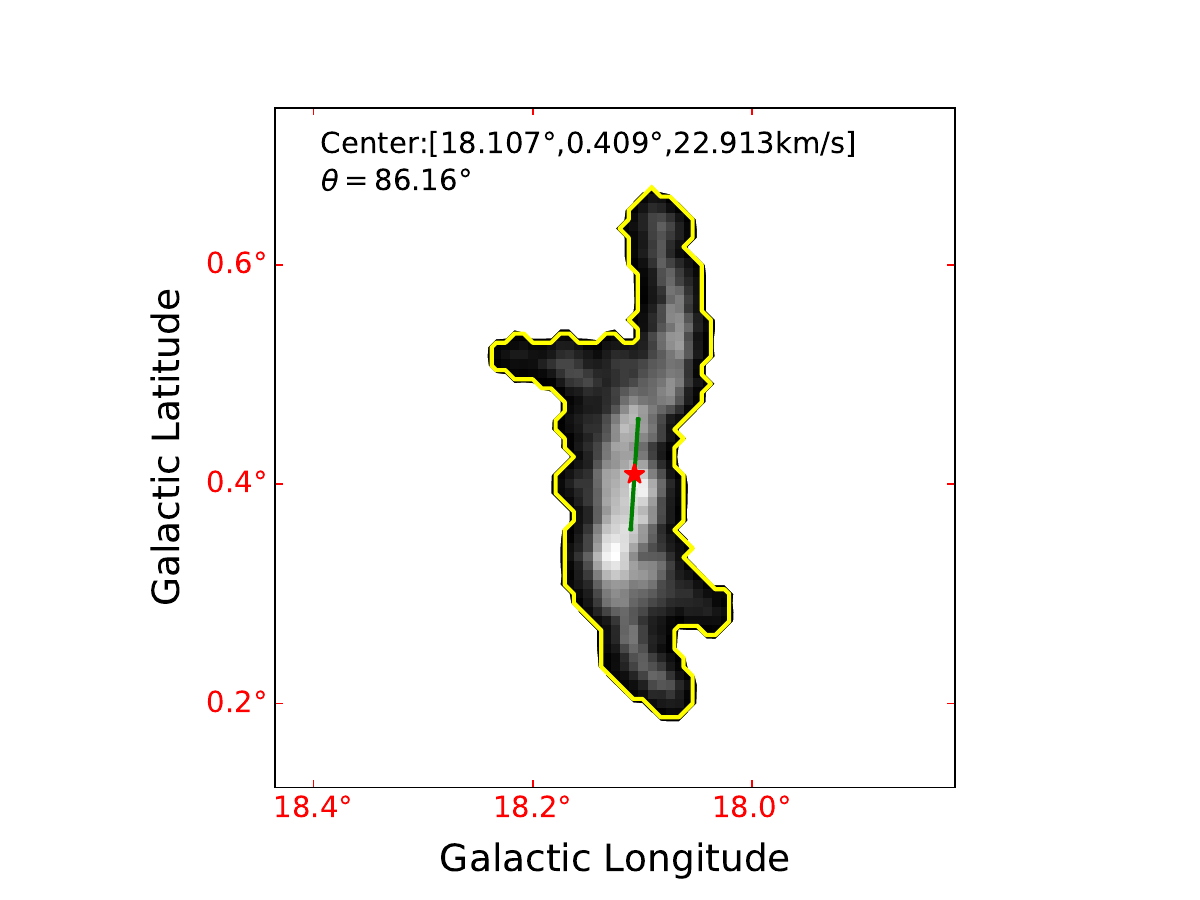}}
\end{minipage}%
\begin{minipage}[t]{0.3\textwidth}
    \centering
    \centerline{\includegraphics[width=3.2in]{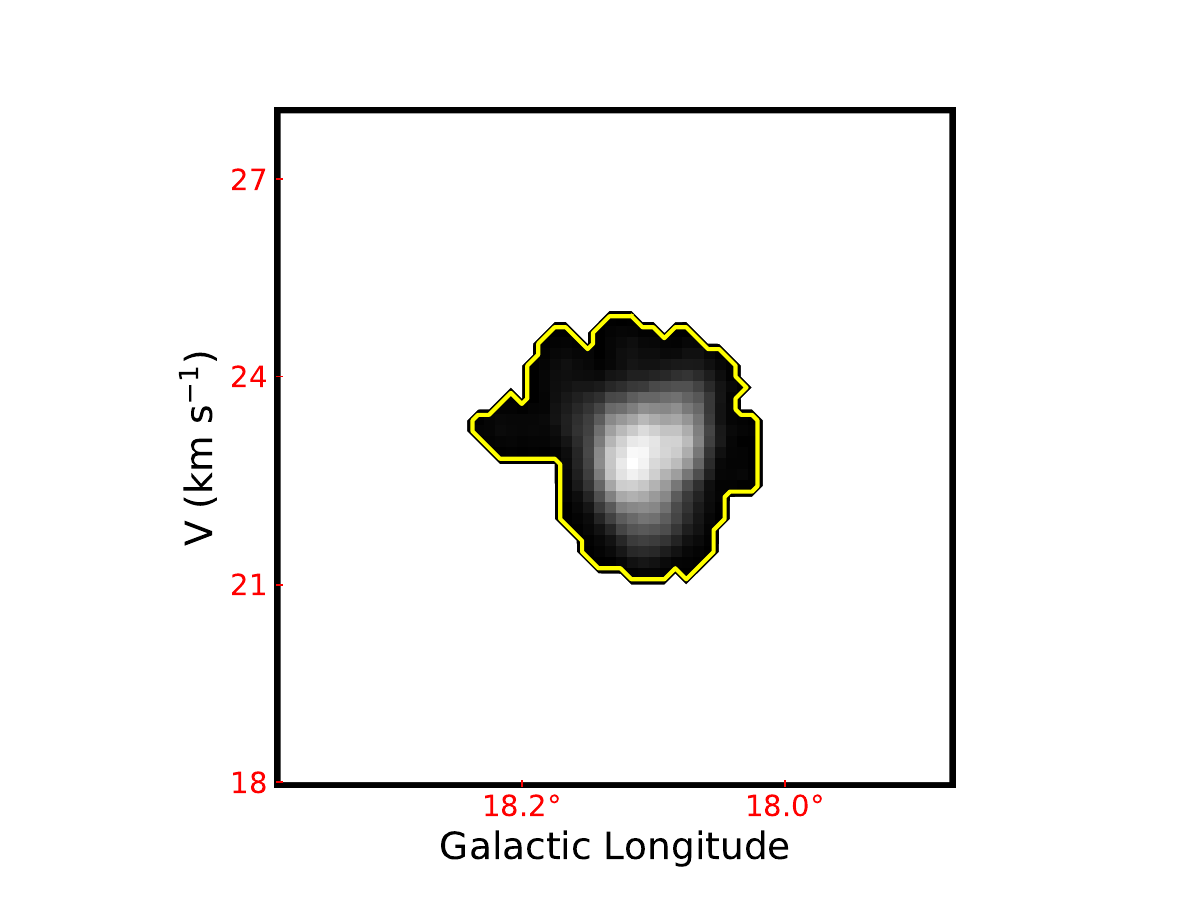}}
\end{minipage}%
\begin{minipage}[t]{0.3\textwidth}
    \centering
    \centerline{\includegraphics[width=3.2in]{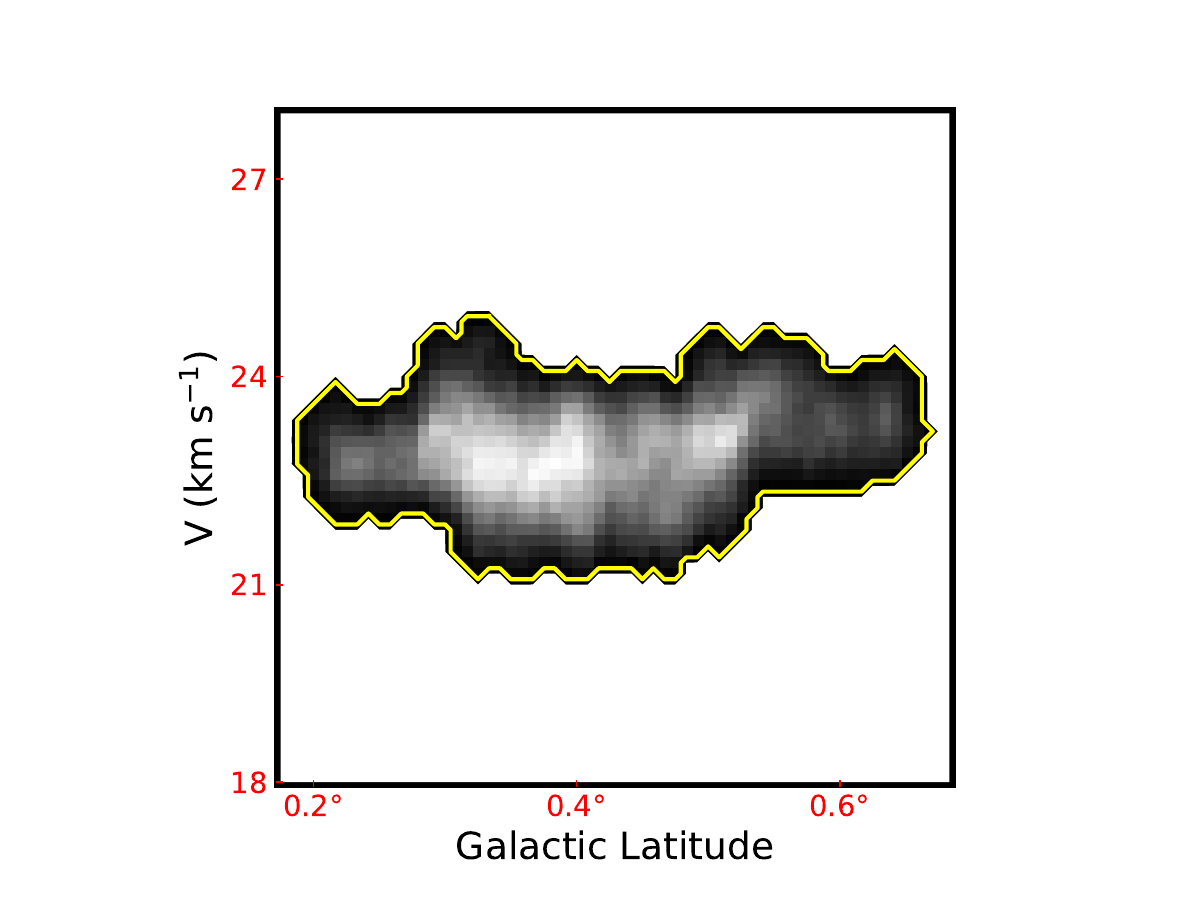}}
\end{minipage}%
\caption{Integrated intensity maps of a filament in different directions. Left panel: velocity-integrated intensity map. The central position of the filament, denoted by a red asterisk, has coordinates $(l,b,v)$ = (18.11$^{\circ}$, 0.41$^{\circ}$, 22.91 km s$^{-1}$), and its principal axis derived through principal component analysis is represented by a green line with an angle of $86.16^{\circ}$. The yellow contours delineate the boundaries of the filament region. Middle panel: latitude-integrated intensity map. Right panel: longitude-integrated intensity map. }
\label{Filament_Example_Item}
\end{figure*}

\subsubsection{Estimate Consistency between Clumps and Filament}\label{EvaluateConsistency}
The identification of filaments is based on the information of molecular clumps, which includes their spatial directions, spatial positions, masks, and the connectivity among clumps. The objective is to accurately recognize the filaments as defined here from PPV astronomical observational data. 

Figure \ref{Filament_Detect_Description} provides an illustrative example of how to assess the consistencies. To begin, start with any clump A (e.g., clump 1 in Figure \ref{Filament_Detect_Description}) and search for its neighboring clumps set S1. For each neighboring clump B in S1 (e.g., clump 2), connect the spatial positions of clump A and B to form a line $L_{AB}$ (e.g., the cyan line in Figure \ref{Filament_Detect_Description}). Next, calculate the angles $\Delta \theta_A$ and $\Delta \theta_B$ between the directions of clump A and clump B with respect to this line. If both $\Delta \theta_A$ and $\Delta \theta_B$ are smaller than the specified angle tolerance (parameter $TolAngle$), then these two neighboring clumps are considered to have directional consistency. Record all clumps that have directional consistency with clump A. 

When clump B exhibits directional consistency with clump A, search for the neighboring clumps set S2 of clump B, and merge it with S1 to obtain the neighboring clumps set S31 for the structure composed of clumps A and B. For each clump C1 (e.g., clump 3) in S31, calculate the intersection point of the perpendicular line passing through clump C1 and the line $L_{AB}$. If the distance between the intersection point and clump C1 is smaller than the specified distance tolerance (parameter $TolDistance$), then clump C1 is considered to have positional consistency with clumps A and B. If clump C1 has positional consistency and the intersection point is outside the line segment $LS_{AB}$, then search for the neighboring clumps set S32 of clump C1. For each clump C2 in S32 (e.g., clump 5), determine whether C2 has positional consistency with $LS_{AB}$. Record all clumps that have positional consistency with clumps A and B. The clumps situated at the edges are not considered during the evaluation of directional consistency, but they are included in the assessment of distance consistency. 

After performing the consistency assessment described above on all clumps, update the clump record list, and merge lists that contain the same clumps to obtain new ID lists representing filaments. As shown in Figure \ref{Filament_Detect_Description}, clumps numbered 1 and 2, as well as clumps numbered 4 and 6, exhibit directional consistency. Clumps numbered 3 and 5 have positional consistency with clumps 1 and 2, while clump 3 has positional consistency with clumps 4 and 6. Since clump 3 is the common clump, the two sublists will be merged together to form a filament. Whether it is ultimately confirmed as a filament requires a judgment based on the aspect ratio, as described in Section \ref{AspectRatio}.

The task of regional segmentation for filaments is challenging, and the choice of different region delineations can largely impact the radius profile of the filament and other physical properties \citep{Profile_Region_1, Profile_Region_2,C18O_1}. Figure \ref{Filament_Example_Item} shows the integrated maps of a filament identified using the consistency-based method in different integration directions. The yellow contour outlines the region of the filament, which is inherited from the regional information of the associated clumps. Due to the relatively accurate segmentation of the clump regions, the filament regions obtained through this method have higher precision compared to other commonly used approaches, such as extracting regions within a specific width along the filament skeletons.

\begin{figure*}
\centering
\vspace{0cm}
\begin{minipage}[t]{0.3\textwidth}
    \centering
    \centerline{\includegraphics[width=2.5in]{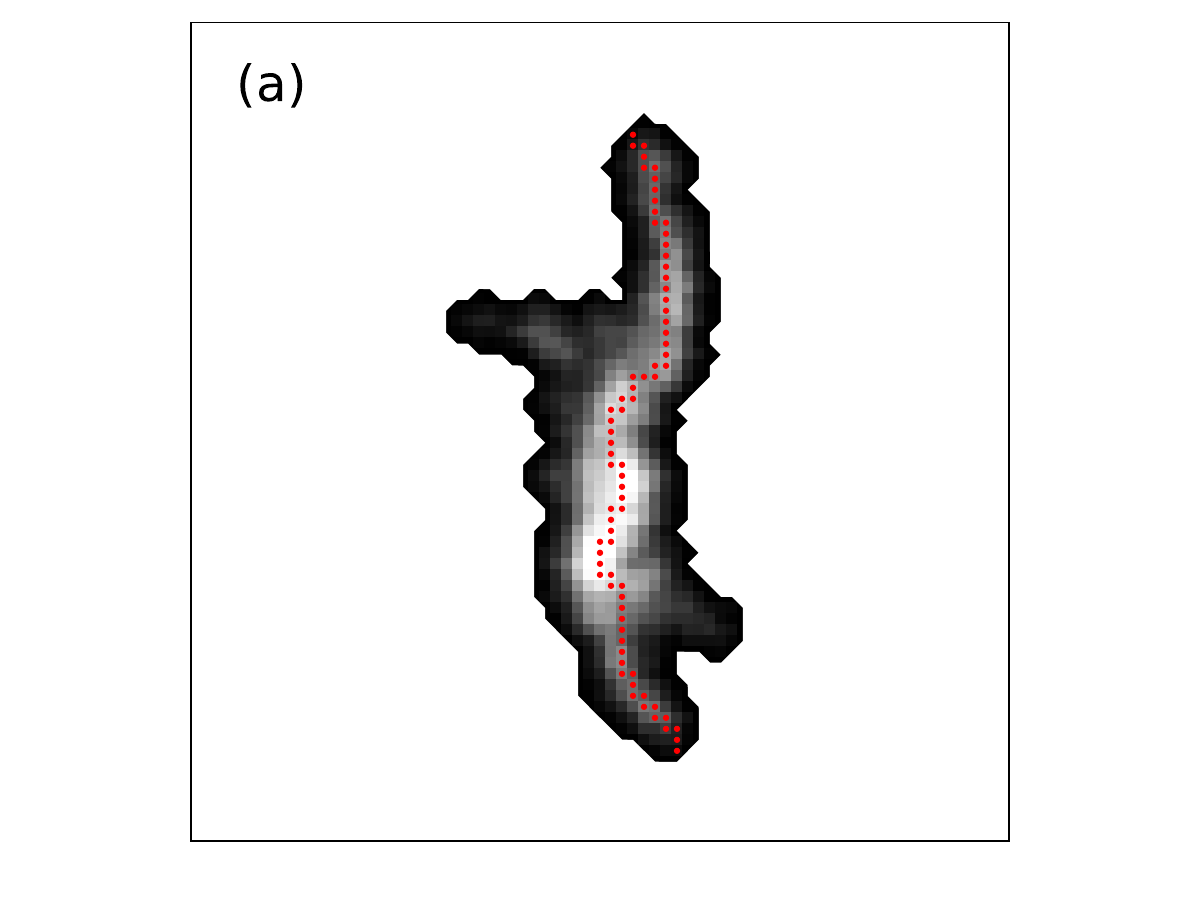}}
\end{minipage}%
\begin{minipage}[t]{0.3\textwidth}
    \centering
    \centerline{\includegraphics[width=2.5in]{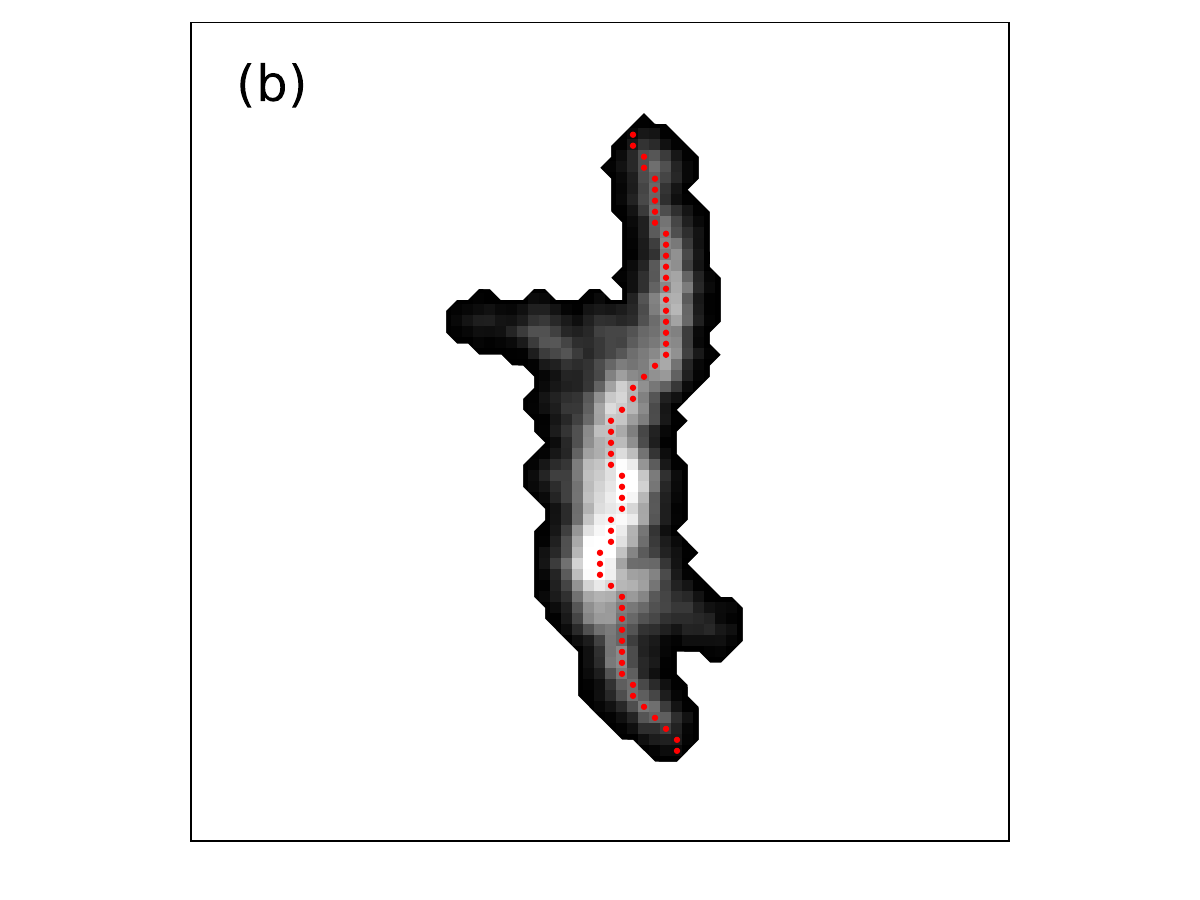}}
\end{minipage}%
\begin{minipage}[t]{0.3\textwidth}
    \centering
    \centerline{\includegraphics[width=2.5in]{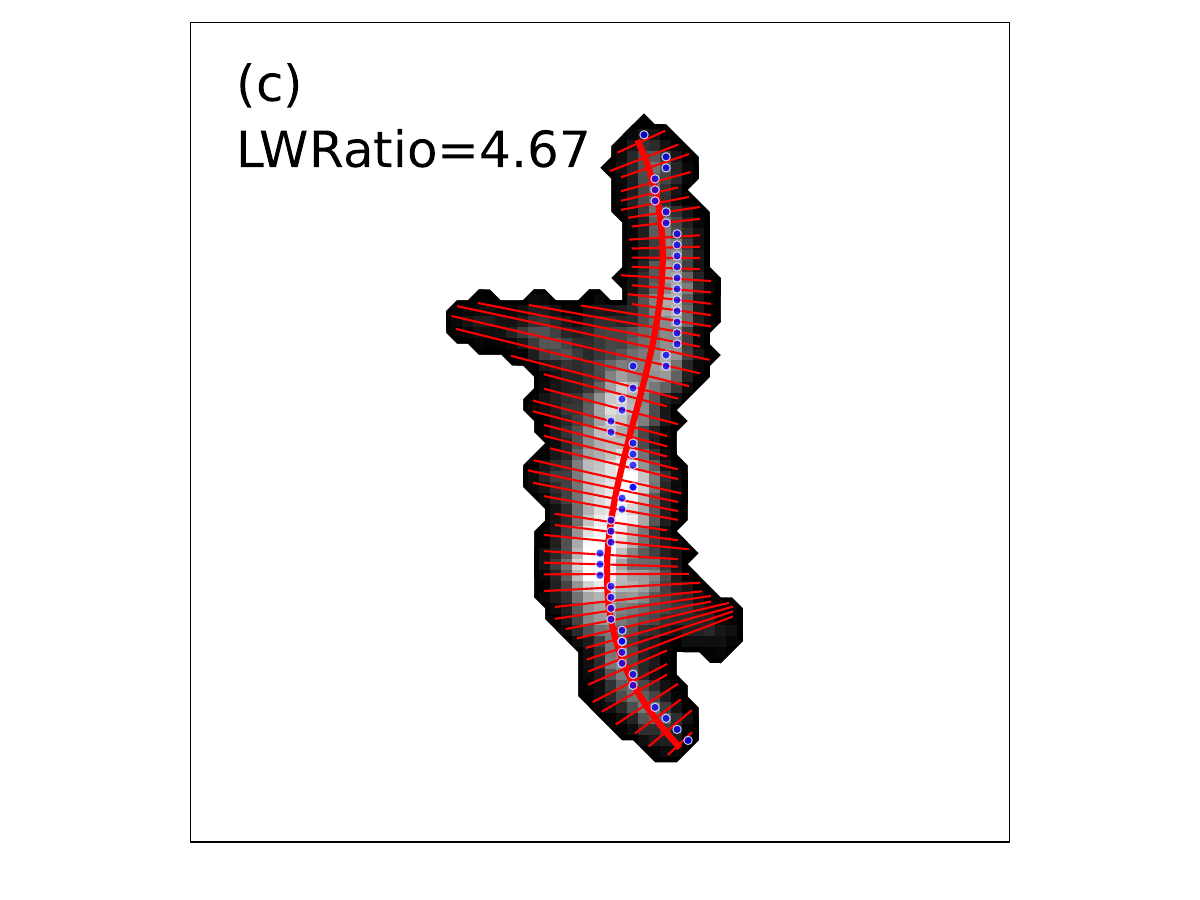}}
\end{minipage}%
\caption{Intensity skeleton analysis. (a) The initial intensity skeleton extracted using the graph-based skeletonization method; (b) the final thinned intensity skeleton; (c) the fitted intensity skeleton and profiles. The thicker red curve represents the smoothed intensity skeleton obtained through B-spline fitting, while the thinner red straight lines represent the profiles. The blue scatter points denote the peak intensity pixels on the profiles. The aspect ratio is 4.67. }
\label{Filament_Intensity_Point}
\end{figure*}

\subsection{The Graph-based Skeletonization Method}\label{Intensity_Spine}
The mathematical morphological skeleton extraction methods, such as the medial axis-based skeletonization method \citep{FilFinder}, revolve around processing binary images to extract their medial axes, while ignoring the intensity values within the image. However, filaments inherently contain intensity information and exhibit partial asymmetry \citep[e.g.][]{Filament_Formation_9}. Hence, we have developed a graph-based\footnote{\href{https://networkx.org/}{https://networkx.org/}} skeletonization method to extract the intensity skeleton of the filament. To generate the filament profiles, we employ spline interpolation to fit the skeleton and derive its B-spline curve. Afterwards, by leveraging the first derivatives of the knots on the B-spline curve, we are equipped to construct the profiles of the filament. 

The geometric minimum spanning tree (GMST) is a fundamental concept in graph theory extensively employed in various fields of astronomy. At its core, GMST revolves around the selection of edges with the lowest weights to construct a tree that spans all vertices. This process guarantees that the resulting tree connects all vertices with the minimum weight possible, rendering it an optimal solution for specific graph-related problems. By avoiding cycles, the GMST ensures the absence of redundant edges, facilitating the creation of an efficient and acyclic structure. 

\subsubsection{The Intensity Skeleton}\label{Skeleton}
Due to the asymmetry in the intensity of filaments, the intensity skeleton should be a crucial element in analyzing filament properties such as length, characteristic radius, and line mass. 

To obtain the intensity skeleton of a filament, we propose a graph-based skeletonization method. All points within the spatial mask of a filament are considered as nodes in a graph. Each point is connected only to its neighboring points by edges, with the weight of each edge determined by a specific formula (Equation \ref{WeightSK_1}). This procedure creates a weighted undirected graph, denoted as G1. The GMST algorithm is then utilized on G1 to derive the tree T1, which includes all points within the mask. The weight of tree T1 is subsequently updated using Equation (\ref{WeightSK_2}). Leaf nodes situated on the boundary of the mask within T1 are searched. Next, we calculate the shortest paths between every pair of these leaf nodes and select the path with the greatest sum weights among all shortest paths. This yields the initial version of the desired intensity skeleton, as shown in Figure \ref{Filament_Intensity_Point}(a). 

\begin{equation}\label{WeightSK_1}
\begin{split}
WeightSK_{Gij}=\frac{DistLB_{ij}}{I_i+I_j}
\end{split}
\end{equation}

\begin{equation}\label{WeightSK_2}
\begin{split}
WeightSK_{Tij}=\frac{I_i+I_j}{DistLB_{ij}}
\end{split}
\end{equation}

\noindent where $WeightSK_{Gij}$ represents the wight of two neighboring coordinates within a filament region in the PP space in the graph G2, $WeightSK_{Tij}$ represents the updated weight of two neighboring coordinates in the tree T1. $DistLB_{ij}$ is the Euclidean distance between these neighboring coordinates, where the possible values are either 1 or $\sqrt{2}$. $I_i$ and $I_j$ are the integrated intensity of coordinate $i$ and coordinate $j$, respectively. 

The weights $WeightSK$ we designed encourage the desired path to predominantly follow neighboring points at a distance of 1, which promotes a straighter skeleton with higher intensity. However, this design compromises the slenderness of the skeleton, leading to less precise length calculations. To address this, we thin the initial skeleton by retaining the three most distant skeleton points within each 3×3 neighborhood. The final intensity skeleton, following the thinning process, is depicted in Figure \ref{Filament_Intensity_Point}(b). 

\subsubsection{The Profile and Aspect Ratio}\label{AspectRatio}
The coordinate set of the intensity skeleton is ordered, arranged from one end to the other, ensuring that adjacent points in space are also adjacent in the coordinate set. This orderly arrangement allows these coordinates to serve as control points for the B-Spline curve. By utilizing cubic basis functions, the control points can be fitted, generating a smooth and continuous representation of the skeleton, depicted as the thicker red curve in Figure \ref{Filament_Intensity_Point}(c). 

To acquire the filament profiles, the B-Spline curve can be uniformly sampled at pixel intervals, generating a series of sampling points. Subsequently, by calculating the first derivative for each sampling point, a line perpendicular to that particular point can be determined. The segments of these lines that fall within the filament mask range correspond to the profiles, represented by the thinner red straight lines in Figure \ref{Filament_Intensity_Point}(c). For detailed instructions on building filament profiles, please refer to the RadFil \citep{RadFil}. 

The aspect ratio of a filament is calculated as the length divided by the width. The length is determined by the number of coordinates along the intensity skeleton, while the width is derived from the median length of the profiles. Structures with smaller aspect ratios (parameter $LWRatio$) are deemed nonfilamentary and will be filtered out.

\begin{figure}
\centering
\centerline{\includegraphics[width=4in]{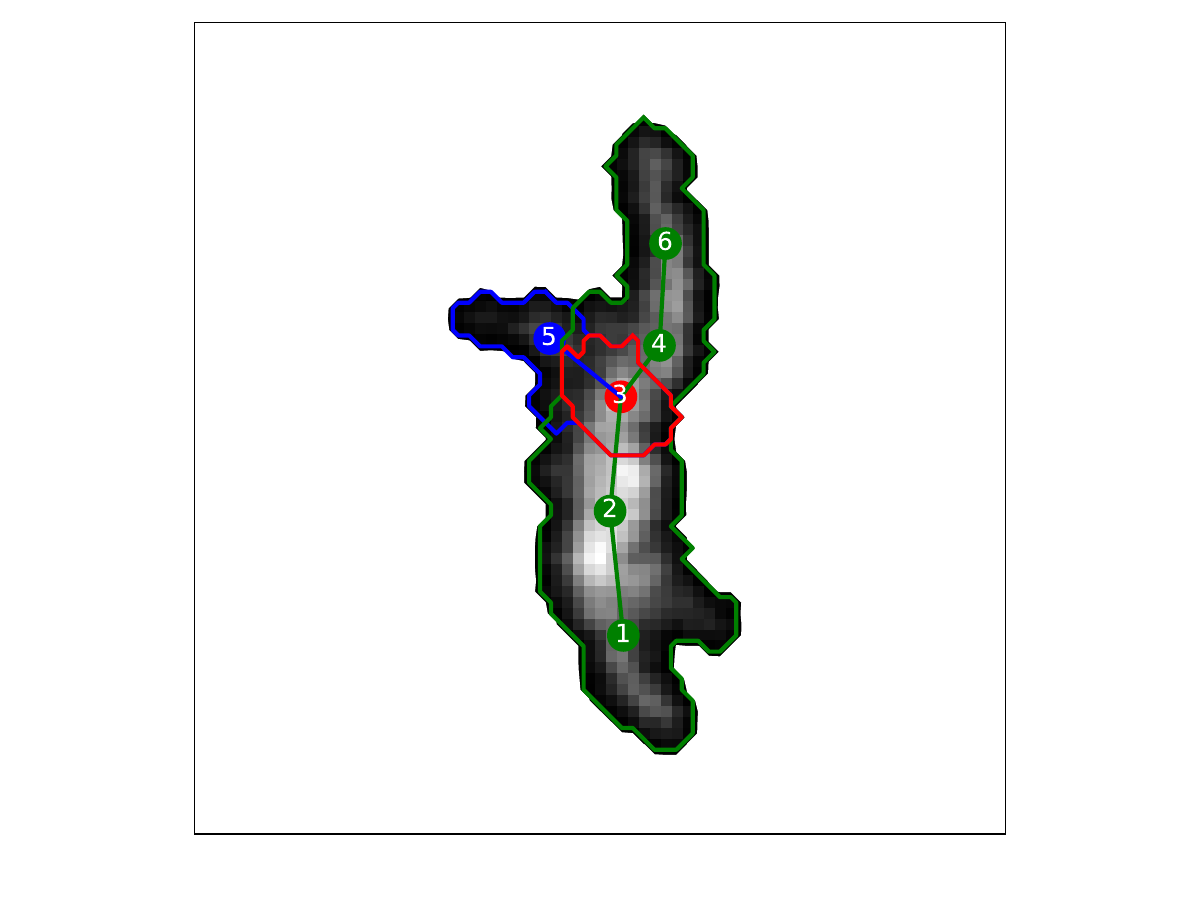}}
\caption{Substructures of the filament. The numbered green and red circles connected by green lines represent the positions of clumps within the first sub-filament, while the green contour represents the boundary of this substructure. Similarly, the numbered blue and red circles connected by blue lines represent the positions of clumps within the second sub-filament, while the blue contour represents the boundary of this substructure. The red clump corresponds to a shared node between the two sub-filaments, with its boundary outlined by the red contour. }
\label{Skeleton_Sub_Graph}
\end{figure}

\begin{figure*}[htbp]
\centering
\begin{minipage}[b]{0.3\textwidth}
    \centering
    \begin{minipage}[b]{\textwidth}
        \centering
        \includegraphics[width=2.1in]{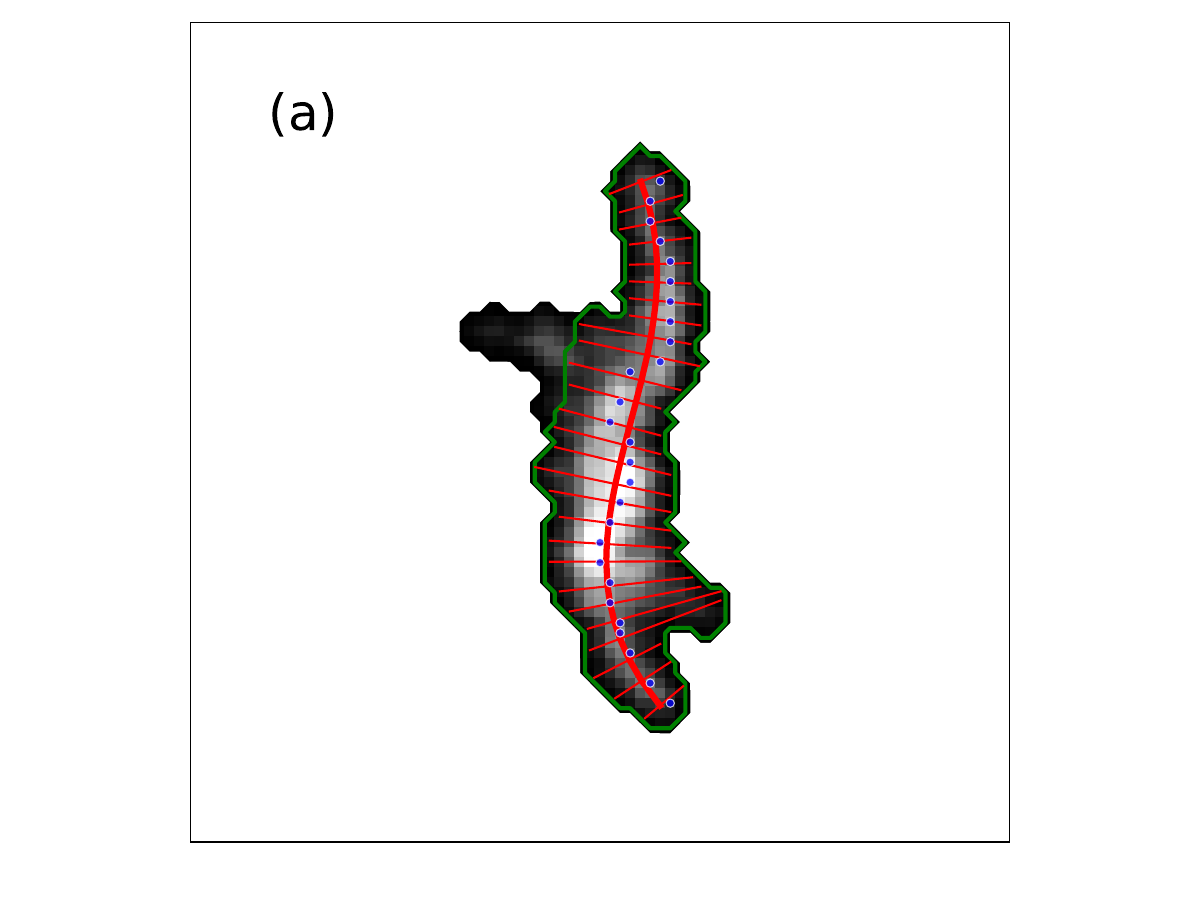}
    \end{minipage}
    \begin{minipage}[b]{\textwidth}
        \centering
        \includegraphics[width=2.1in]{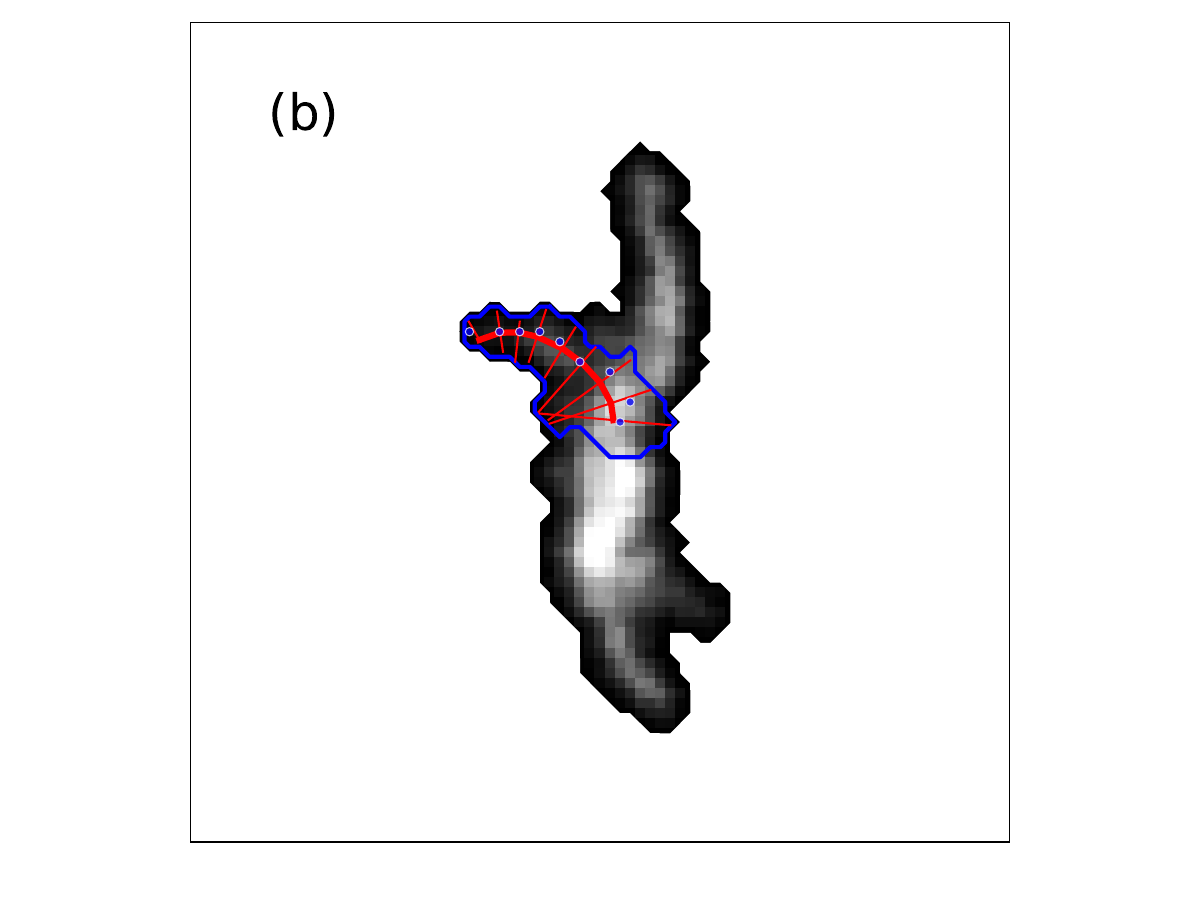}
    \end{minipage}
\end{minipage}
\begin{minipage}[b]{0.5\textwidth}
    \centering
    \includegraphics[width=4.5in]{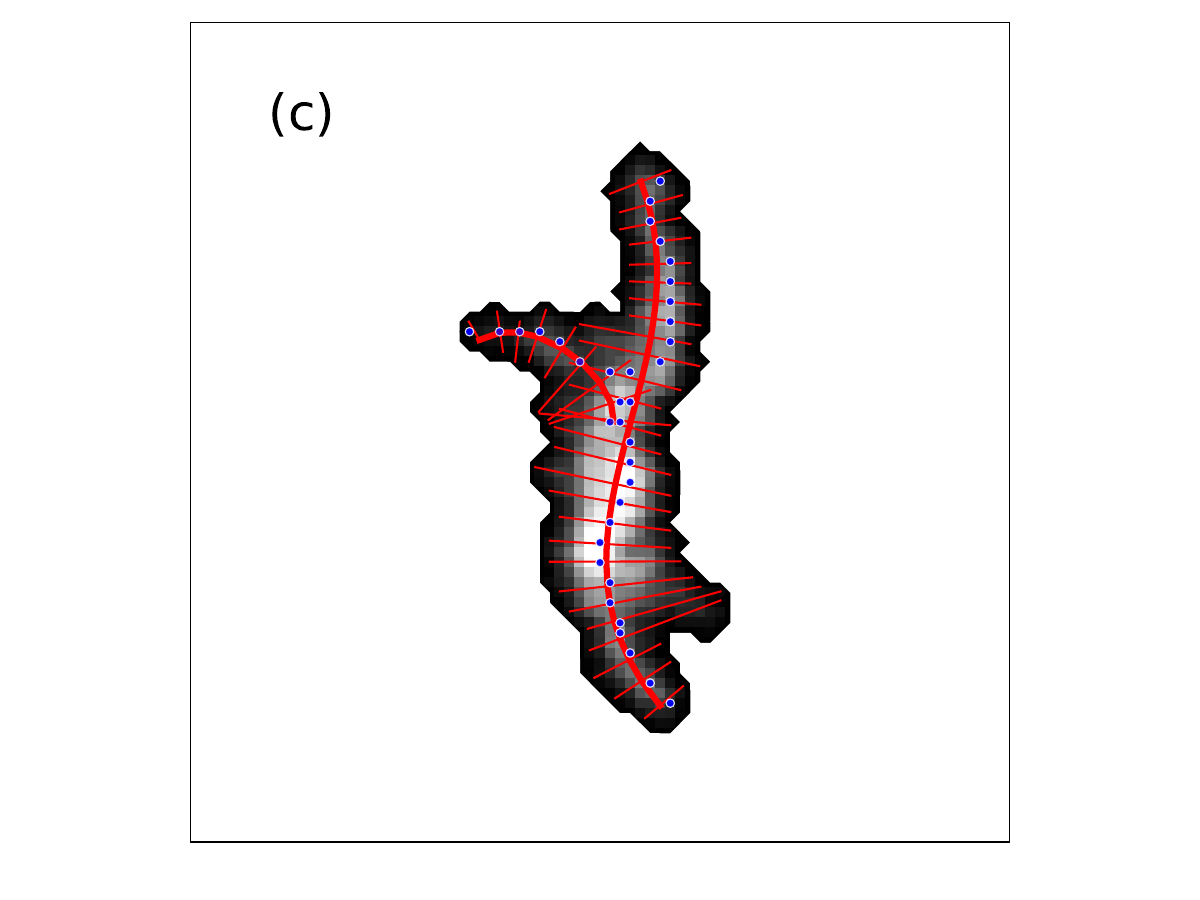}
\end{minipage}
\caption{(a) The fitted intensity skeleton and profiles of the first sub-filament. The green contour outlines the region of this sub-filament. (b) The fitted intensity skeleton and profiles of the second sub-filament. The blue contour outlines the region of this sub-filament. (c) The overall fitted intensity skeletons and profiles.}
\label{Filament_Intensity_Spine_Total}
\end{figure*}

\begin{figure*}
\centering
\vspace{0cm}
\begin{minipage}[t]{0.45\textwidth}
    \centering
    \centerline{\includegraphics[width=3in]{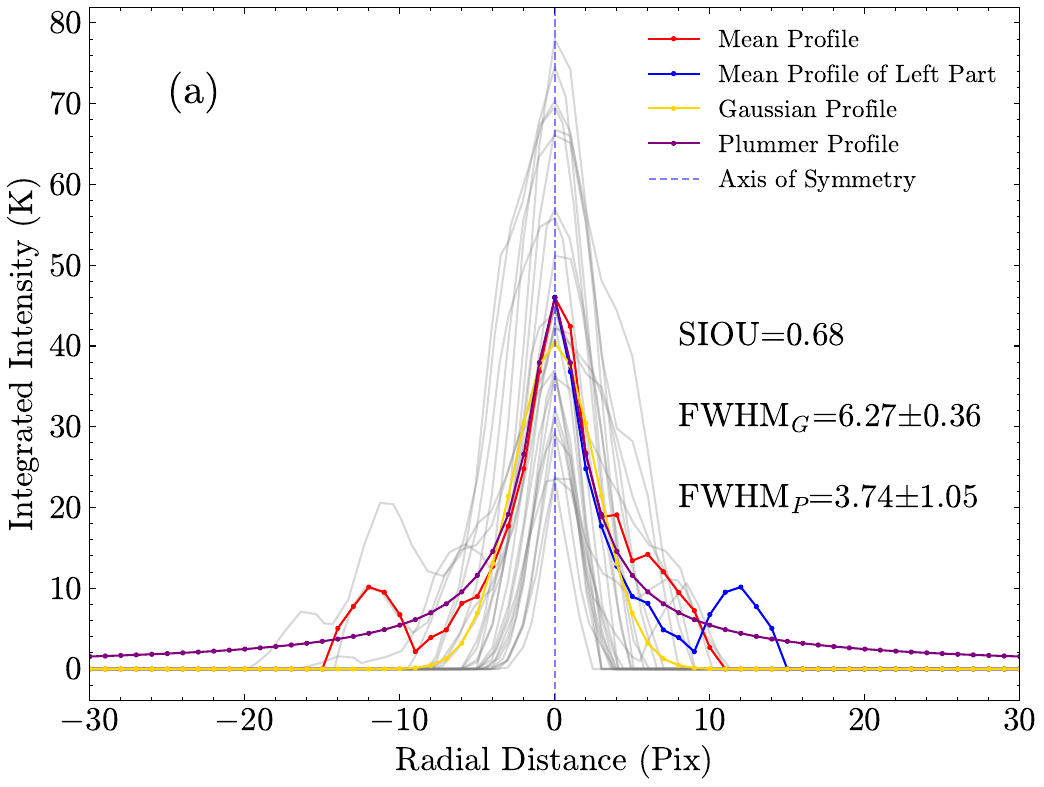}}
\end{minipage}%
\begin{minipage}[t]{0.45\textwidth}
    \centering
    \centerline{\includegraphics[width=3in]{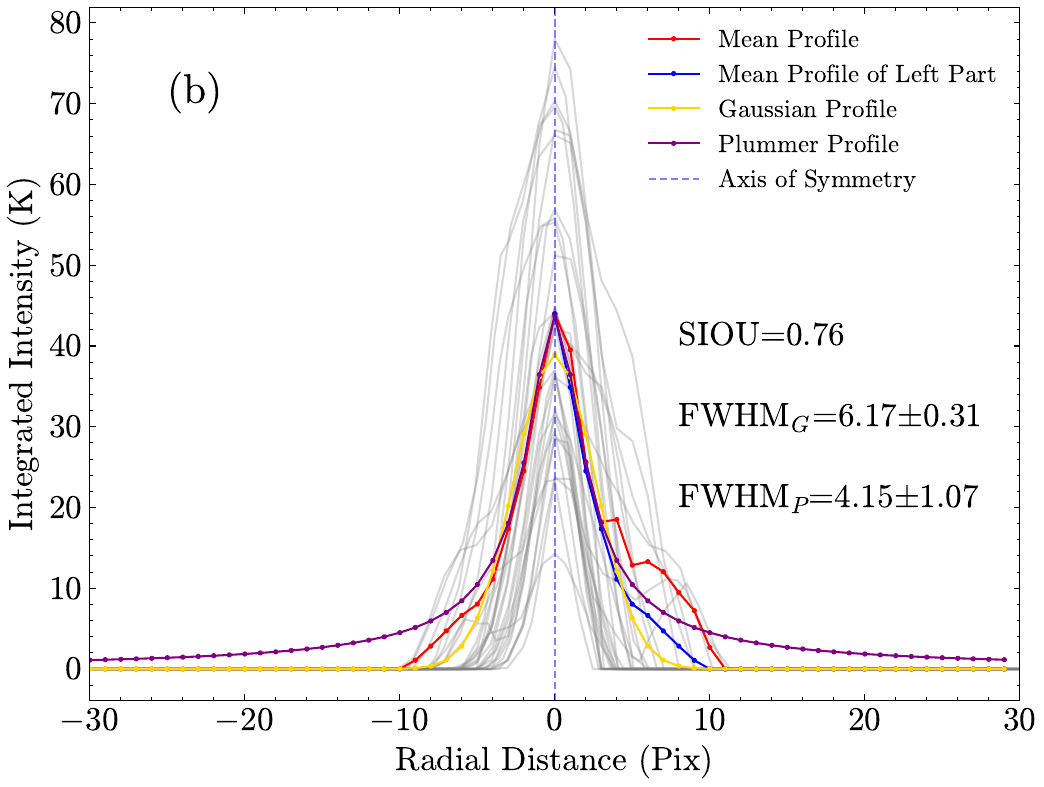}}
\end{minipage}%
\caption{(a) Intensity profiles of Figure \ref{Filament_Intensity_Point}(c). The gray lines are the profiles, the red line is the average line of the profiles after filtering out abnormal profiles (see the manual), and the blue line is the reflection of the left average line along the dashed symmetry axis. The golden profile is obtained by fitting the profiles with a Gaussian function, while the purple profile is obtained by fitting the profiles with a Plummer function. The symmetry of the average profile is 0.68. The FWHM obtained from the Gaussian fitting is FWHM$_G=6.27\pm0.36$ pixels, while the FWHM from Plummer-like fitting is FWHM$_P=3.74\pm1.05$ pixels and its power index is $p=2.2\pm0.28$. (b) Intensity profiles of Figure \ref{Filament_Intensity_Spine_Total}(c). The symmetry of the average profile is 0.76. The FWHM obtained from the Gaussian fitting is FWHM$_G=6.17\pm0.31$ pixels, while the FWHM from Plummer-like fitting is FWHM$_P=4.15\pm1.07$ pixels and its power index is $p=2.3\pm0.32$.}
\label{Intensity_Profile_R_Mean}
\end{figure*}

\subsection{The Graph-based Sub-structuring Method}\label{Substructure}
Filament formation and fragmentation are complex processes, characterized by tangled, interconnected sub-structures and complicated kinematics, on both large and small scales \citep{Subfilaments,MapPPVPPP}. Studying the substructures within intricate filament systems and analyzing the connections and interactions among individual sub-filaments can provide observational insights into the processes of filament formation and fragmentation. Increasingly, studies suggest that the intersection points of sub-filaments are preferred environments for the formation of high-mass stars and star clusters \citep{Filament_Core_1,InterHighMassStar,Core_Formation_Filaments_6}.

\subsubsection{The Substructure}\label{Substructure_1}

To obtain the substructure of a filament, we treat the positions of the clumps within the filament as nodes in a graph. The weight, as indicated in Equation (\ref{WeightST_1}), is calculated between any two neighboring clumps. These weights are then utilized to establish connections and build a weighted undirected graph, denoted as G2. The GMST algorithm is applied to G2 to obtain the tree T2 which includes all the clumps that constitute the filament. Then, the weight of tree T2 is updated by Equation (\ref{WeightST_2}).

\begin{equation}\label{WeightST_1}
\begin{split}
WeightST_{Gij}=\frac{DistLB_{ij}\times DistV_{ij}}{\frac{1}{N}\sum_{1}^{N}I_k}
\end{split}
\end{equation}

\begin{equation}\label{WeightST_2}
\begin{split}
WeightST_{Tij}=DistLB_{ij}\times DistV_{ij}\times \frac{1}{N}\sum_{1}^{N}I_k
\end{split}
\end{equation}

\noindent where $WeightST_{Gij}$ represents the wight of two neighboring clumps $i$ and $j$ in the graph G2, and $WeightST_{Tij}$ represents the updated wight of two neighboring clumps in the tree T2. $DistLB_{ij}$ is the Euclidean distance between the neighboring clumps in the spatial direction, while $DistV_{ij}$ is the difference in the velocity channel. $I_k$ is the intensity of coordinate $k$ along the segment between the central coordinates of clumps $i$ and $j$ in the PPV space, with $N$ representing the number of voxels in this segment. 

When neighboring clumps are close in distance, the tree should prioritize connecting clumps with higher intensity, while larger substructures need to have both greater length and higher intensity. $WeightST$ can robustly achieve these goals. We have devised the following recursive function to decompose any complex filament: 

\setenumerate[1]{label=} 
\begin{enumerate}
\itemsep=2pt
\item $Operation 1$: Retrieve and record the longest shortest path of T2. The "longest" here is defined as having the maximum sum of weights.
\item $Operation 2$: Remove the edges in the longest shortest path in T2, and examine the corresponding nodes. If the degree of a node becomes 0, remove it as well.
\item $Operation 3$: If there are remaining nodes, generate all the subtrees T2i by utilizing the remaining nodes and edges, with edge weights determined using Equation (\ref{WeightST_1}). 
\item $Operation 4$: Replace T2 with each subtree T2i and update the weights using Equation (\ref{WeightST_2}). Then, recursively execute operations 1-3.
\end{enumerate}

The recursion terminates when there are no nodes in T2. The nodes along different longest shortest paths correspond to the clumps within different sub-filaments. As shown in Figure \ref{Skeleton_Sub_Graph}, the filament can be decomposed into two sub-filaments, with one containing five clumps and the other containing two clumps. 

These sub-filaments share a common intersection clump, and each includes at least one additional clump beyond the shared one. Each sub-filament has distinct local high-density structures, which allows for a better examination of stellar formation along both the primary and secondary filaments, as well as the possible variations in stellar activity at the intersections.

\subsubsection{The Symmetry and Radius}\label{SymmetryRadius}
Here, we elaborate on how the sub-filament analysis method further improves the accuracy of the profile, which is crucial for calculating the characteristic radius of a filament. 

To extract the intensity skeleton of the secondary substructures, first obtain the tree T1 as described in Section \ref{Substructure_1} for the region. Next, compute the shortest paths solely between existing skeleton points on the intersection clumps and the leaf nodes on the regional boundary. Finally, select the shortest path with the maximum weight as the intensity skeleton. It is worth noting that the profiles of the substructures are restricted within the masked area of the clumps where each skeleton point is located, as well as their neighboring clumps. This prevents excessively long and abnormal profiles that can arise from curved structures. 

In Figure \ref{Filament_Intensity_Spine_Total}, we present the fitted intensity skeletons and profiles of the two sub-filaments shown in Figure \ref{Skeleton_Sub_Graph}, as well as the overall fitted intensity skeletons and profiles. We align the profiles based on the intensity peak on each profile and show the profiles of Figures \ref{Filament_Intensity_Point}(c) and \ref{Filament_Intensity_Spine_Total}(c) in Figure \ref{Intensity_Profile_R_Mean}. By comparing Figure \ref{Filament_Intensity_Point}(c), Figure \ref{Skeleton_Sub_Graph}, and Figure \ref{Intensity_Profile_R_Mean} together, it can be observed that the accuracy and symmetry of the profiles decrease due to the influence of clump 5. 

To accurately evaluate the symmetry of the profile, we introduce a metric called symmetry intersection over union (SIOU). It is defined as the intersection of two regions divided by their union. The first region is the area enclosed by the blue average profile shown in Figure \ref{Intensity_Profile_R_Mean} and the radial distance axis within the same range. The second region is the area enclosed by the right half of the red average profile shown in Figure \ref{Intensity_Profile_R_Mean} and the radial distance axis within the same range. The symmetry of the profile obtained through the sub-filament analysis method improves by 0.08, indicating that the new method can achieve more accurate profiles. 

To describe the characteristic radius, we employ both the Gaussian model and the Plummer-like model \citep{Arzoumanian_2011,Profile_Region_1,Large_Scale_Galactic_Filaments,C18O_1} to fit the profiles. This approach enables us to determine the FWHM, which serves as a measure of the characteristic radius. The FWHM is calculated using the formula $FWHM=2\sqrt{2log(2)}\sigma$, where $\sigma$ represents the standard deviation of the Gaussian function or the flattening radius of the Plummer-like function \citep{Musca,RadFil}. The fitted curves and calculated FWHM values are displayed in Figure \ref{Intensity_Profile_R_Mean}.

\begin{table}
\centering
\caption{The Input Parameters of DPConCFil for Identifying Filaments.}
\begin{tabular}{p{1.8cm}p{4.8cm}p{0.8cm}}\hline\hline
    Parameters&Explanation&Value\\\hline
    $TolAngle$&The angle tolerance that indicates the presence of directional consistency between two neighboring clumps, in degrees. &30\\\hline
    $TolDistance$&The distance tolerance that indicates the presence of positional consistency between a clump and local filament axis, in pixels. &4 (2$\arcmin$)\\\hline
    $LWRatio$&The minimum aspect ratio of a filament. &2.5\\\hline 
\end{tabular}
\label{InputPar}
\end{table}

\begin{figure*}
\centering
\vspace{0cm}
\begin{minipage}[t]{0.35\textwidth}
    \centering
    \centerline{\includegraphics[width=2.8in]{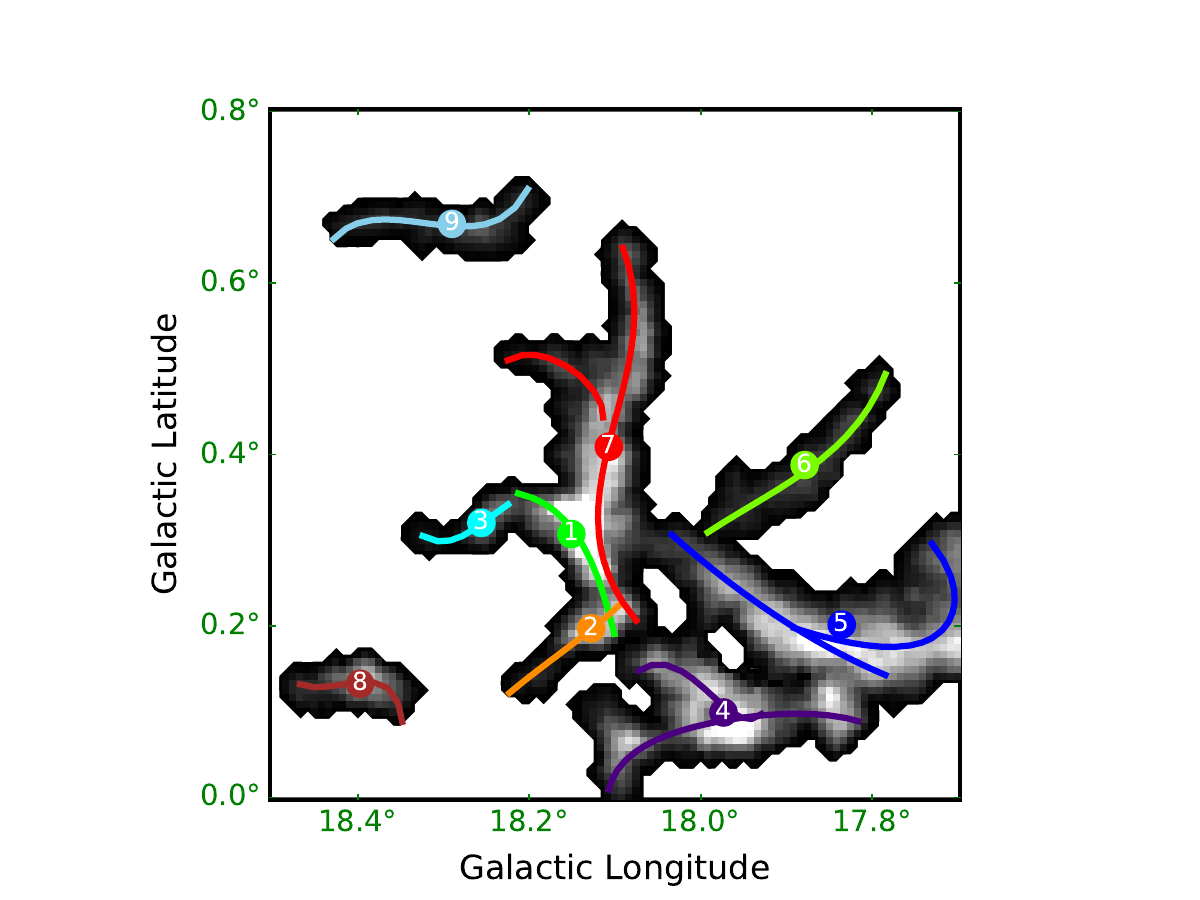}}
\end{minipage}%
\begin{minipage}[t]{0.35\textwidth}
    \centering
    \centerline{\includegraphics[width=2.8in]{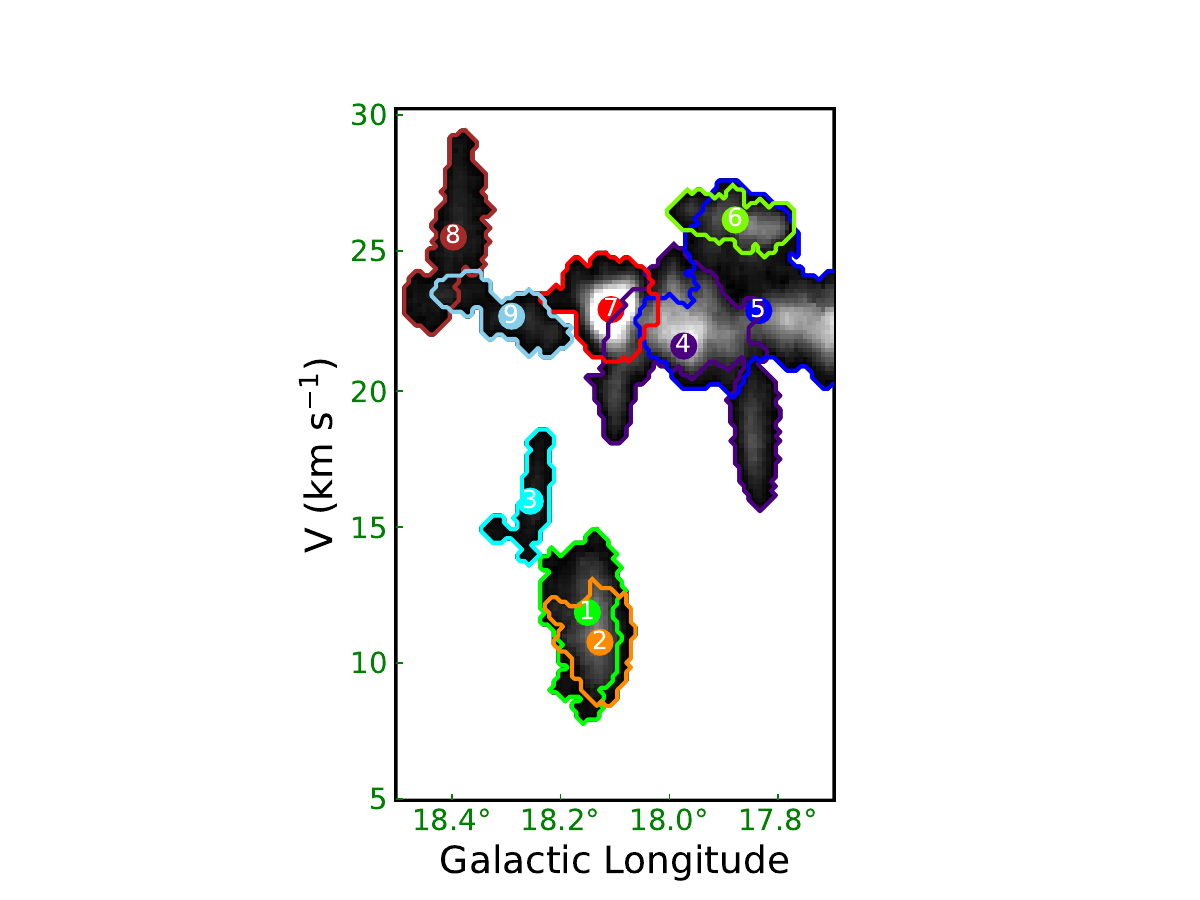}}
\end{minipage}%
\begin{minipage}[t]{0.20\textwidth}
    \centering
    \centerline{\includegraphics[width=2.8in]{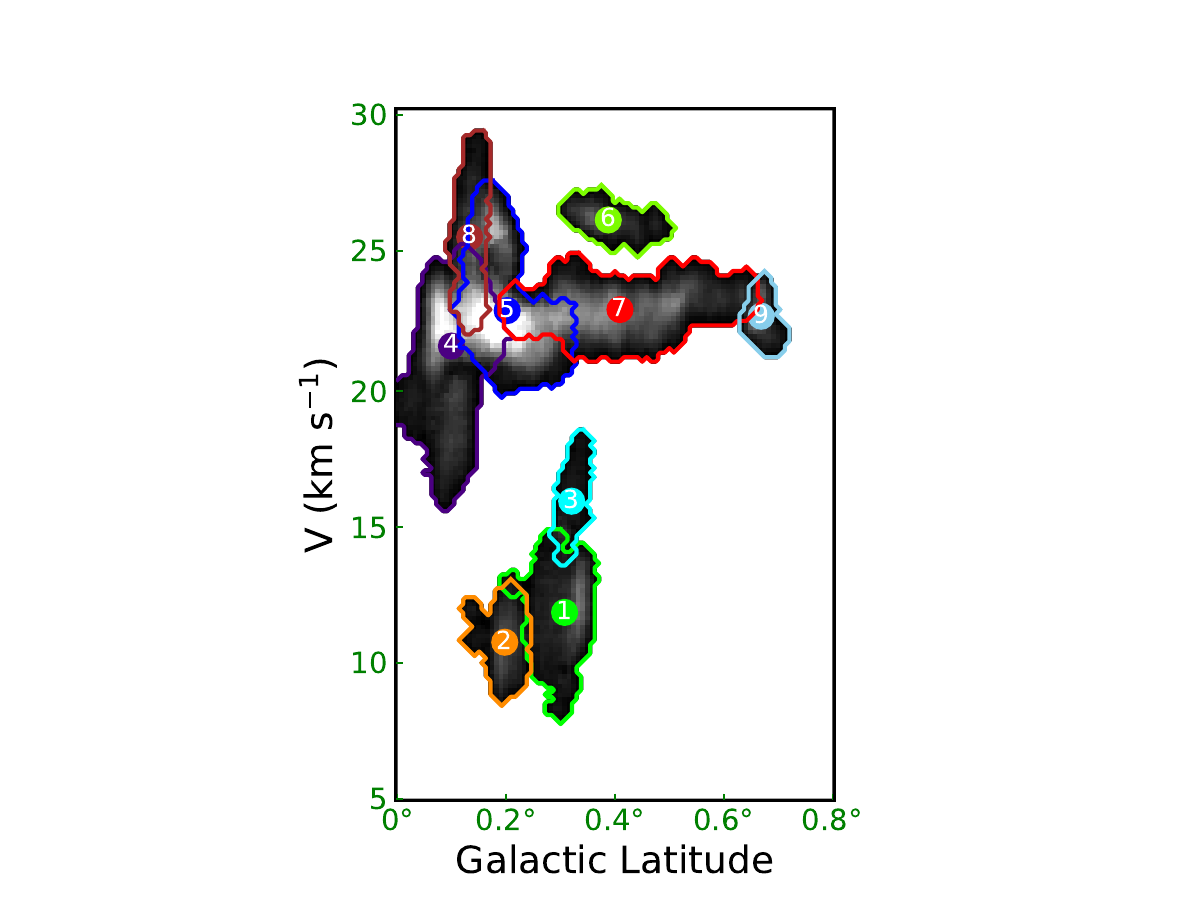}}
\end{minipage}%
\caption{All filaments identified from Figure \ref{Clumps_Infor}. Left panel: velocity-integrated intensity map. The curves of different colors represent the B-spline curves of the intensity skeletons of different filaments, while the numbered circles of different colors indicate the positions of different filaments and retain the same meanings across the three panels. Middle panel: latitude-integrated intensity map. The contours, marked in different colors, illustrate the specific regions of various filaments. Right panel: longitude-integrated intensity map. Except for Filament 7, which has been shown in the previous text, the detailed plots of other filaments are displayed in Figure \ref{Filament_Fit_Spine_Items} of Appendix \ref{FilamentItems}. }
\label{Filaments_All}
\end{figure*}

\begin{table*}
\centering
\caption{\centering A Catalog of Filaments of the Example Data.}
\begin{tabular}{c|cccccccccccc}\hline\hline
    Fils&$L_{cen}$&$B_{cen}$&$V_{LSR}$&Length&Area&LWRatio&Angle&$V_{Grad}$&SIOU&FWHM$_{G}$&$N_{Cl}$&$N_{Sf}$\\
    ID&(deg)&(deg)&(km s$^{-1}$) &(arcmin)&(arcmin$^{2}$)&&(deg)&(km s$^{-1}$arcmin$^{-1}$)&&(arcmin)&&\\
    (1)&(2)&(3)&(4)&(5)&(6)&(7)&(8)&(9)&(10)&(11)&(12)&(13)\\\hline		
    1&18.151&0.307&11.812&13.0&50.25&3.9&-51.41&0.176&0.92&2.37±0.15&5&1\\ 
    2&18.128&0.197&10.718&9.0&35.75&3.4&36.25&0.168&0.80&2.21±0.12&3&1\\ 
    3&18.256&0.320&15.888&7.0&21.25&3.5&24.11&0.436&0.78&1.72±0.08&4&1\\ 
    4&17.973&0.099&21.575&18.5&137.25&3.4&3.52&0.292&0.74&4.00±0.21&7&2\\ 
    5&17.836&0.202&22.875&28.5&172.00&4.2&-9.65&0.110&0.53&3.82±0.14&8&2\\ 
    6&17.879&0.388&26.198&14.5&60.50&4.5&39.44&0.059&0.96&2.35±0.12&3&1\\ 
    7&18.107&0.409&22.913&27.5&167.00&4.7&86.19&0.034&0.76&3.08±0.15&6&2\\ 
    8&18.397&0.133&25.570&9.0&33.50&2.7&-1.68&0.429&0.86&2.33±0.28&3&1\\ 
    9&18.290&0.669&22.670&14.5&49.00&5.8&1.01&0.127&0.92&2.04±0.27&4&1\\ 
    
    \hline
\end{tabular}
\begin{tablenotes}
    \item \textbf{Note.} Column (1): ID of flaments. Columns (2)-(4): flux-weighted longitude, latitude, and LSR velocity. Column (5): the angular length. Column (6): the projected angular area in the spatial direction. Column (7): the ratio between length and width. Column (8): the angle. Column (9): the velocity gradient. Column (10): the symmetry of the profile. Column (11): the FWHM of Gaussian fitting. Column (12): the number of clumps within flaments. Column (13): the number of sub-filaments. 
\end{tablenotes}
\label{Catalogue_Table}
\end{table*}

\begin{figure*}
\centering
\vspace{0cm}
\begin{minipage}[t]{0.3\textwidth}
    \centering
    \centerline{\includegraphics[width=5in]{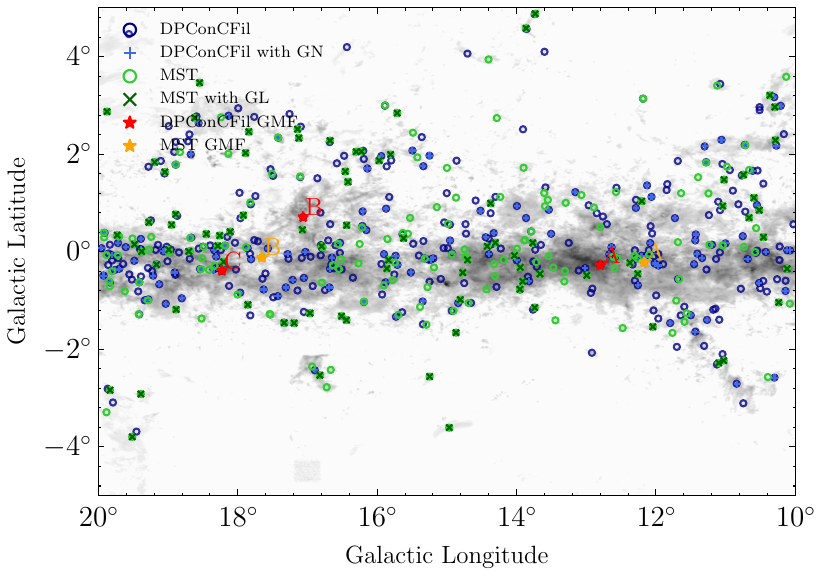}}
    \begin{minipage}[t]{0.3\textwidth}
        \centering
        \centerline{\includegraphics[width=5in]{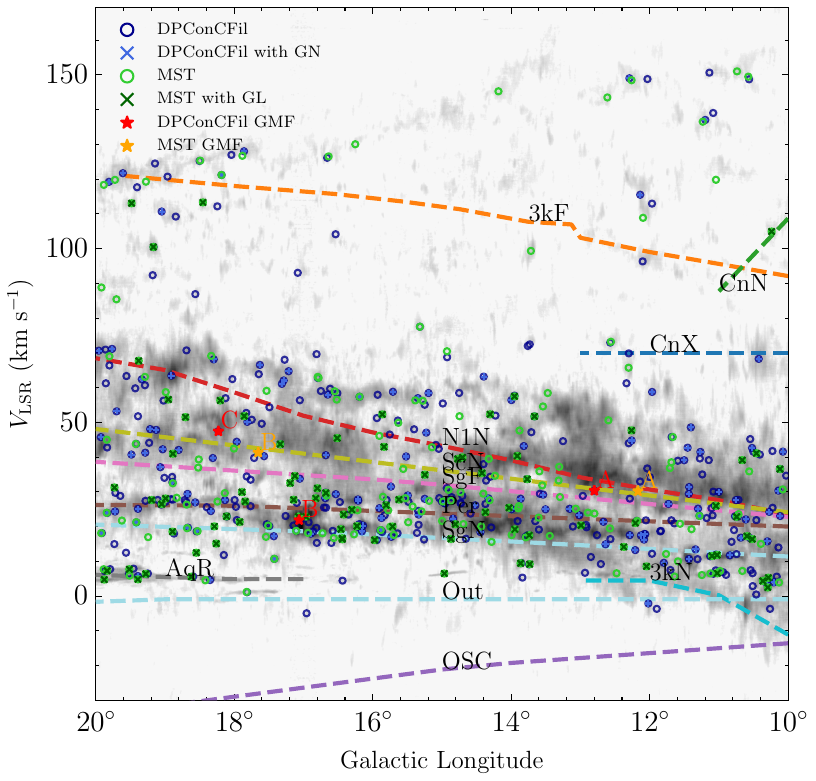}}
    \end{minipage}
\end{minipage}
\caption{Distribution of all filaments in the Galactic longitude-latitude and Galactic longitude-velocity planes in the application data. The background map illustrates the distribution of molecular gas traced by the integrated $^{13}$CO emission. Dark blue circles denote filaments identified by DPConCFil, and royal blue plus signs indicate those with a greater number of clumps. Lime green circles represent structures isolated by MST, and dark green crosses highlight filaments with greater linearity. Red asterisks mark giant filaments identified by DPConCFil, and orange stars denote those identified by MST; details are in Appendix \ref{ApplicationImages}. Different colored dashed lines illustrate various spiral arms from the model by \cite{Distance_1}, with their names labeled.}
\label{Distribution_LBV}
\end{figure*}


\subsection{The Parameters of DPConCFil}\label{Parameters_DPConCFil}
The explanation and default values of the input parameters of DPConCFil for identifying filaments are listed in Table \ref{InputPar}. The default values are determined based on extensive experimental results, and the parameters can be fine-tuned to suit specific applications and requirements. If the consistency is considered to be weak, such as indicated by the PP clumps shown in Appendix \ref{ClumpsInfor_2D}, one can increase the angle and distance tolerances.

DPConCFil is capable of retrieving all the information pertaining to filaments as presented in the preceding text. The essential elements include the IDs of clumps associated with the filaments and the specific regions that the filaments occupy. The regions are identified via a mask array, where each filament is assigned a unique index (starting from one) that corresponds to the same number in the mask. Additional parameters, such as centroid and angle, can be deduced through direct computation based on the filament regions. By employing the analytical methods outlined in DPConCFil, the filament original data and regions can be processed to acquire the skeleton and profiles. 

The filaments identified from Figure \ref{Clumps_Infor} are displayed in Figure \ref{Filaments_All}. DPConCFil demonstrates the ability to completely identify all visually possible filamentary structures in Figure \ref{Example_Data}. Filaments 1 and 7 overlap in spatial direction, but they have a substantial velocity difference of approximately 11 km s$^{-1}$. The regions associated with Filaments 4, 5, and 7 are connected and demonstrate comparable velocities, measuring less than 2 km s$^{-1}$. Nonetheless, DPConCFil correctly recognizes these distinct filaments. 

\section{Application and Statistics}\label{ApplicationAnalysis}
We apply the identification sub-method of DPConCFil to the application data described in Section \ref{Data_CO}, successfully identifying a batch of filaments that span various scales. Using the filaments from the example data as an illustration, we compile a filament catalog that includes detailed explanations of its basic parameters and additional parameters derived through the analysis sub-methonds of DPConCFil. We choose a subset of these parameters for conducting histogram statistics. Statistical and physical analyses with multiple distance measurements will be explored in a subsequent paper (Jiang et al. in prep). The application to the simulated molecular clouds is detailed in Appendix \ref{Simulation}.

\subsection{Catalog}\label{Catalogue} 
Table \ref{Catalogue_Table} presents the parameters for the nine filaments identified in the example data. The parameters of the identified filament in the application data are listed in a table similar to that of Table \ref{Catalogue_Table}. Although the range of the application data encompasses that of the example data, not all filaments listed in Table \ref{Catalogue_Table} are included in the application data table due to differences in edge clumps between the example and application datasets.


The index ID of each filament, as returned by DPConCFil, is displayed in column (1). Columns (2) - (4) are the intensity-weighted center of the filament in spatial coordinates, and the intensity-weighted radial velocity with respect to the local standard of rest (LSR). The length, which is calculated by multiplying the number of intensity skeleton points by the angular size (0.5 arcmin) of a pixel, is presented in column (5), while the angular area can be found in column (6). The area value is obtained by multiplying the projected area in spatial direction by the angular area (0.25 arcmin$^2$) of a pixel \citep{Visual_Based_Definition_Yuan}. Column (7) displays the aspect ratio, and its definition can be found in Section \ref{AspectRatio}. Column (8) contains the angle, which is determined by diagonalizing the moment of inertia matrix \citep{FacetClumps}. This angle represents the orientation of the major axis of a filament relative to the negative direction of $l$, ranging from $-90^\circ$ to $90^\circ$. The velocity gradient is shown in column (9), calculated as the difference between the maximum and minimum velocities on the intensity skeleton of the velocity field divided by the length of the skeleton. Column (10) is used to quantify the symmetry of the profile, with its specific definition shown in Section \ref{Substructure}. Column (11) displays the FWHM obtained from Gaussian model fitting and its uncertainty range. It should be noted that this value has not been deconvolved based on beamwidth. The number of clumps within a filament and sub-filaments derived through the graph-based sub-structuring method for each filament can be found in columns (12) and (13), respectively. 

\begin{figure}
\centering
\centerline{\includegraphics[width=3.3in]{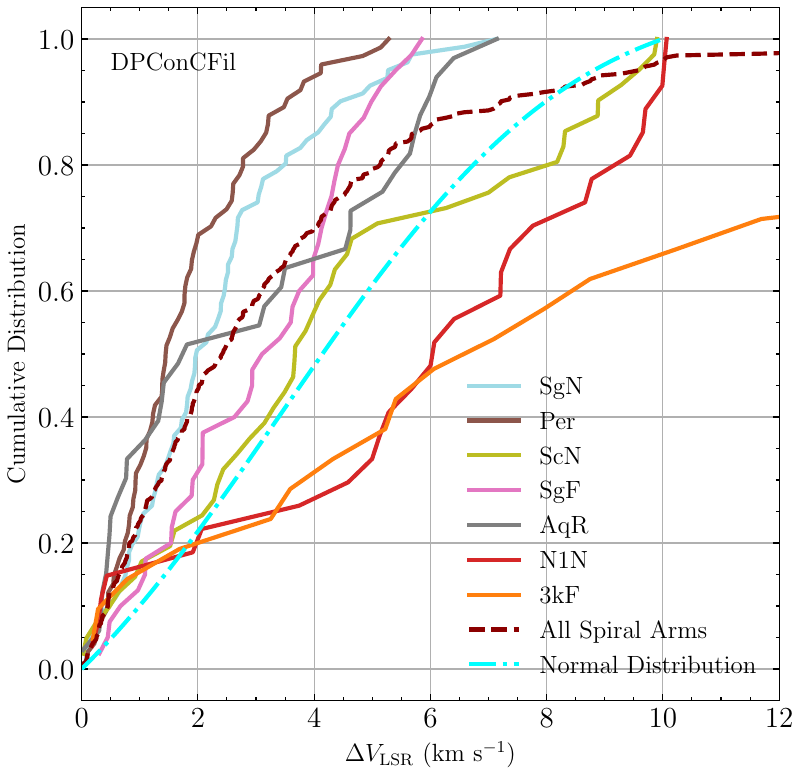}}
\caption{Cumulative distribution of the velocity separation between filaments identified by DPConCFil and their nearest spiral arm along the line of sight. The spiral arm velocities are depicted in Figure \ref{Distribution_LBV}. The curves in different colors illustrate the cumulative distributions of velocity differences for the  spiral arms. The dark red dashed line represents the cumulative distribution of all velocity differences, while the cyan dashed line depicts the cumulative distribution function of a truncated normal distribution with the same mean and variance as the overall velocity difference.}
\label{Distribution_CDF}
\end{figure}

\begin{figure}
\centering
\centerline{\includegraphics[width=3.3in]{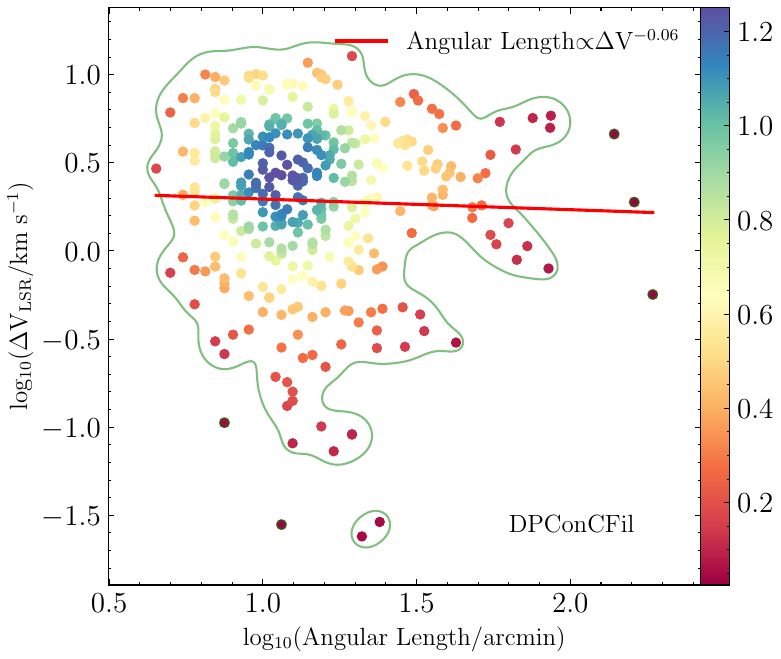}}
\caption{The angular length of filament, as certified by DPConCFil, in relation to the velocity separation between filaments and their nearest spiral arm. Points of different colors represent varying density estimates from the kernel density estimation, with the red line indicating the fitted relationship.}
\label{RelationLenV}
\end{figure}

\subsection{Analysis}\label{StatisticalAnalysis}
The coherence-based identification methods in DPConCFil and MST, designed specifically for detecting filaments in the PPV space, thus apply to the observational data. An introduction to MST and a comparison of algorithms can be found in Appendices \ref{Comparison_MST} and \ref{ApplicationImages}. Due to the study of large-scale filaments, MST requires a minimum of five clumps. Therefore, filaments identified by DPConCFil with five or more clumps are categorized as DPConCFil with a greater number (GN). Additionally, filaments with lower linearity identified by MST are excluded by the algorithm, so those with higher linearity are categorized as MST with a greater linearity (GL). The identification results from DPConCFil will be used to compile the catalog, so the emphasis is placed on these results.

\begin{figure*}[h]
\centering
\vspace{0cm}
\begin{minipage}[t]{0.3\textwidth}
    \centering
    \centerline{\includegraphics[width=2.2in]{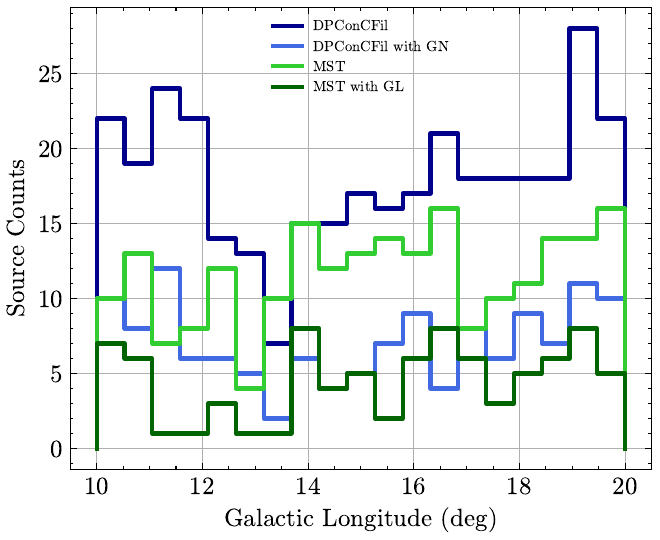}}
    \centerline{(a) Galactic Longitude}
\end{minipage}
\begin{minipage}[t]{0.3\textwidth}
    \centering
    \centerline{\includegraphics[width=2.2in]{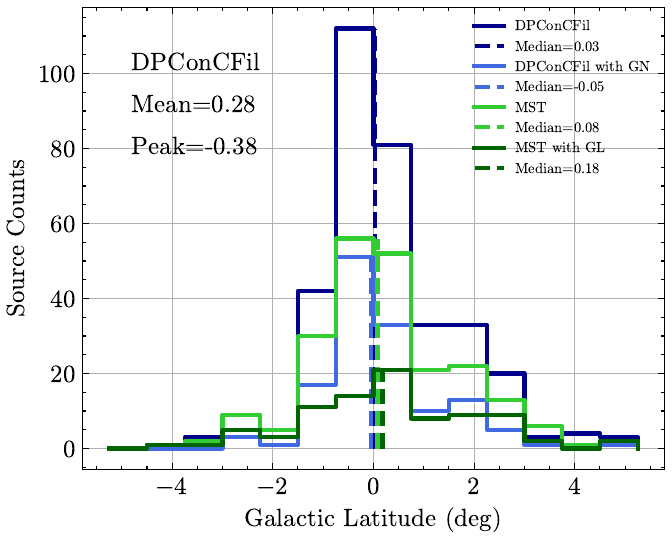}}
    \centerline{(b) Galactic Latitude}
\end{minipage}
\begin{minipage}[t]{0.3\textwidth}
    \centering
    \centerline{\includegraphics[width=2.2in]{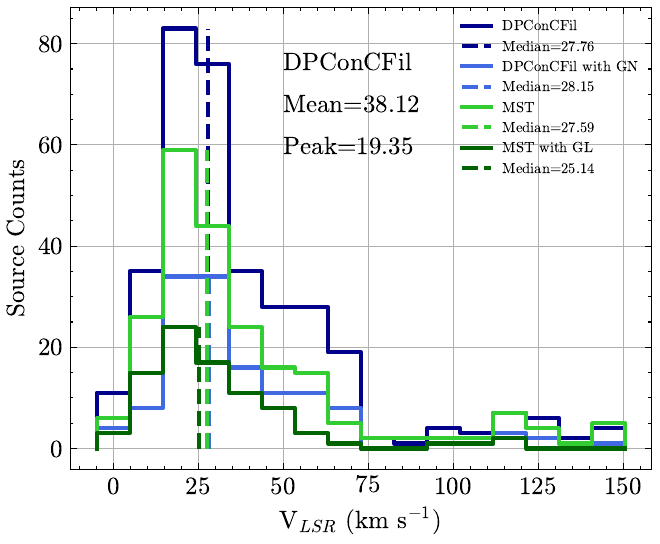}}
    \centerline{(c) Velocity }
\end{minipage}
\caption{The histogram statistics of Galactic distribution. Panels (a)-(c) display the distribution of filament positions, including Galactic longitude, Galactic latitude, and velocity. Dark blue denotes filaments identified by DPConCFil, while royal blue indicates those with a greater number of clumps. Lime green represents structures isolated by MST, and dark green highlights filaments with greater linearity. The dashed lines represent the median, and the mean and peak of the distribution for DPConCFil are also calculated.}
\label{Statistical_Compare_1}
\end{figure*}

\begin{figure*}
\centering
\vspace{0cm}	
\begin{minipage}[t]{0.3\textwidth}
    \centering
    \centerline{\includegraphics[width=2.2in]{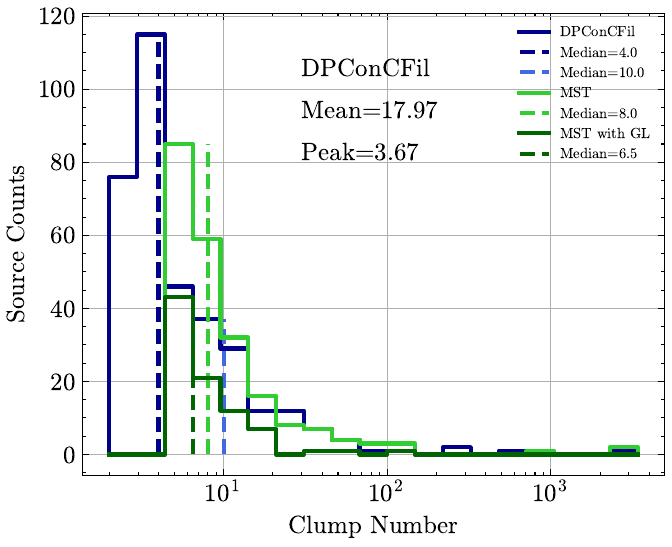}}
    \centerline{(d) Clumps Number}
\end{minipage}
\begin{minipage}[t]{0.3\textwidth}
    \centering
    \centerline{\includegraphics[width=2.2in]{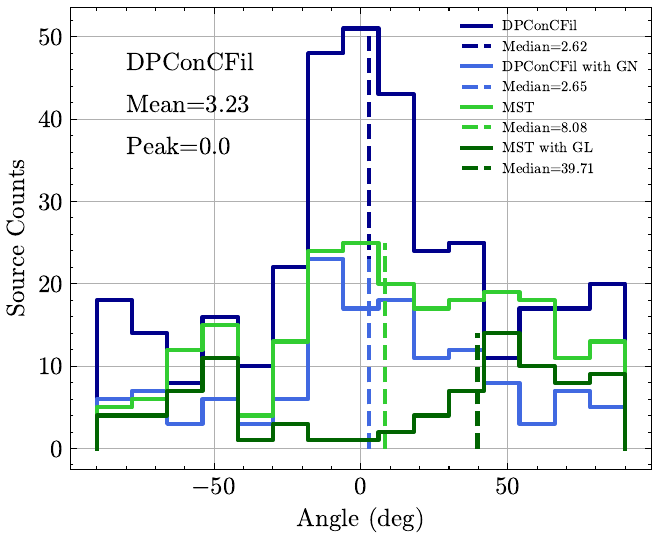}}
    \centerline{(e) Angle }
\end{minipage}
\begin{minipage}[t]{0.3\textwidth}
    \centering
    \centerline{\includegraphics[width=2.2in]{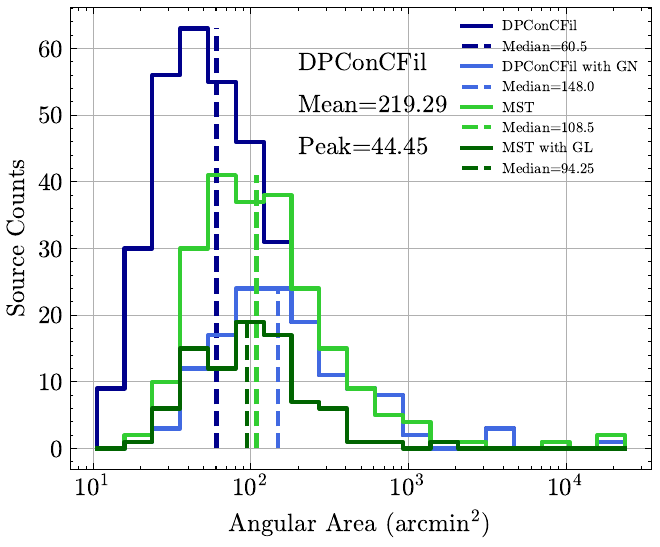}}
    \centerline{(h) Angular Area }
\end{minipage}

\begin{minipage}[t]{0.3\textwidth}
    \centering
    \centerline{\includegraphics[width=2.2in]{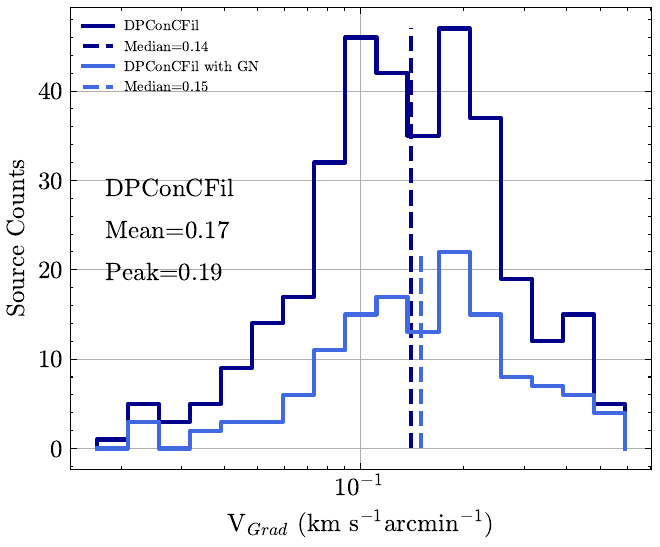}}
    \centerline{(d) Velocity Gradient }
\end{minipage}
\begin{minipage}[t]{0.3\textwidth}
    \centering
    \centerline{\includegraphics[width=2.2in]{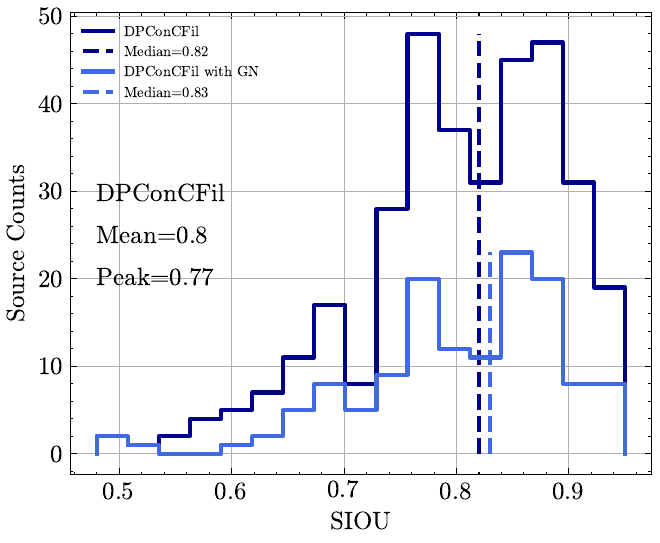}}
    \centerline{(e) SIOU }
\end{minipage}
\begin{minipage}[t]{0.3\textwidth}
    \centering
    \centerline{\includegraphics[width=2.2in]{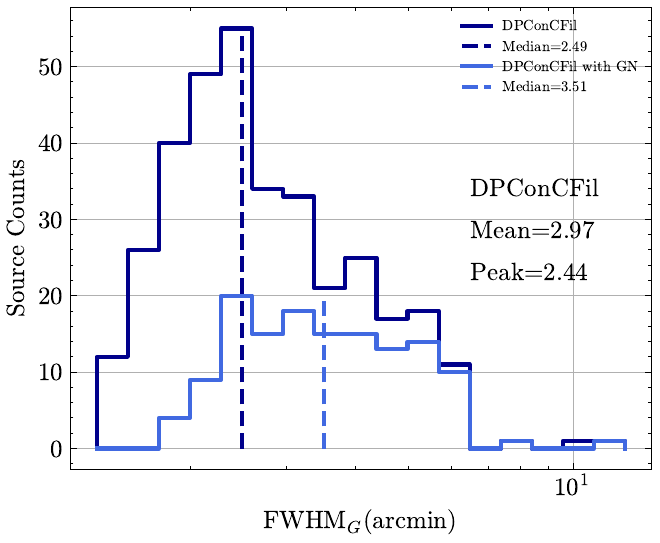}}
    \centerline{(f) Angular FWHM of Gaussian Fitting}
\end{minipage}

\caption{The histogram statistics of properties. Panels (a)-(c) display the number of clumps within the filament, the angle, and the angular area for DPConCFil and MST. Panels (d)-(f) show the velocity gradient, SIOU, and angular FWHM from Gaussian fitting for DPConCFil.}
\label{Statistical_Compare_2}
\end{figure*}

\subsubsection{The Galactic Distribution}\label{Distribution}
In Figure \ref{Distribution_LBV}, we show the distribution of filaments in the Galactic longitude-latitude and Galactic longitude-velocity planes, overlaid on the integrated $^{13}$CO emission map from MWISP. The longitude and velocity of the spiral arms are from the model suggested by \cite{Distance_1} using trigonometric parallaxes of Galactic high-mass star-forming regions. Asterisks denote the positions of giant molecular filaments \citep[GMF;][]{GiantFilament_1,GiantFilament_3, Large_Scale_Galactic_Filaments}, with images of these filaments illustrated in Appendix \ref{ApplicationImages}.

The spiral arms are the primary structures of the Milky Way, and large-scale filaments tend to be distributed along these arms. \citep[e.g.][]{MST,LatitudePeak,Large_Scale_Galactic_Filaments,MST_Apply_1}. To quantitatively assess the relationship between filament velocities and spiral arm velocities, we plot the cumulative distribution function of the velocity separation between filaments identified by DPConCFil and the nearest arm velocities in Figure \ref{Distribution_CDF}. The dark red dashed line represents the overall cumulative distribution, showing that approximately 95\% of the velocity differences are less than 10 \text{km}~\text{s}$^{-1}$, and 80\% are less than 5 \text{km}~\text{s}$^{-1}$. The cyan dashed line denotes a truncated normal distribution with the same mean and variance as the overall cumulative distribution. This comparison reveals that filaments more closely associated with the arms include Perseus (Per), Sagittarius N (SgN), Aquarius R (AqR), and Sagittarius F (SgF), with the most closely associated filaments being with Per, where all velocity separations are less than 5 \text{km}~\text{s}$^{-1}$. Scutum N (ScN) approximates a random distribution, while Norma 1 N (N1N) and 3 kpc Flare (3kF) exhibit weaker associations and also resemble random distributions. The filaments closest to SgN account for the highest proportion, at 25.4\%. 

\cite{GiantFilament_4} find no difference in the observable length of large-scale filaments across different galactic environments. We do not distinguish between arm-filaments and interarm-filaments but instead analyze the relationship between angular length and velocity separation. The density estimates obtained from kernel density estimation, along with the correlation fit line, are presented in Figure \ref{RelationLenV}. The main density distribution is circular, and the fitted relationship is $Angular Length \propto \Delta V^{-0.06}$, which further indicates no significant correlation between the scale of the filaments and the spiral arms. 

Figure \ref{Statistical_Compare_1} (a)-(c) show the intensity-weighted Galactic longitude, latitude, and velocity of filaments identified by different methods. Filaments are discretely distributed across the Galactic longitude, with fewer in the 12$^{\circ}$-15$^{\circ}$ as the presence of the largest GMFs A, which contains most of the molecular gas in this area. The majority of filaments are found within the latitude range of $-2^{\circ} < b < 3^{\circ}$. The latitude peaks for DPConCFil, DPConCFil with GN, and MST occur at negative latitudes, consistent with the trend reported in \cite{LatitudePeak}. The latitude median for DPConCFil, DPConCFil with GN, and MST differs from the Galactic midplane by less than 0.1$^{\circ}$, while MST with GL shows a larger median. The peak latitude for DPConCFil is $b_{\text{peak}} = -0.38^{\circ}$ (negative latitude), whereas the mean latitude is $b_{\text{mean}} = 0.28^{\circ}$ (positive latitude). This distribution mirrors the molecular gas density, with denser signals found at low negative latitudes and filamentary signals more common at high positive latitudes compared to high negative latitudes. Filament velocities mostly fall within -5 to 75 \text{km}~\text{s}$^{-1}$, with the cantilever velocities also concentrated in this range. The median, mean, and peak velocities for DPConCFil are 27.76, 38.12, and 19.35 \text{km}~\text{s}$^{-1}$, respectively.

\begin{table*}
\centering
\caption{\centering The Statistics.}
\begin{tabular}{c|ccccccccc}\hline\hline
    Method&Fils&M$_{CF}$&R$_{CF}$&Number&Angle&Area&V$_{Grad}$&SIOU&FWHM$_{G}$\\
    (1)&(2)&(3)&(4)&(5)&(6)&(7)&(8)&(9)&(10)\\\hline
    DPConCFil &344& 2484&52.3\% & 4 &2.62& 60.5 & 0.14 & 0.82 & 2.52\\
    DPConCFil with GN &135&2484 & 47\% &10  &2.65& 148 & 0.15 & 0.83 & 3.44 \\
    MST &177&3392& 80.1\% & 8 & 8.08& 108.5 & \textbackslash & \textbackslash &\textbackslash \\
    MST with GL &85&105& 7.2\% & 6.5 & 39.71 & 94.25 & \textbackslash & \textbackslash & \textbackslash\\\hline
\end{tabular}
\begin{tablenotes}
    \item \textbf{Note.} Column (1): different methods. Column (2): the number of filament. Columns (3): the max number of clumps within filaments. Column (4): the number ratio of clumps within filaments. Columns (5)-(10): the median of various statistics as shown in Figures \ref{Statistical_Compare_2}.  
\end{tablenotes}
\label{PropertiesTable}
\end{table*}

\subsubsection{The Properties }\label{Properties}
Histogram statistics for selected parameters are shown in Figure \ref{Statistical_Compare_2}. Panels (a)-(f) illustrate the number of clumps within the filament, the orientation angles in the projected sky, the angular area, the velocity gradient, SIOU, and angular FWHM from Gaussian fitting, respectively. Number information and statistical medians are summarized in Table \ref{PropertiesTable}. The graph-based substructuring and skeletonization methods are suitable for connected structures, whereas MST-identified structures often lack connectivity \citep[see Appendix \ref{Comparison_MST} and][]{Large_Scale_Galactic_Filaments}. Consequently, properties like velocity gradient, SIOU, and FWHM obtained from the analysis methods are not represented for MST.

The total number of filaments identified by DPConCFil is 344, with 60.8\% being small-scale filaments containing fewer than five clumps. Examples of small-scale filaments are shown in Figure \ref{Filament_Fit_Spine_Items}. MST identifies 177 structures, 48\% of which have GL. Approximately 50\% of the clumps are located on filaments identified by DPConCFil, while 80.1\% are on structures isolated by MST. Among these, only 7.2\% of the clumps are on filaments with GL. This is because clumps are mainly concentrated on a few larger MST structures, and filaments with GL contain no more than 105 clumps. Structures with a GN of clumps generally have lower linearity, and the linearity criterion is not restrictive enough to identify elongated features in most cases \citep{Large_Scale_Galactic_Filaments}. 

Figure \ref{Statistical_Compare_2}(b) demonstrates that filaments of DPConCFil exhibit a peak of around $\theta_{peak}=0$ in the angular distribution, with most filaments falling within the range of $-30^{\circ}<\theta<30^{\circ}$. This alignment is consistent with previous research, indicating that elongated molecular clouds tend to be oriented parallel to the Galactic plane \citep[e.g.][]{Angle,LatitudePeak,Large_Scale_Galactic_Filaments,MST_Apply_1,MST_Apply_2}. We have distinguished between positive and negative filament directions, which carry different implications. The mean angle for DPConCFil is 3.23$^{\circ}$, and the median angle for both DPConCFil and DPConCFil with GN is 2.6$^{\circ}$. MST are more widely distributed at positive angles, with a median of 8.08$^{\circ}$, while MST with GL have a median of 39.71$^{\circ}$.

Figure \ref{Statistical_Compare_2}(c) displays the angular areas, with medians of 60.5, 148, 108.5 and 94.25 arcmin$^2$ for DPConCFil, DPConCFil with GN, MST, and MST with GL, respectively. The mean and peak values for DPConCFil are 219.29 and 44.45 arcmin$^2$. For angular areas exceeding 110 arcmin$^2$, the distributions of DPConCFil and DPConCFil with GN are identical.

Figure \ref{Statistical_Compare_2}(d) and (e) show that the distributions of velocity gradient and SIOU for DPConCFil and DPConCFil with GN are similar, each displaying two peaks. The velocity gradient values range from 0.02 to 0.6 \text{km}~\text{s}$^{-1}$arcmin$^{-1}$, with a concentration around 0.14 \text{km}~\text{s}$^{-1}$arcmin$^{-1}$. Filament asymmetry is clearly evident, showing a wide range of SIOU values from 0.5 to 0.95, concentrated around 0.82. As shown in Figure \ref{Statistical_Compare_2}(f), the median, mean, and peak values for the angular FWHM of Gaussian fitting for DPConCFil are 2.49, 2.97, and 2.44 arcmin, respectively, while the median for DPConCFil with GN is 3.51 arcmin. To better understand the velocity gradient and radius of molecular cloud filaments, further analysis in conjunction with distance measurements is required.

\section{Summary}\label{Summary}
We introduce DPConCFil, a suite of innovative algorithms designed for filament identification and analysis. The consistency-based identification method employs clump properties such as directions, positions, and regional masks extracted by FacetClumps to identify elongated filaments that exhibit spatial and velocity continuity in the PPV space. The graph-based skeletonization method derives intensity skeletons from graphs and trees, which are weighted based on the spatial distances and intensities of neighboring points within the filament region. The graph-based sub-structuring method employs a recursive function to decompose the continually updated tree, constructed with weights based on spatial direction and velocity channel distances, as well as the mean line intensity between neighboring clumps, to extract substructures from complex filaments. Each sub-method can be applied independently and efficiently. The consistency-based identification method is especially effective for PPV data. By applying a molecular cloud clump detection algorithm to known filaments to extract clumps, and then utilizing the graph-based substructuring and skeletonization methods, both 2D maps and 3D datacubes can be analyzed.

We apply DPConCFil to MWISP $^{13}$CO emission and successfully identify a batch of filaments across various scales in the PPV space. A catalog containing the basic parameters of these filaments has been generated, and their positions and properties have been subjected to statistical analysis. The main statistical results are as follows:

(i) The velocity separation between these filaments and the Per arm is the smallest, while the region near SgN contains the greatest number of filaments.

(ii) The angular scale of the filaments shows no significant correlation with the Galactic environment.

(iii) Filaments are primarily distributed in latitudes ranging from $2^{\circ}$ to $3^{\circ}$, slightly below the midplane, with a peak at negative latitudes.

(iv) Approximately 50\% of the clumps are located within filaments.

(v) The direction of the filaments tends to align along the Galactic plane, with this trend being more pronounced on smaller scales.

(vi) The typical velocity gradient of a filament is 0.14 km s$^{-1}$arcmin$^{-1}$, and the typical symmetry is 0.82.

By comparing the principles and the visual results of filament images produced by algorithms such as FilFinder, DisPerSE, and MST, we conclude that DPConCFil is better suited for identifying and analyzing filaments of various scales in the PPV space, such as the MWISP molecular gas. In future work, we plan to extend the application of DPConCFil to additional observational datasets and perform a more detailed analysis of the physical parameters of the identified filaments by incorporating distance measurements. The relatively accurate regions and skeletons provided by DPConCFil make it easier and more precise to compute the physical parameters of the filaments. 

Furthermore, we will aim to utilize the intensity skeleton and surrounding profiles obtained through DPConCFil analysis methods, along with the velocity field, gravitational field, and magnetic field, to statistically analyze the relationships between the directions of these different fields within the convex hull of the profiles and the local filamentary skeleton. This will aid in understanding the formation and evolution of filaments.

\section*{Acknowledgements}
We are grateful to the anonymous referee for the invaluable insights and comments, which enabled us to refine and enhance this work. This work is supported by the National Natural Science Foundation of China (grants Nos. U2031202, 12373030, and 11873093) and the National Key R\&D Program of China (grant No. 2023YFA1608001). This research makes use of the data from the Milky Way Imaging Scroll Painting (MWISP) project, which is a multiline survey in $^{12}$CO/$^{13}$CO/C$^{18}$O along the northern galactic plane with PMO-13.7m telescope. We are grateful to all the members of the MWISP working group, particularly the staff members at PMO 13.7m telescope, for their long-term support. MWISP was sponsored by the National Key R\&D Program of China with grants 2023YFA1608000, 2017YFA0402701, and the CAS Key Research Program of Frontier Sciences with grant QYZDJSSW-SLH047. 

$Software$: Numpy \citep{Numpy}, Astropy \citep{Astropy}, Matplotlib \citep{Matplotlib}, Scikit-learn \citep{scikit-learn}, Scikit-image \citep{scikit-image}, Networkx \citep{Networkx}, FacetClumps \citep{FacetClumps}, RadFil \citep{RadFil}.

\bibliography{Filament_Identification_Analysis.bib}{}
\bibliographystyle{aasjournal}

\appendix

\section{Filaments in the Example Data}\label{FilamentItems}
\begin{figure*}
\centering
\vspace{0cm}
\begin{minipage}[t]{0.24\textwidth}
    \centering
    \centerline{\includegraphics[width=2.1in]{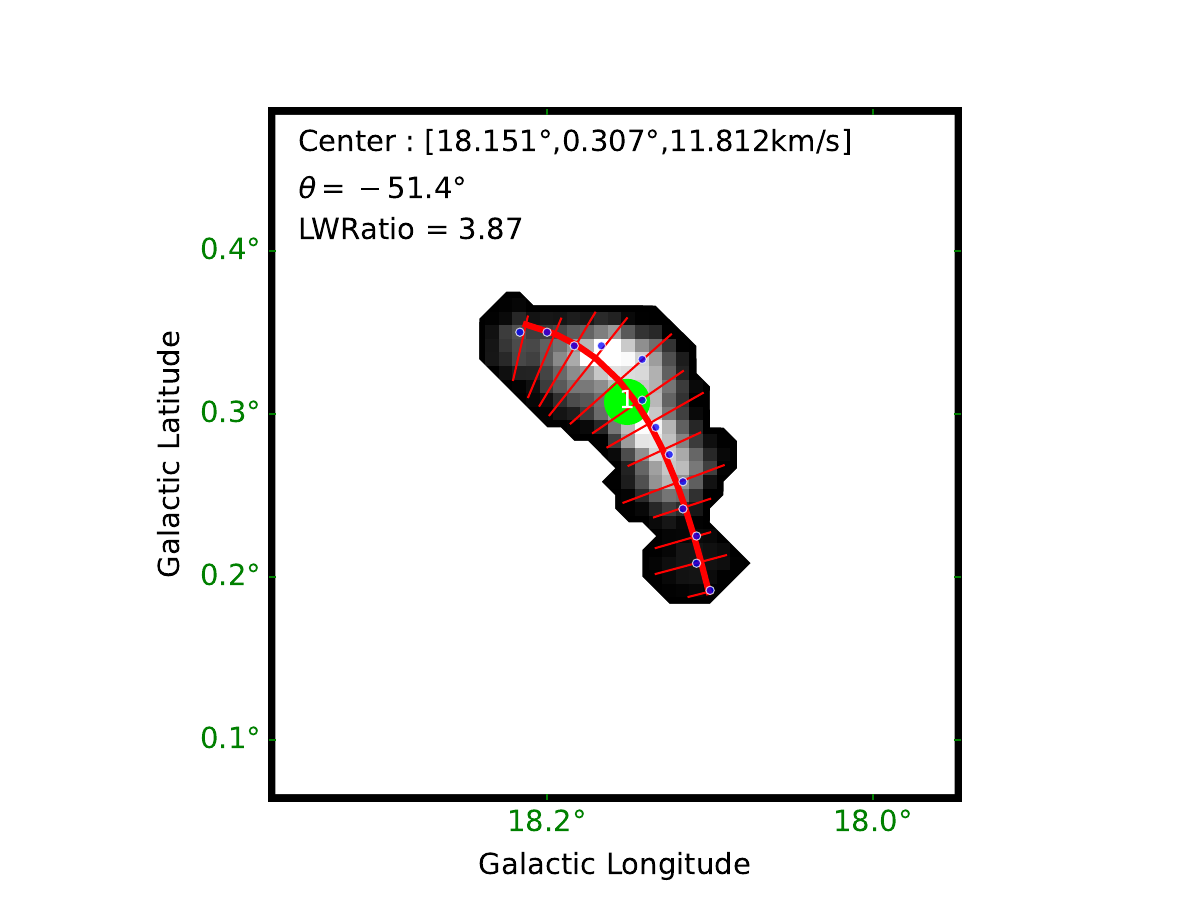}}
\end{minipage}%
\begin{minipage}[t]{0.24\textwidth}
    \centering
    \centerline{\includegraphics[width=2.1in]{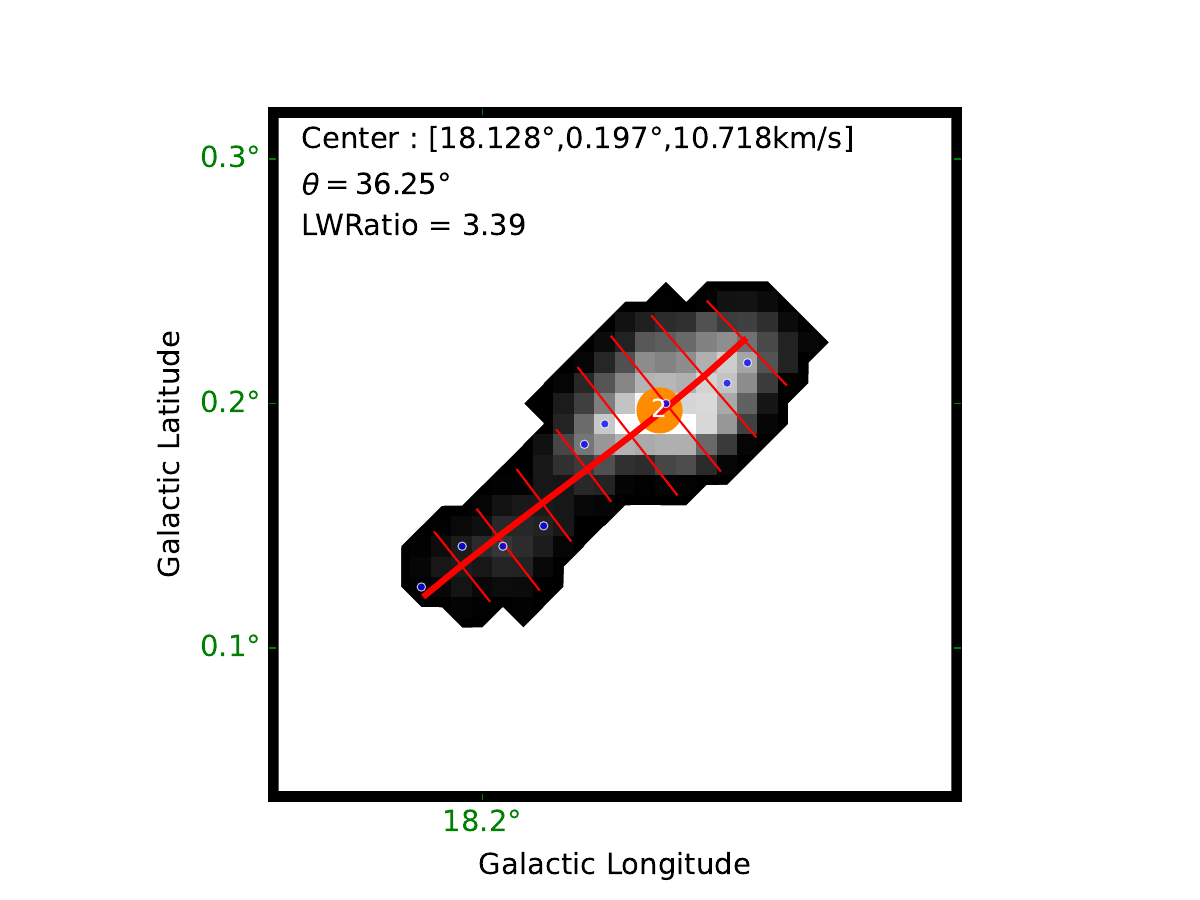}}
\end{minipage}%
\begin{minipage}[t]{0.24\textwidth}
    \centering
    \centerline{\includegraphics[width=2.1in]{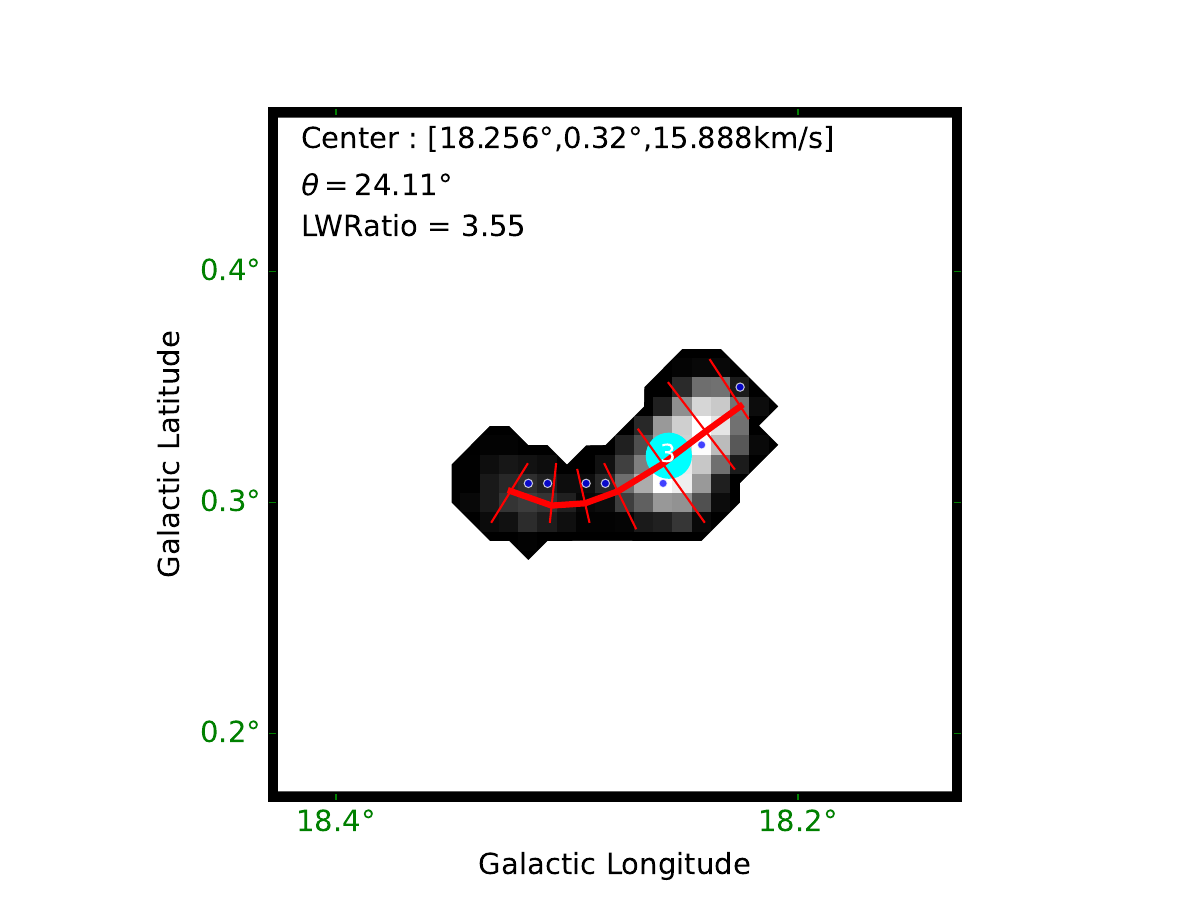}}
\end{minipage}%
\begin{minipage}[t]{0.24\textwidth}
    \centering
    \centerline{\includegraphics[width=2.1in]{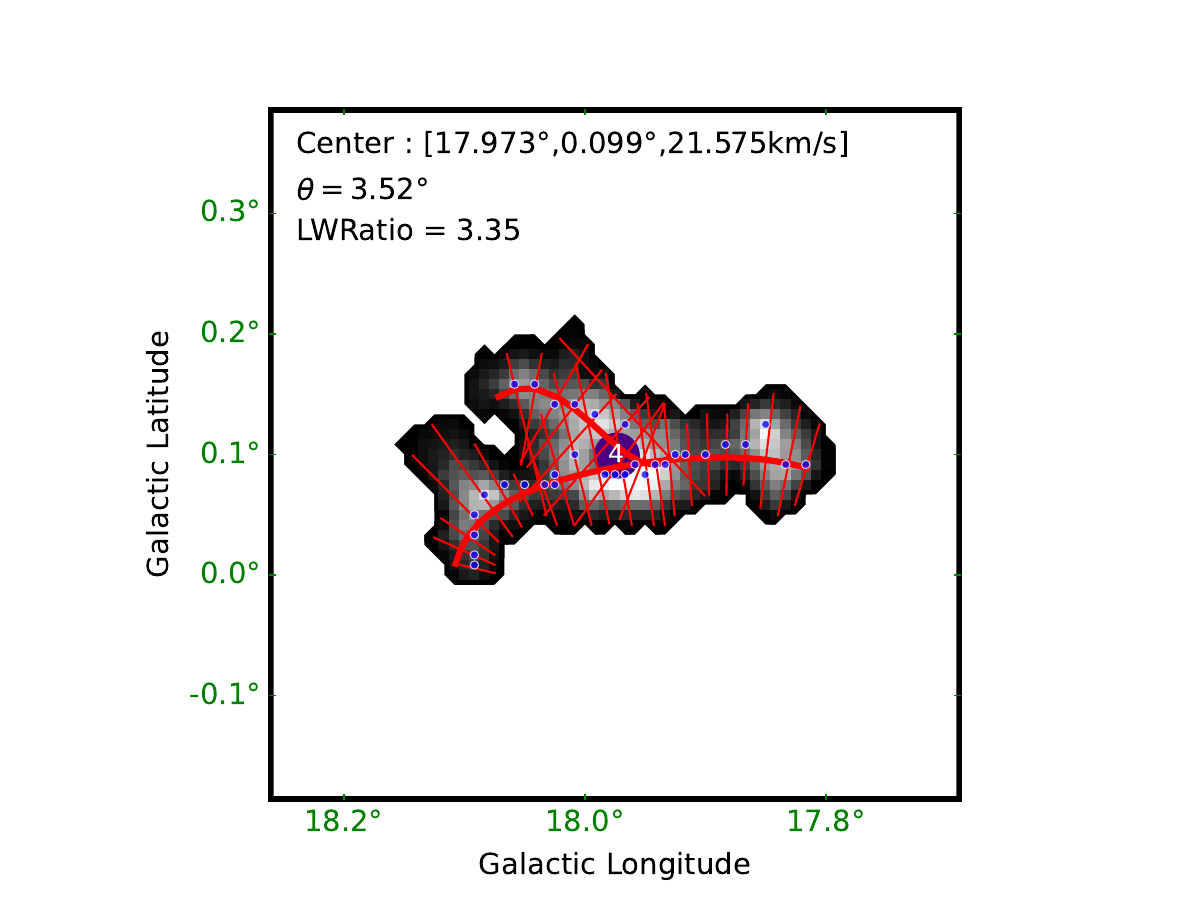}}
\end{minipage}%

\begin{minipage}[t]{0.24\textwidth}
    \centering
    \centerline{\includegraphics[width=2.1in]{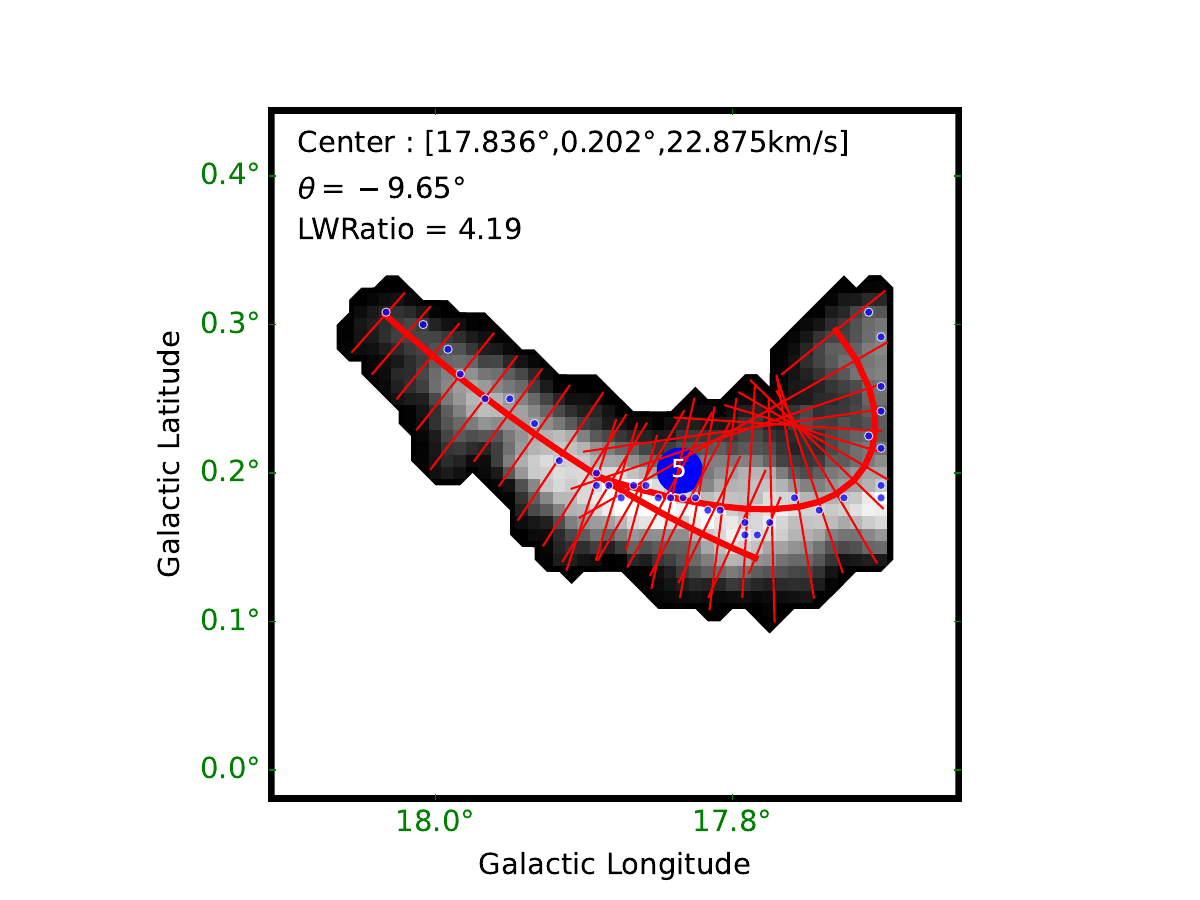}}
\end{minipage}%
\begin{minipage}[t]{0.24\textwidth}
    \centering
    \centerline{\includegraphics[width=2.1in]{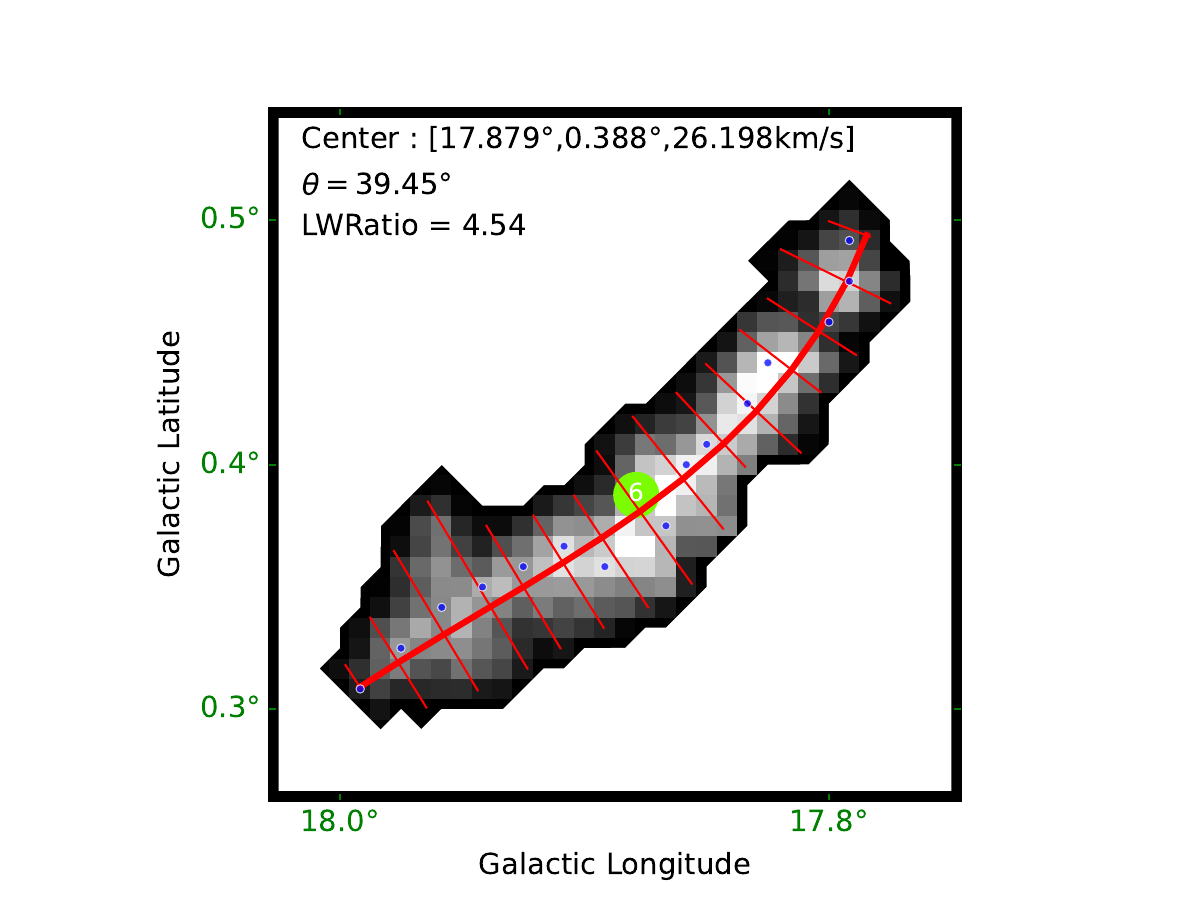}}
\end{minipage}%
\begin{minipage}[t]{0.24\textwidth}
    \centering
    \centerline{\includegraphics[width=2.1in]{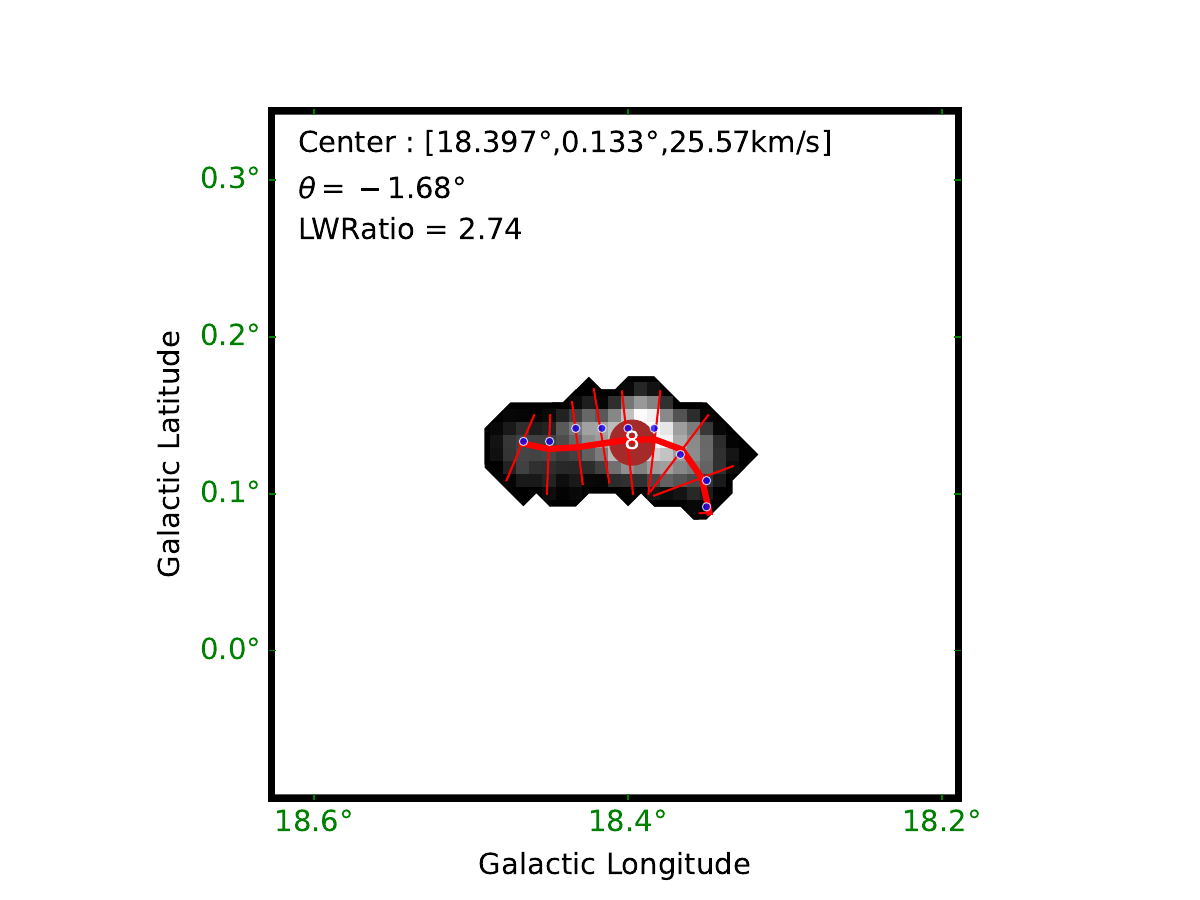}}
\end{minipage}%
\begin{minipage}[t]{0.24\textwidth}
    \centering
    \centerline{\includegraphics[width=2.1in]{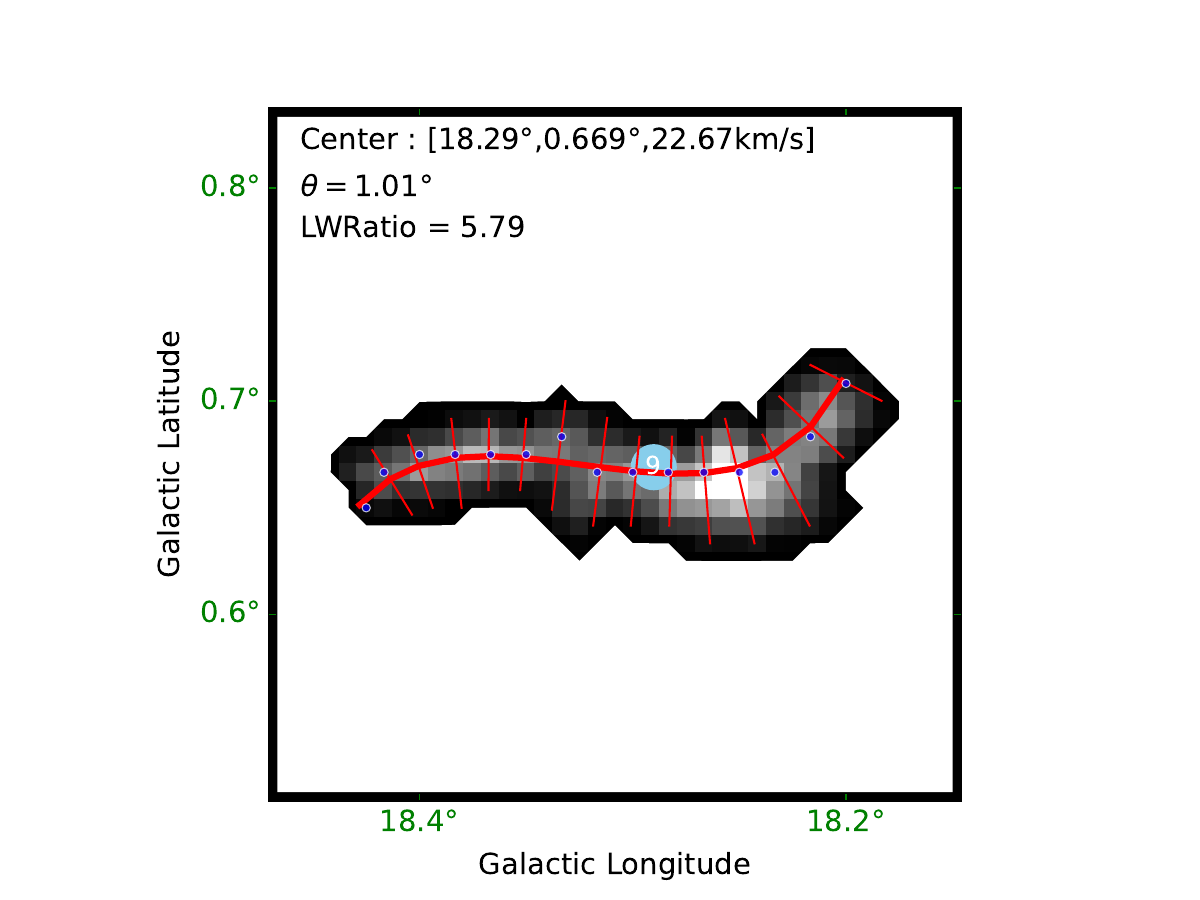}}
\end{minipage}%
\caption{Velocity-integrated intensity images of filaments in Figure \ref{Filaments_All}. The thicker red curves represent the fitted intensity skeletons, while the thinner red straight lines represent the profiles. The blue scatter points denote the peak intensity pixels on the profiles. The numbered circles of different colors indicate the positions of different filaments, and are consistent with the labels in Figure \ref{Filaments_All}. Filaments 2, 3, 6, 8, and 9 are small scale.}
\label{Filament_Fit_Spine_Items}
\end{figure*}

We showcase the substructures, fitted intensity skeletons, and profiles of the filaments identified in the example data in Figure \ref{Filament_Fit_Spine_Items}. In addition, the figures also display the central coordinates of the filaments in the PPV space, as well as the angles and aspect ratios in the PP space. Filaments 2, 3, 6, 8, and 9 are small-scale filaments with fewer than five clumps. 

\begin{figure*}
\centering
\centerline{\includegraphics[width=4in]{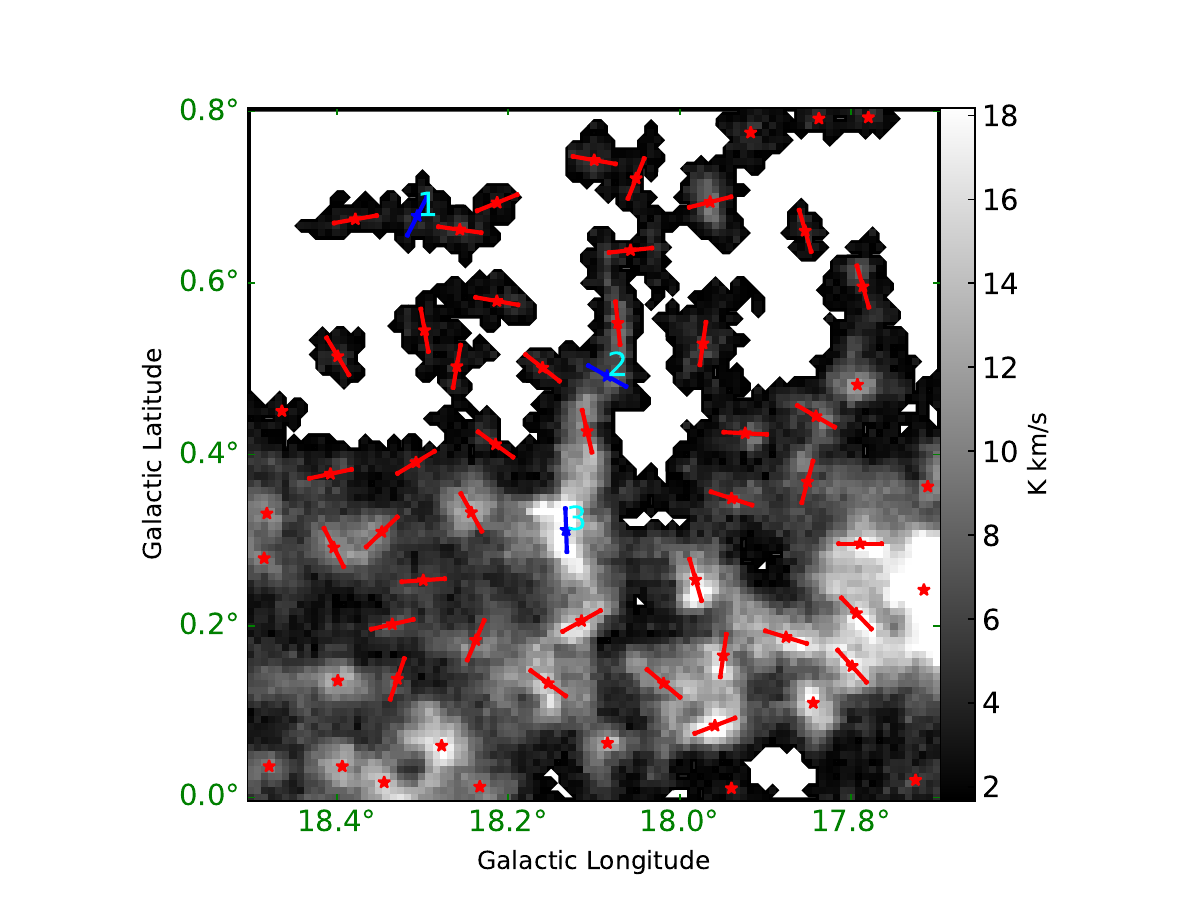}}
\caption{The clumps detected by FacetClumps in the velocity-integrated map. The intensity threshold is 1.6 K km s$^{-1}$. The asterisks denote the positions of the clumps, the lines denote the direction of the principal axis of the clumps that have not touched the edge. Clumps 1, 2, and 3 are examples to illustrate the large shifts in direction and position.}
\label{ClumpsInfor_2D_Img}
\end{figure*}

\section{Clumps of the Example Data in the PP Space}\label{ClumpsInfor_2D}
To assess the impact of integration effects on the direction and position of clumps, we use FacetClumps for clump detection on the velocity-integrated intensity maps, as shown in Figure \ref{ClumpsInfor_2D_Img}. Clumps 1 and 2 are located similarly to the clumps depicted in Figure \ref{Clumps_Infor} for Filaments 9 and 7, as illustrated in Figures \ref{Filaments_All} and \ref{Filament_Fit_Spine_Items}. However, noise has caused their orientations to diverge from the local axes of the filaments. Clump 3 corresponds to the combined signal of Filaments 1 and 7 at different velocities, and its position also deviates from the local axis.

The clumps in Figure \ref{Clumps_Infor} and Figure \ref{GiantFilaments} indicate that consistency between the direction and position of clumps and the local axis of the filaments in PPV space is prevalent, regardless of whether the filaments are large scale or small scale. Because of the integration effect of signals at different velocity channels, the consistency in the PP space is significantly reduced. Consequently, the consistency-based identification method will need to accommodate larger angle and distance tolerances in order to accurately identify the filaments.

\section{Comparison: The Structures Isolated by FilFinder, DisPerSE, and MST}\label{Comparison}

\begin{figure*}
\centering
\vspace{0cm}
\begin{minipage}[t]{0.4\textwidth}
    \centering
    \centerline{\includegraphics[width=2.7in]{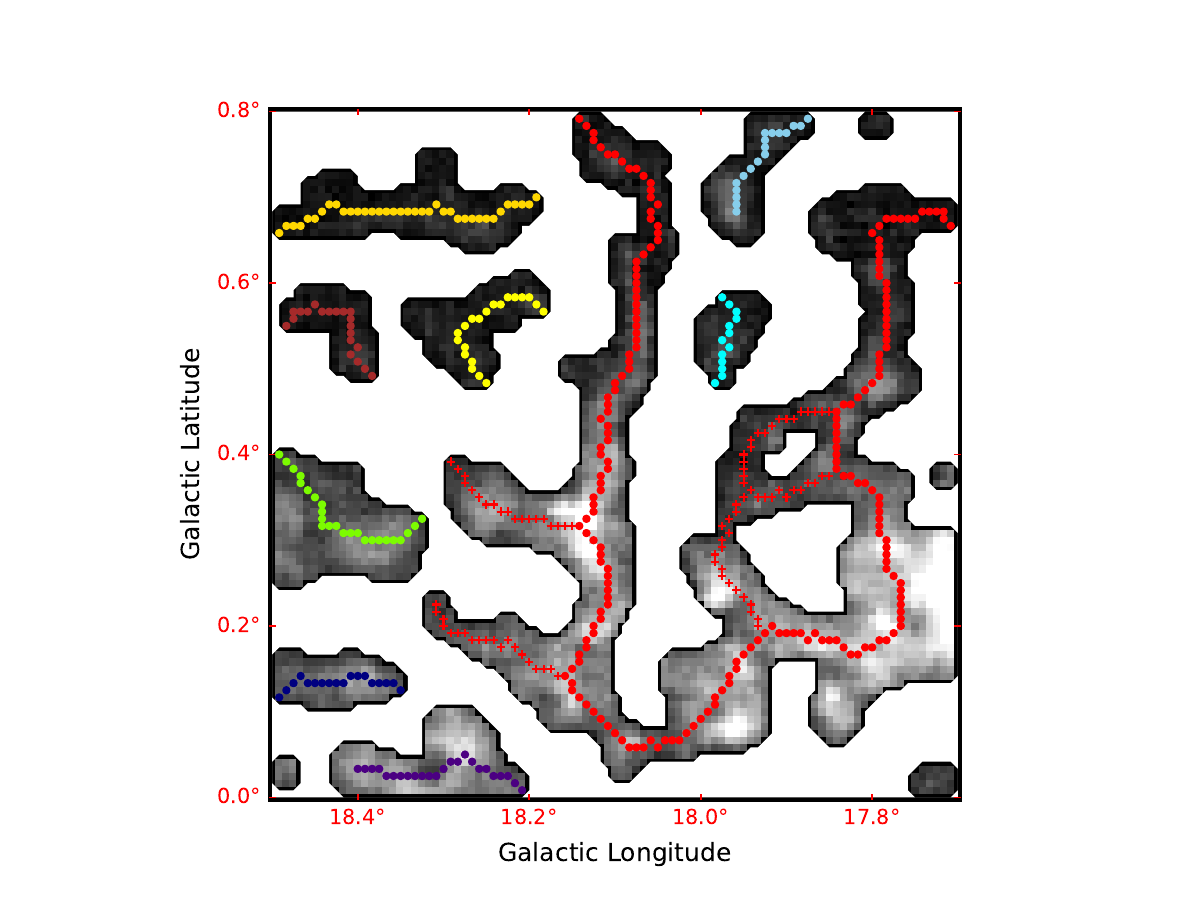}}
\end{minipage}
\begin{minipage}[t]{0.4\textwidth}
    \centering
    \centerline{\includegraphics[width=2.7in]{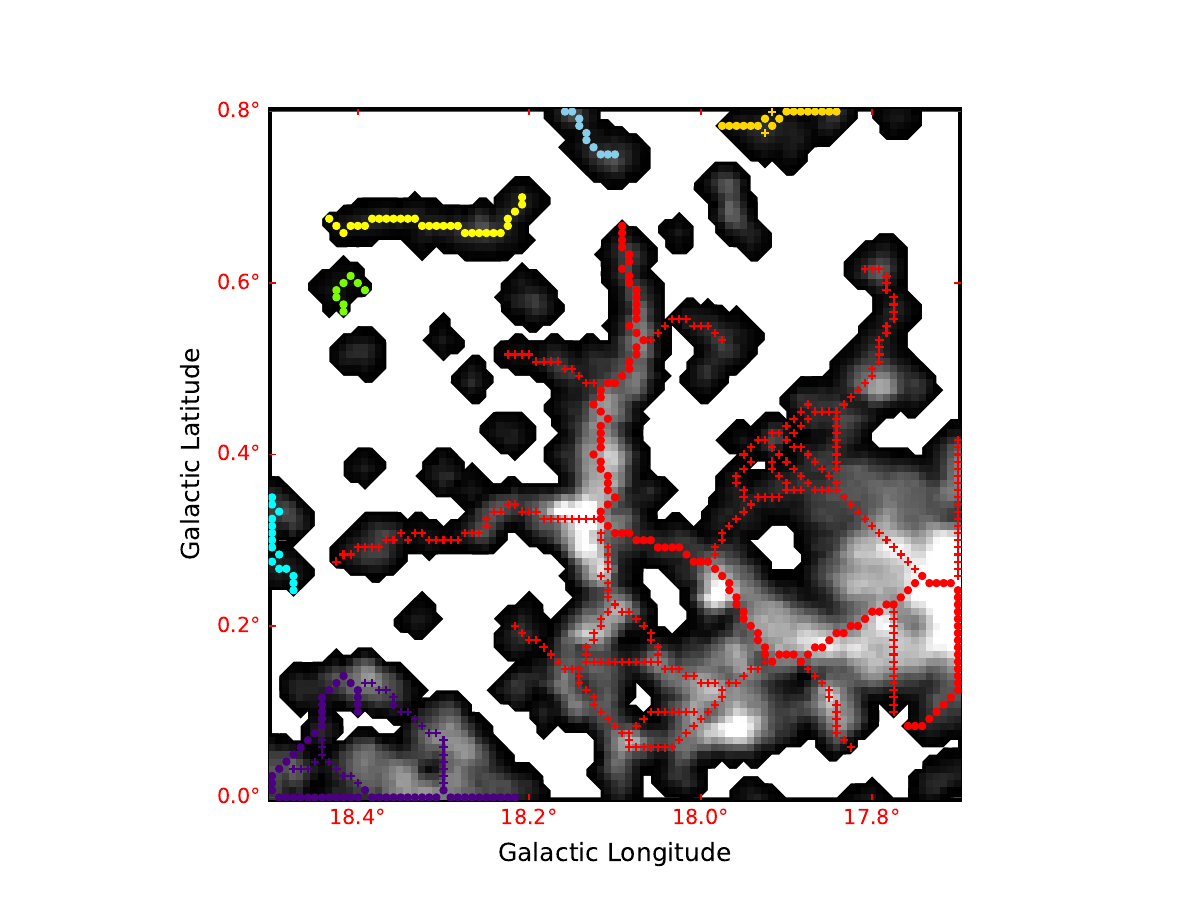}}
\end{minipage}
\caption{The structures isolated by the FilFinder algorithm. The left panel shows the skeletons extracted from the mask obtained by FilFinder on the velocity-integrated map, with a global threshold of 0.8 K km s$^{-1}$. The right panel shows the skeletons extracted from the mask of the integrated clumps. Different colors denote different skeletons, with circles indicating the longest skeleton and plus signs denoting branches. The minimum length threshold for both the skeletons and branches is set to 10 pixels.}
\label{FilFinder_Structures}
\end{figure*}

\subsection{The Algorithm in the PP Space: FilFinder}\label{Comparison_FilFinder}
FilFinder \citep{FilFinder} first flattens the image using an arctangent transform, determining the mean and standard deviation in the transformed space by fitting a log-normal distribution to the brightness data. The flattened data are then smoothed with a Gaussian (generally using an empirical FWHM, 1 pixel). An adaptive threshold method is applied to create a mask, requiring the central pixel's intensity in the smoothed data to exceed the median of its neighborhood. This mask is combined with a globally thresholded mask to remove regions below the noise level. FilFinder then reduces each structure within the mask to a skeleton using a medial axis transform, which is further pruned to highlight the dominant features. 

FilFinder is a commonly used algorithm that can be employed for filament identification in 2D data. The structures isolated by FilFinder in the PP space are shown in Figure \ref{FilFinder_Structures}. As a result of integration effects, filaments with different velocity components, such as Filaments 1 and 7 in Figure \ref{Filaments_All}, are inevitably identified as the same structure by FilFinder. Furthermore, the skeleton of FilFinder, extracted from masks without intensity information, is distributed along the central axis of the region and may not accurately represent the skeleton of asymmetrical filaments.

\begin{figure*}
\centering
\vspace{0cm}
\begin{minipage}[t]{0.35\textwidth}
    \centering
    \centerline{\includegraphics[width=2.8in]{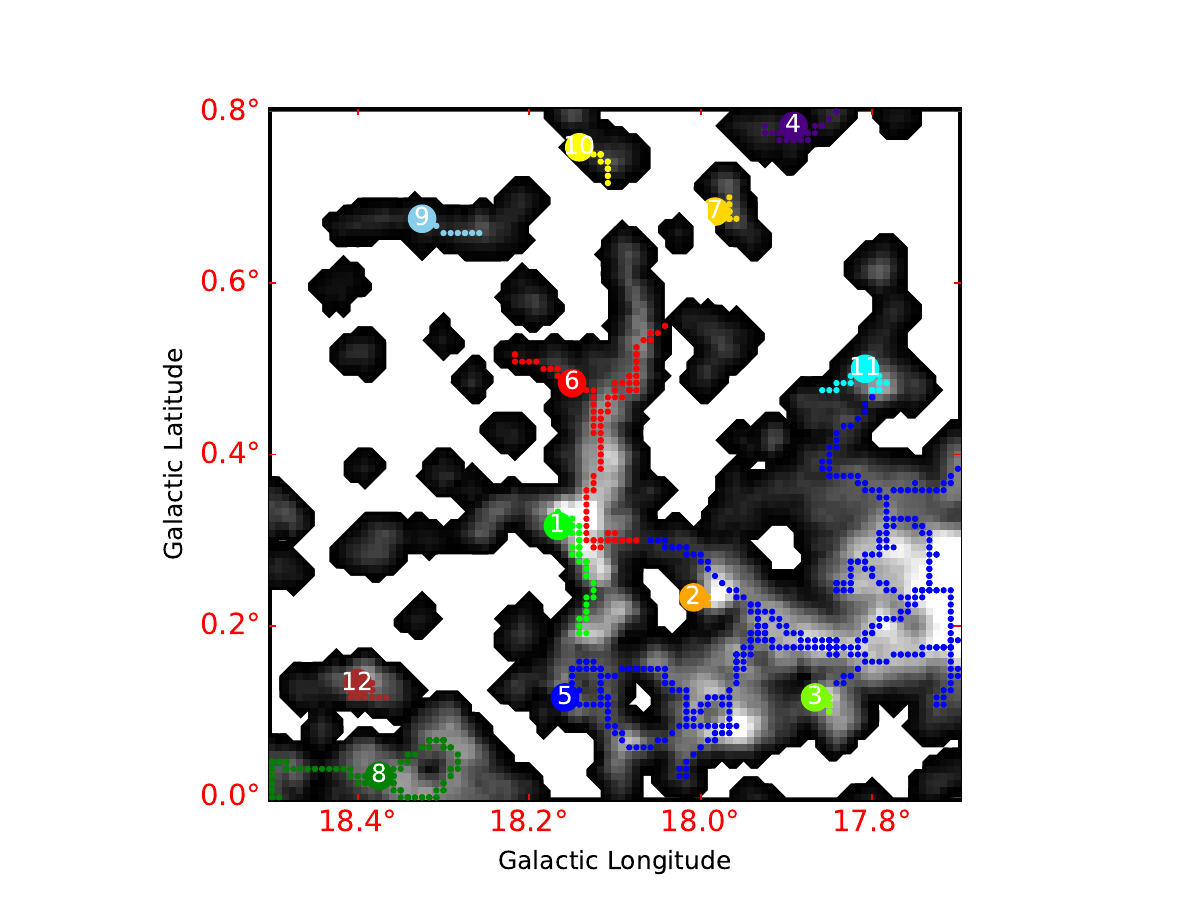}}
\end{minipage}
\begin{minipage}[t]{0.35\textwidth}
    \centering
    \centerline{\includegraphics[width=2.8in]{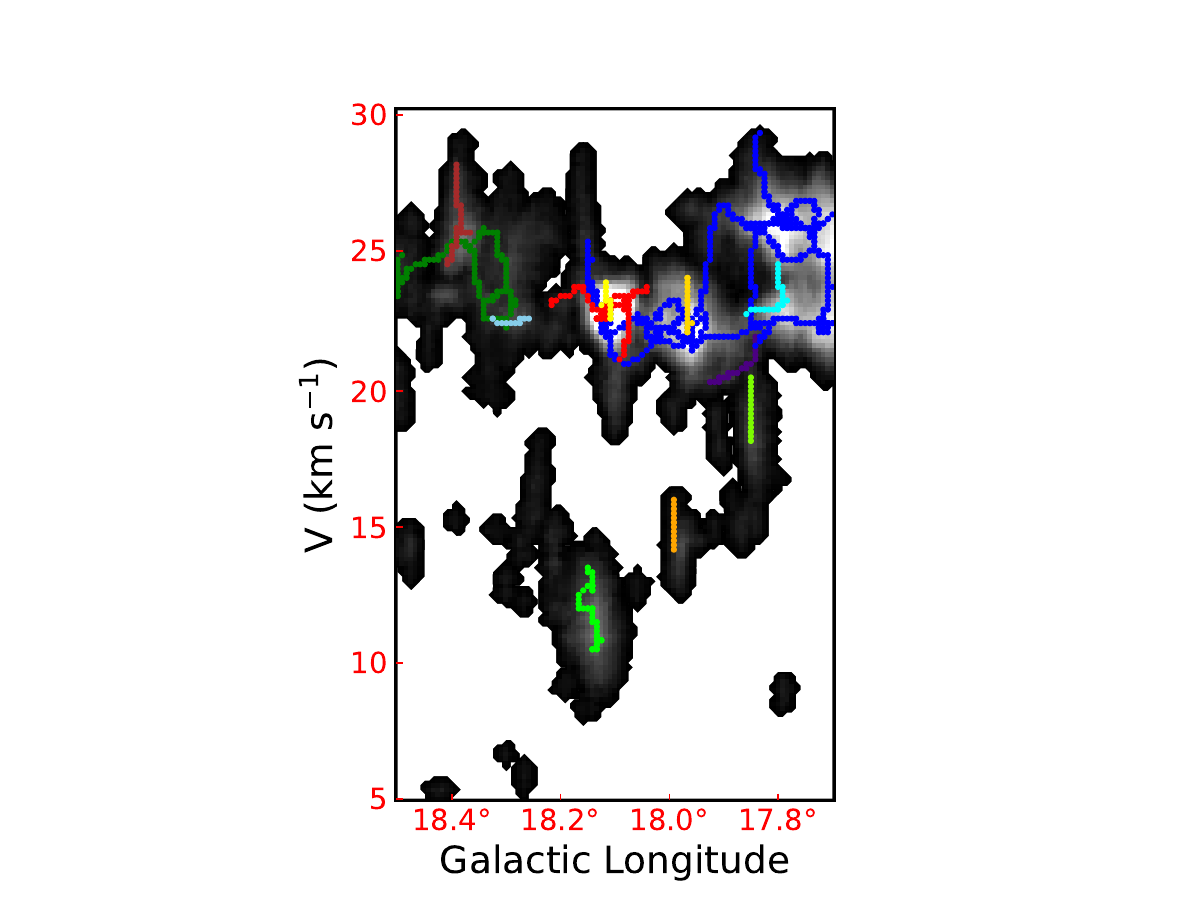}}
\end{minipage}%
\begin{minipage}[t]{0.20\textwidth}
    \centering
    \centerline{\includegraphics[width=2.8in]{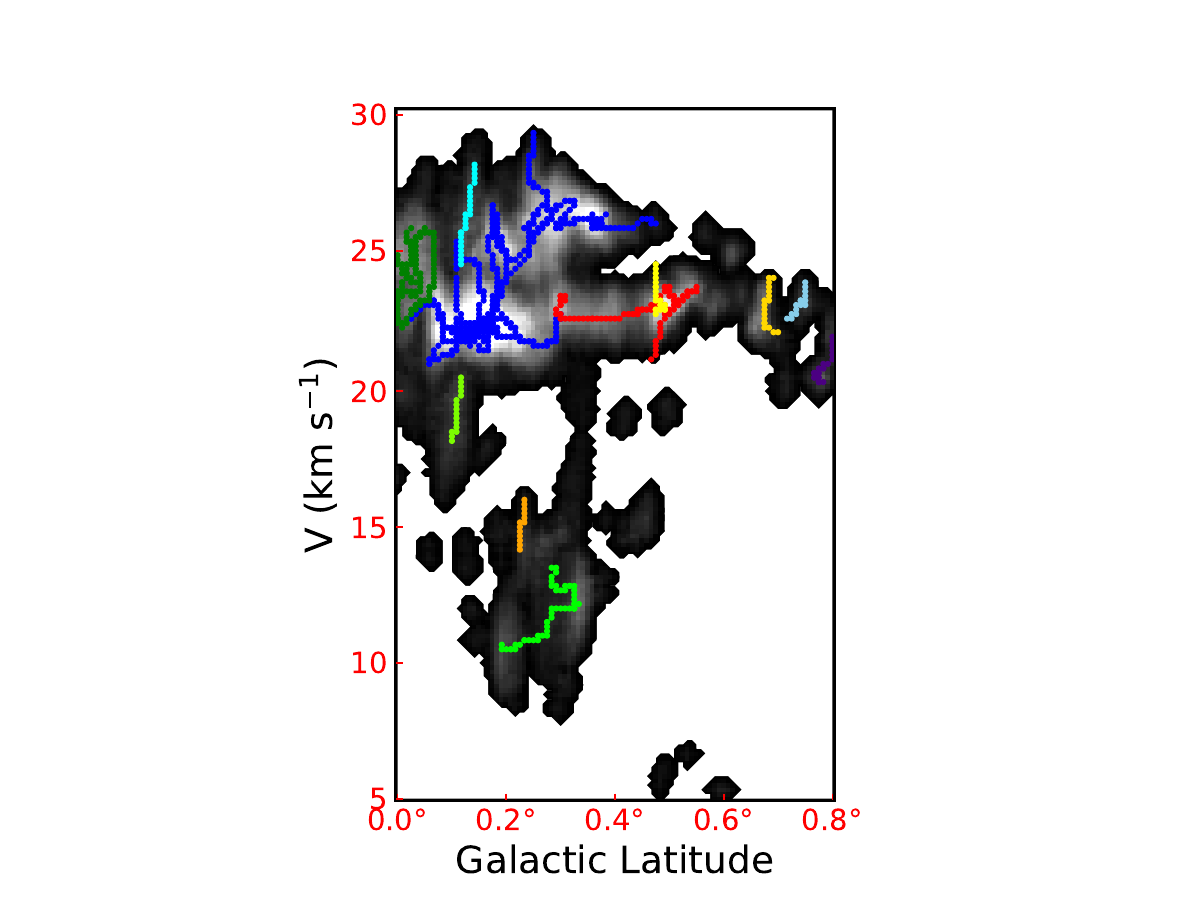}}
\end{minipage}%
\caption{The structures isolated by the DisPerSE algorithm. Left panel: velocity-integrated intensity map. Curves in different colors denote different skeletons, each labeled with a corresponding number in the same color. Middle panel: latitude-integrated intensity map. Right panel: longitude-integrated intensity map. The minimum length threshold for skeletons is set to 10 pixels.}
\label{DisPerSe_Structures}
\end{figure*}

\subsection{The Algorithm in the PPP Space: DisPerSE}\label{Comparison_DisPerSe}
DisPerSE \citep{DisPerse,DisPerse_2} leverages discrete Morse theory to analyze a density field by treating it as a topological terrain with peaks, valleys, and saddle points. It first constructs a Delaunay triangulation to provide a precise mesh of points for evaluating gradients. Critical points such as maxima (peaks) and saddle points (intermediate structures) are identified, and filaments are formed by connecting these points with smooth curves. DisPerSE uses persistence to differentiate between significant structures and noise by comparing intensity differences between critical points, and assesses robustness to ensure that detected filaments are well defined against the background.

DisPerSE is originally developed for analyzing the PPP structures of cosmological simulations and has since been widely adopted to identify ISM filaments in 2D maps, such as those of infrared data, dust continuum images, and molecular line integrations. However, it has not been widely applied to PPV data on account of its limitation in distinguishing between spatial and velocity axes. We apply DisPerSE to the cube after filtering out noise with the mask of clumps, with both persistence and robustness thresholds set at 1.1 K (equivalent to $5\times RMS$). This identification is considered to be conducted in the PPP space, and the results are presented in Figure \ref{DisPerSe_Structures}. 

Since DisPerSE outputs a one-dimensional filamentary skeleton, additional processing is required to obtain the region of filament. When there are great various emission strengths, DisPerSE tends to extract skeletons with lower integrity in faint emission regions \citep{Confusion_PPV,FilFinder}. Additionally, because the high velocity resolution of MWISP, the filamentous structures observed in the velocity channels may not represent real features. As illustrated by Filaments 2, 3, 7, and 12 (where 12 corresponds to Filament 8 in Figure \ref{Filament_Fit_Spine_Items}), DisPerSE may detect false filaments primarily dominated by stretching in the velocity direction.

\subsection{The Algorithm in the PPV Space: MST}\label{Comparison_MST}
\begin{figure*}
\centering
\vspace{0cm}
\begin{minipage}[t]{0.22\textwidth}
    \centering
    \centerline{\includegraphics[width=2.7in]{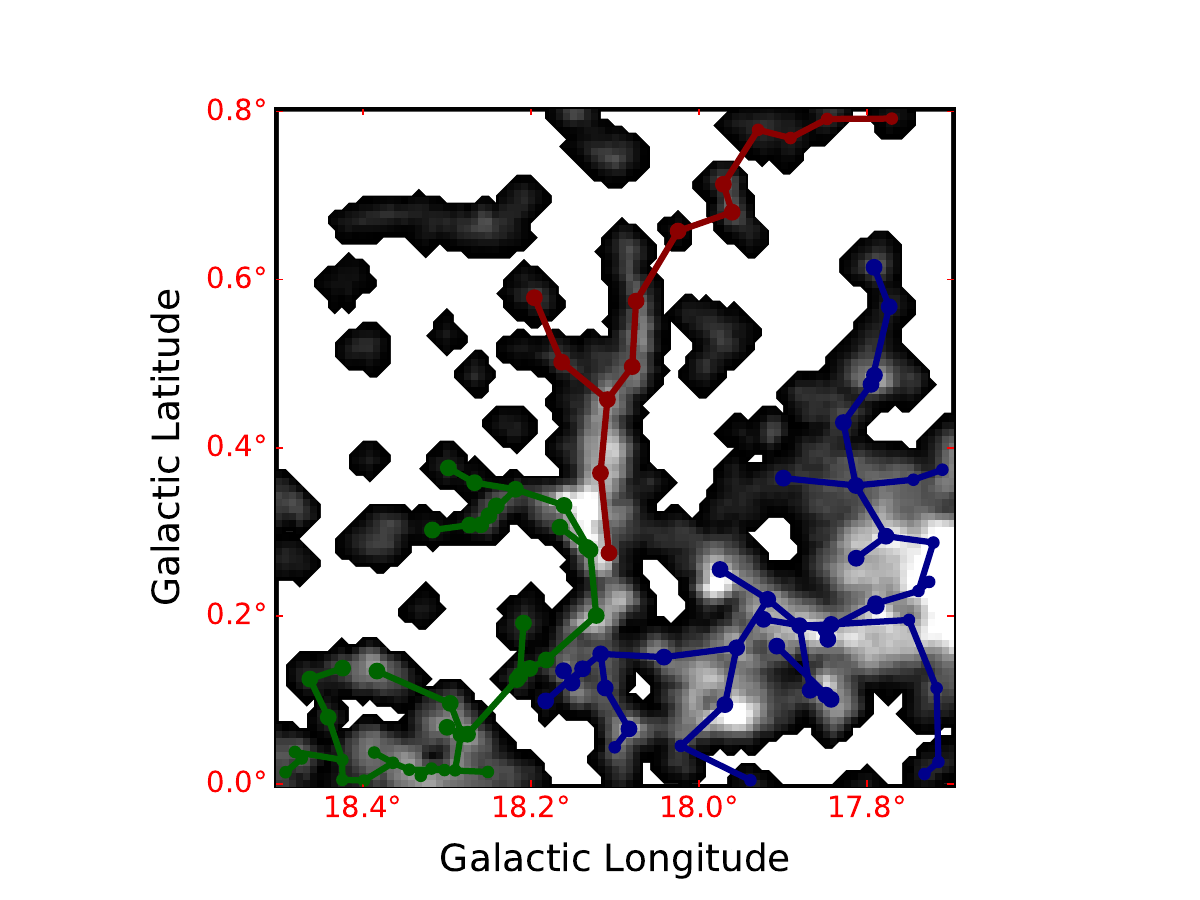}}
\end{minipage}
\begin{minipage}[t]{0.22\textwidth}
    \centering
    \centerline{\includegraphics[width=2.7in]{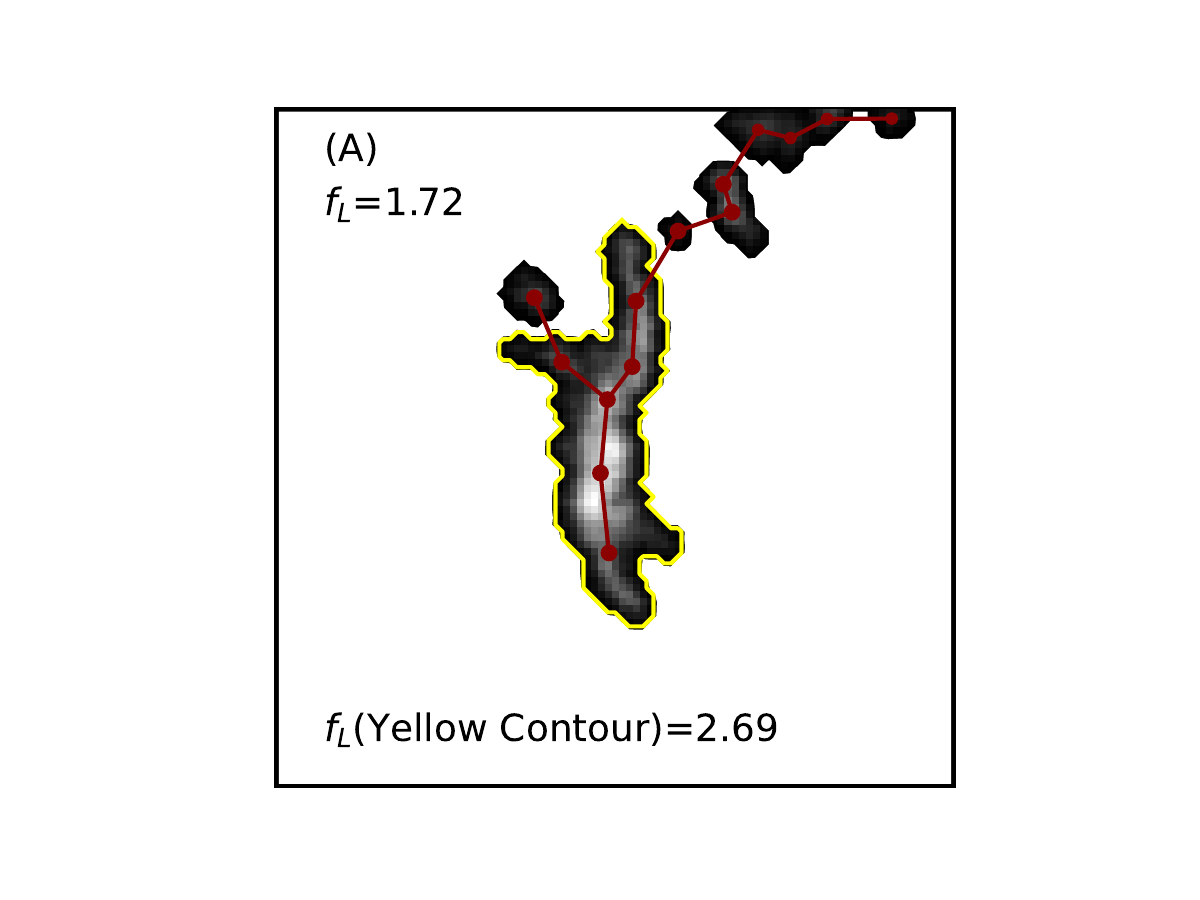}}
\end{minipage}
\begin{minipage}[t]{0.22\textwidth}
    \centering
    \centerline{\includegraphics[width=2.7in]{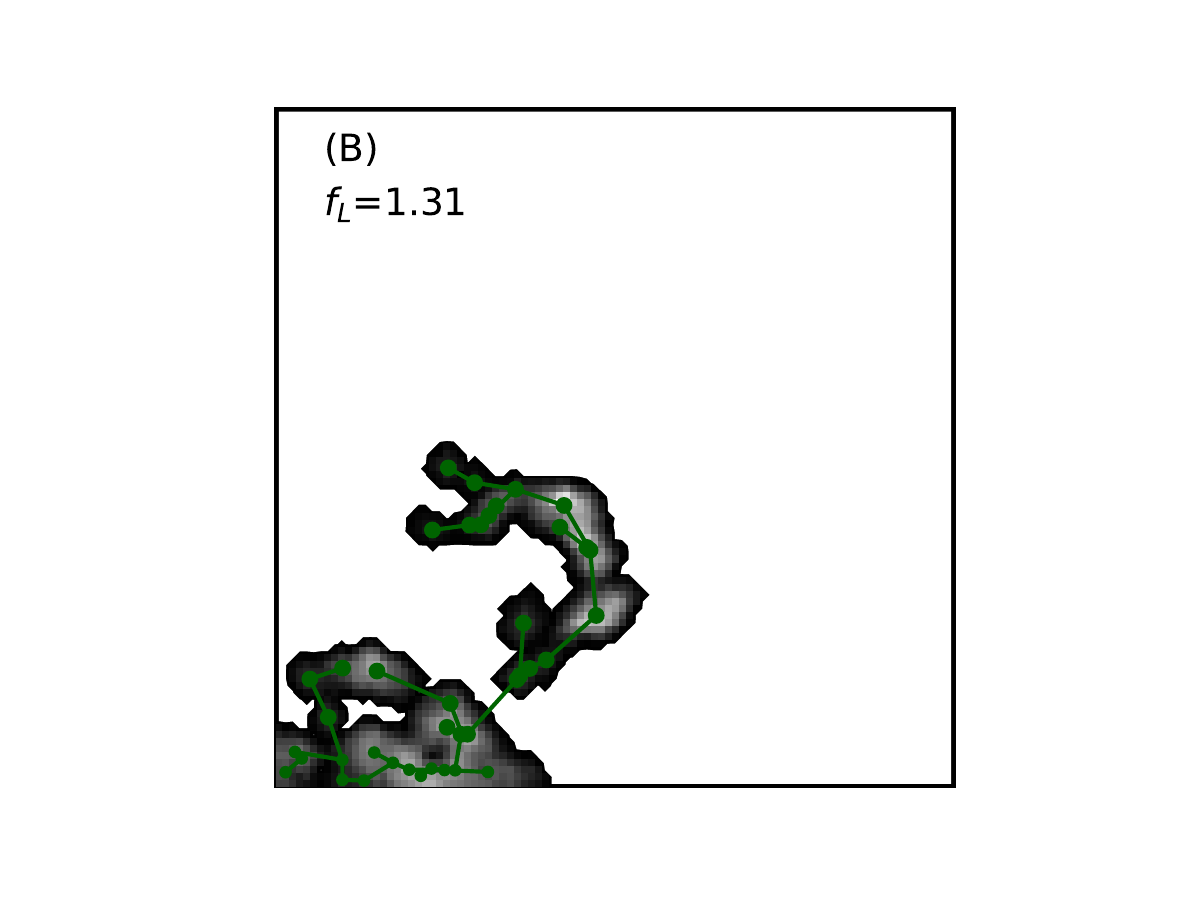}}
\end{minipage}
\begin{minipage}[t]{0.22\textwidth}
    \centering
    \centerline{\includegraphics[width=2.7in]{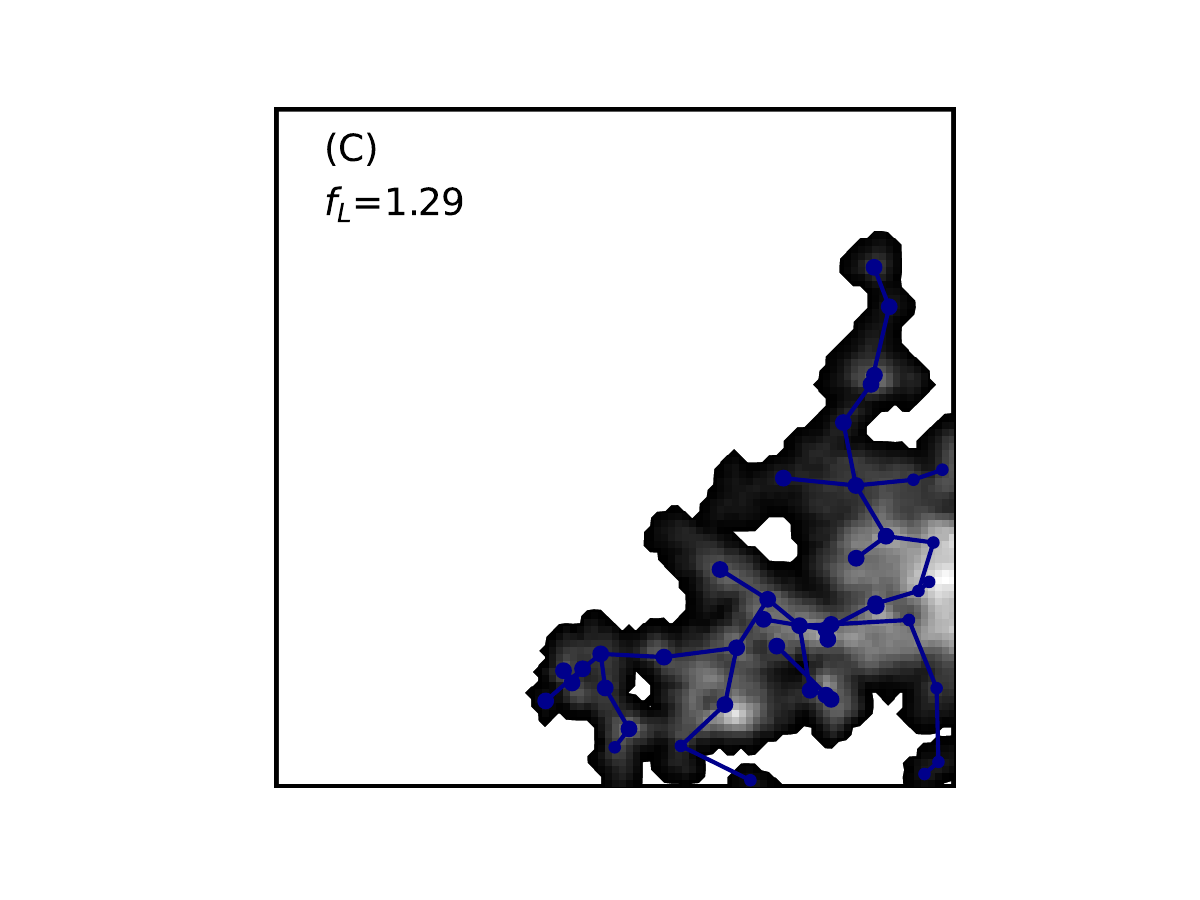}}
\end{minipage}
\caption{The structures isolated by the MST algorithm. In the left panel, lines of various colors interconnect distinct coherent structures, while circles denote the spatial positions of clumps within these structures. The background is the velocity-integrated intensity map of all clumps. Panels (A), (B), and (C) respectively depict three coherent structures alongside their respective linearities. In Panel (A), the linearity of the filament outlined in yellow is also computed.}
\label{MST_Structures}
\end{figure*}

MST \citep{MST} first connects all clumps in the catalog according to their positions in the PPV space, and then filters out widely separated clumps to isolate coherent filaments. Here, "coherence" refers to being "close proximity" in both spatial and velocity direction, meaning that the spatial distance difference between any two conjoint clumps is less than $\Delta L$ (0.1$^\circ$), and the velocity difference in the velocity channel is less than $\Delta V$ (2 km s$^{-1}$). All coherent structures identified by MST in the example data with default parameters are shown in Figure \ref{MST_Structures}. A further linearity check is conducted, and structures (B) and (C) with low linearity ($f_L<1.5$) will be excluded. 

Structures identified by MST are usually disconnected, which may be disadvantageous for the skeleton and profile-related analysis. Additionally, in Figure \ref{MST_Structures} (A), we compute the linearity ($f_L=2.69$) of the example filament outlined by the yellow contour mentioned earlier, and find it to be greater than the linearity ($f_L=1.72$) of the structure identified by MST. Linearity cannot effectively assess filaments that exhibit curvature, and disconnected subcomponents may reduce the linearity of structures in densely overlapping areas, thereby tending to exclude those that contain genuine filaments. 

\section{Filaments in the Application Data}\label{ApplicationImages}
\begin{figure}
\centering
\vspace{0.5cm}
\begin{minipage}[t]{1\textwidth}
    \centering
    \centerline{\includegraphics[width=9.5in]{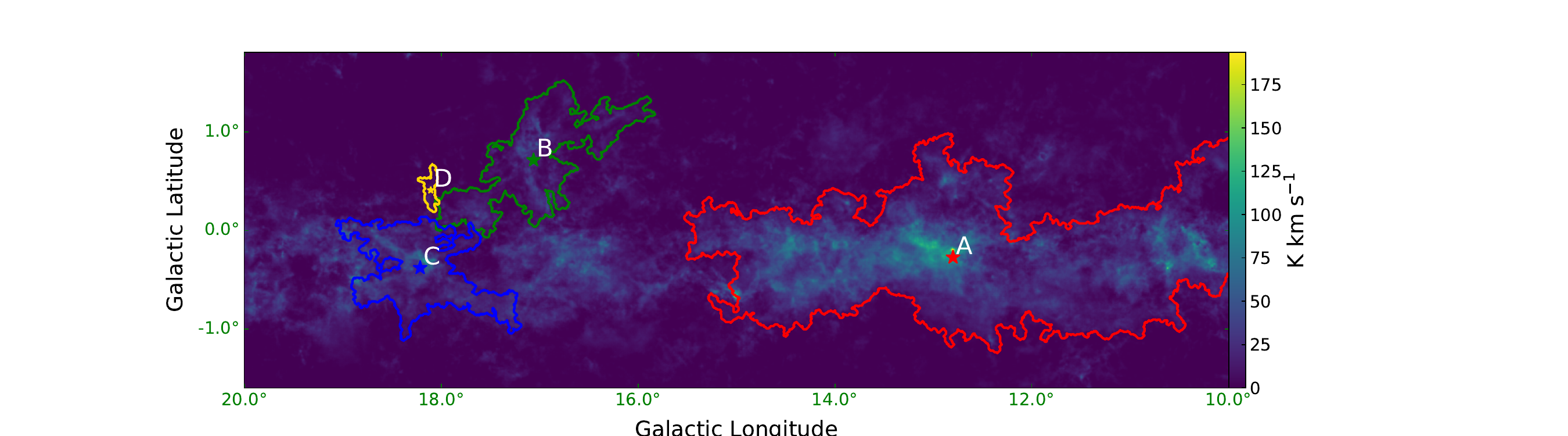}}
    \vspace{-4cm}/
\end{minipage}
\caption{The area of the largest molecular cloud and filaments identified by DPConCFil in the application data. The map is the $^{13}$CO emission of MWISP within $10^{\circ} \leq l \leq 20^{\circ}$, $-1.6^{\circ} \leq b \leq 1.8^{\circ}$ and -25 km s$^{-1}$ $\leq v \leq$ 90 km s$^{-1}$, and is preprocessed employing the signal region extraction method \citep{FacetClumps}. GMF A is the largest giant filament. GMFs B and C are illustrated in Figure \ref{GiantFilaments}, while Filament D is a common filament present in both the example and application data, as previously shown in Figure \ref{Filament_Example_Item}. The central positions of these filaments are denoted by asterisks of different colors, and the boundaries are delineated by contours of different colors. } 
\label{Application_Data} 
\end{figure}

By employing the signal region extraction method proposed by \cite{FacetClumps}, we  have obtained a contiguous giant molecular cloud in the application data, spanning an approximately $10^{\circ} \leq l \leq 20^{\circ}$, $-1.5^{\circ} \leq b \leq 1.7^{\circ}$ and -20 km s$^{-1}$ $\leq v \leq$ 85 km s$^{-1}$ area. The velocity-integrated intensity map for this area, which extends slightly beyond the defined range, is shown in Figure \ref{Application_Data}. To intuitively feel the difference between filaments of various angular scales, we have marked the three significantly elongated GMFs and one of the common filaments found in both the example data and application data in Figure \ref{Application_Data}. 

\begin{figure*}
\centering
\vspace{0cm}
\begin{minipage}[t]{0.4\textwidth}
    \centering
    \centerline{\includegraphics[width=3.5in]{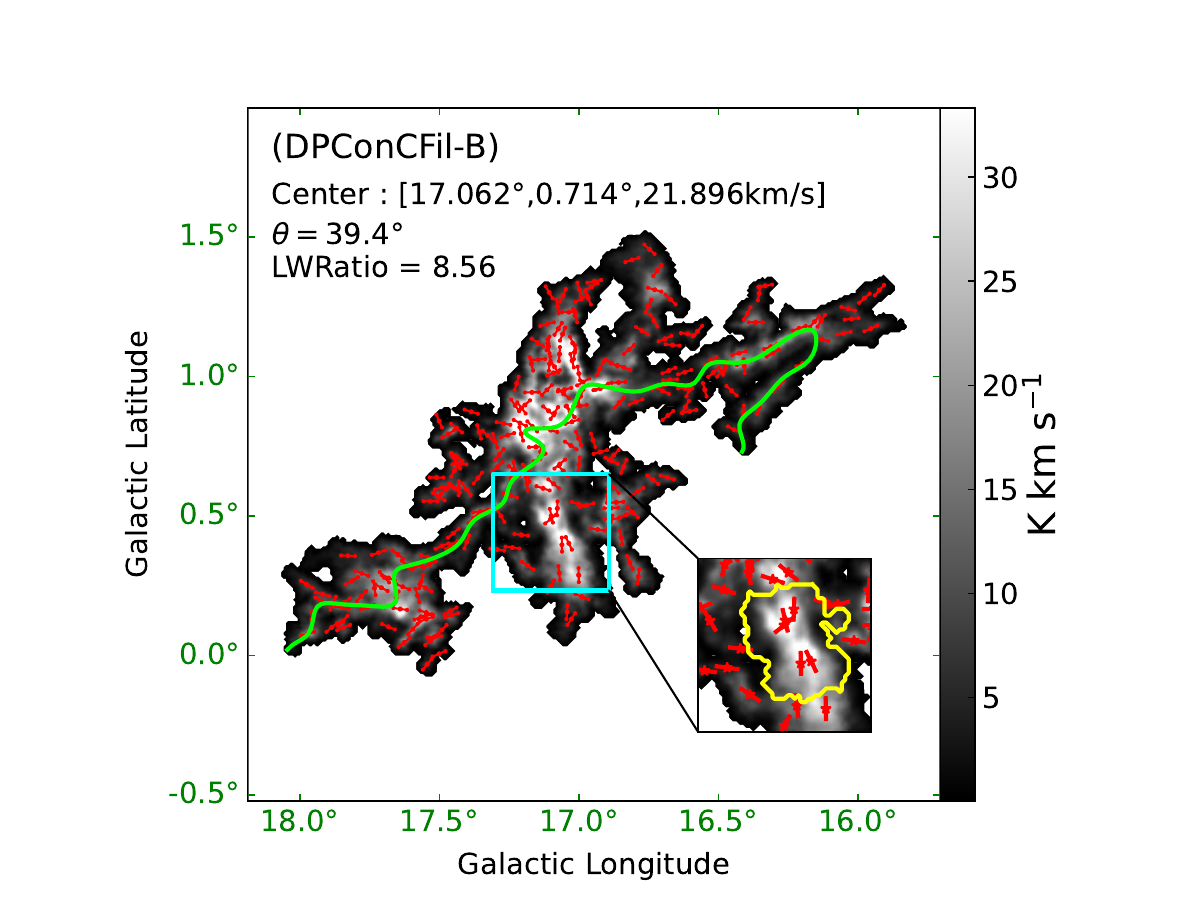}}
\end{minipage}
\begin{minipage}[t]{0.45\textwidth}
    \centering
    \centerline{\includegraphics[width=3.5in]{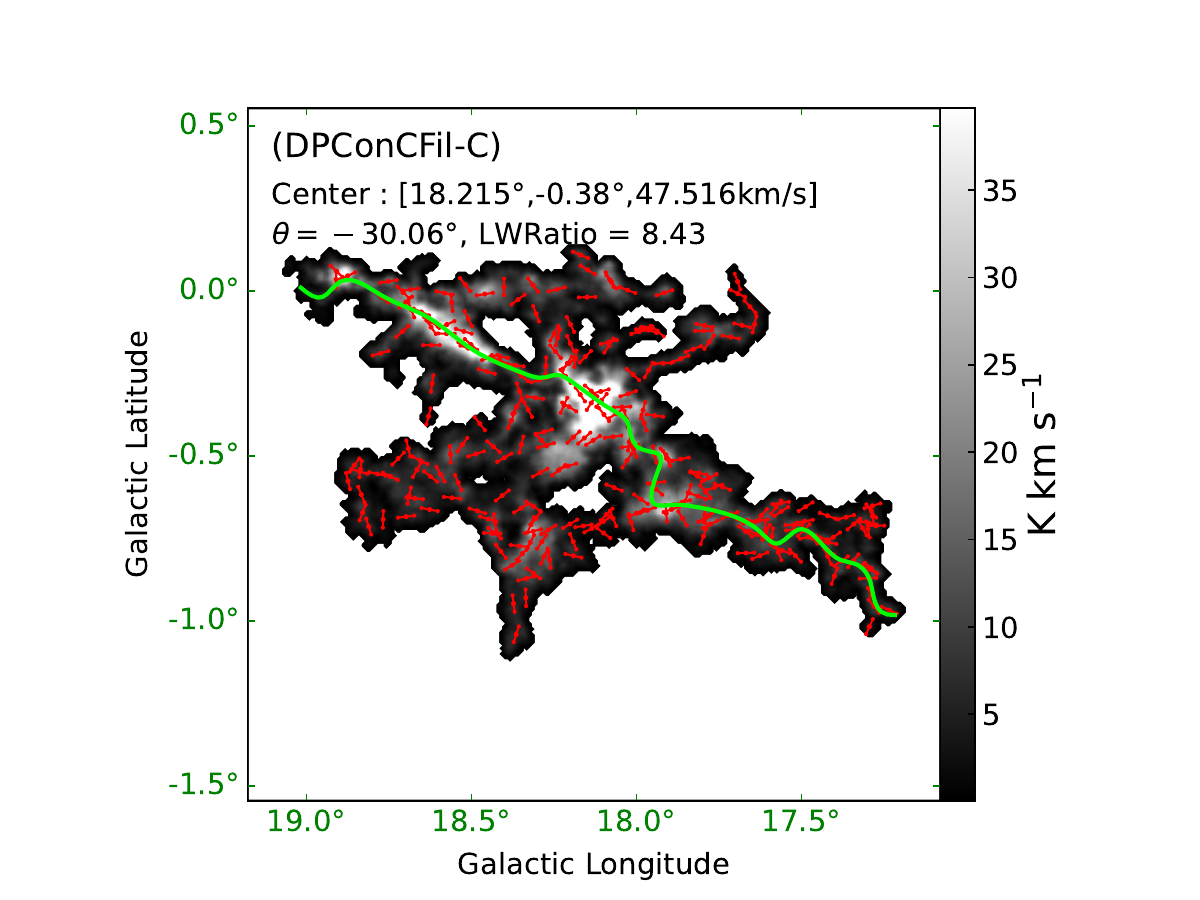}}
\end{minipage}

\begin{minipage}[t]{0.5\textwidth}
    \centering
    \centerline{\includegraphics[width=5in]{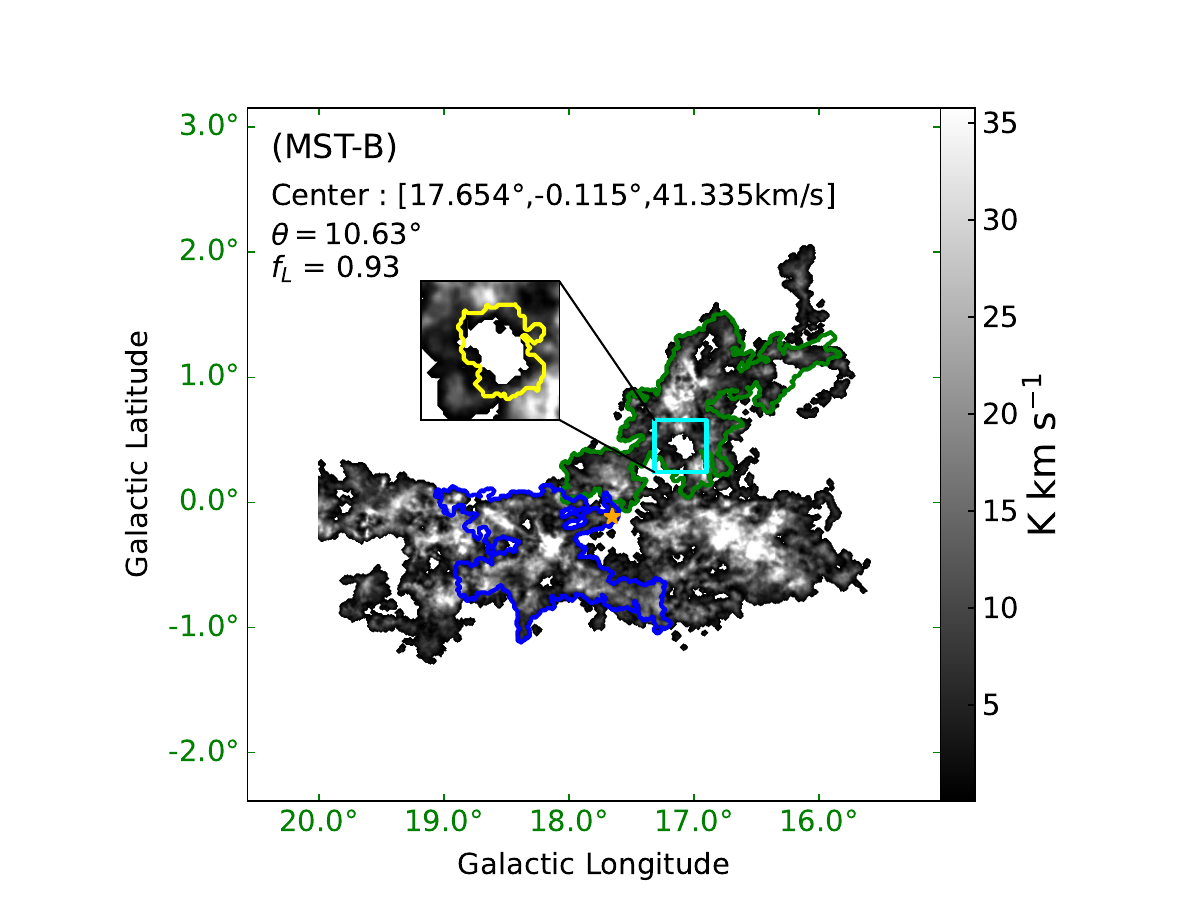}}
\end{minipage}
\caption{Giant structures isolated by DPConCFil and MST from the application data. DPConCFil-B, DPConCFil-C, and MST-B refer to the same filaments marked in Figure \ref{Distribution_LBV}. The number of clumps within the filaments are 240, 255, and 2525, respectively. In DPConCFil-B and DPConCFil-C, lime curves represent the longest fitted intensity skeletons, while red asterisks and line segments denote the positions and orientations of the clumps. In MST-B, orange asterisk denotes the intensity-weighted central position of the structure, and green and blue contours correspond to the boundaries of filaments in DPConCFil-B and DPConCFil-C, respectively. In DPConCFil-B and MST-B, cyan rectangles outline the same areas and provide a zoomed-in view, with yellow contours marking the boundaries of the clumps missing in MST-B within the rectangle.}
\label{GiantFilaments}
\end{figure*}

To visually compare the differences between large-scale filaments identified by DPConCFil and MST, we present GMF B and C from Figure \ref{Application_Data} in Figure \ref{GiantFilaments}, alongside the structures identified by MST that include these two GMFs. The central velocity difference between DPConCFil-B and DPConCFil-C is approximately 25 km s$^{-1}$. Figure \ref{Distribution_LBV} shows that these two GMFs are close to different spiral arms and exhibit a weak spatial connection, which justifies treating them as separate GMFs. A local zoom-in view of DPConCFil-B and MST-B suggests that, owing to constraints in spatial and velocity distance differences, the large-scale structures of MST may miss considerable internal clumps that are distant. Expanding the distance limitations, however, might connect more evidently disconnected substructures. Due to factors such as magnetic fields and density disturbances, the branch structures (e.g., striations or fibers) in DPConCFil-B and DPConCFil-C exhibit weaker alignment with the direction of the main skeleton, but these clumps have directional or positional consistency with the local filament axis and can converge to the main skeleton through consistency with clumps on it. Although DPConCFil may overlook some clumps at the boundaries, it is generally more complete. 

The experiments demonstrate that the consistency-based identification method is effective in identifying filaments of various scales in the PPV data, while the graph-based skeletonization method can robustly extract intensity skeletons even for large-scale filaments. However, when dealing with giant filaments that exhibit intricate interlacing of multiple sub-filaments, such as the GMF A \citep[with multiple velocity components,][]{MultipleVelocity} and DPConCFil-B \citep{GiantFilament_2}, accurately describing these complex structures using a single skeleton becomes challenging. If more skeleton analysis is required, the graph-based sub-structuring method proves to be valuable. 

\section{Filaments in the Simulation Data}\label{Simulation}
The positional consistency between clumps and the local filament axis has been corroborated through astronomical observations in different wavelengths \citep[e.g.][]{Review_2}. In simulations, the directional consistency is also widely observed, with MHD simulations showing a more pronounced trend of alignment compared to hydrodynamical ones \citep{Filament_Inertia_Matrix}. To further validate the applicability of the methods, we apply DPConCFil to a synthetic PPV datacube. 

The original simulated molecular clouds are samples from \cite{Simulation_1} used to study the evolution of filaments in a galactic environment, generated by the galactic-scale ISM suite simulation known as "The Cloud Factory" \citep{Simulation_2}. The Cloud Factory employs a version of the AREPO code \citep{Simulation_3,Simulation_4} to simulate and examine the ISM in spiral galaxies at various scales, from entire galaxies to individual filaments and clumps, while accounting for factors such as gravitational potential, CO and hydrogen chemistry, ultraviolet extinction, and supernova feedback \citep{Simulation_2,Simulation_6}.

We utilize the Polarized Radiation Simulator (POLARIS) code \citep{Simulation_5,Simulation_7} to generate synthetic CO (J=1-0) line emission for the molecular clouds. POLARIS employs the Monte Carlo method to trace the path of light rays and solve the radiative transfer problem. It incorporates a sub-pixeling technique that enhances the resolution of small-scale structures while maintaining reasonable computation times and ensuring a high signal-to-noise ratio. Energy levels and transitions are sourced from the Leiden Atomic and Molecular Database \citep{Simulation_8}. For level population calculations, we adopt the large velocity gradient method, which is developed in the context of idealized velocity fields \citep{Simulation_9,Simulation_10}. The resulting synthetic PPV datacube has a velocity resolution of 0.17 km s$^{-1}$ and a physical resolution of 0.5 pc pix$^{-1}$, comparable to MWISP observations.

\begin{figure*}
\centering
\vspace{0cm}
\begin{minipage}[t]{0.4\textwidth}
    \centering
    \centerline{\includegraphics[width=5in]{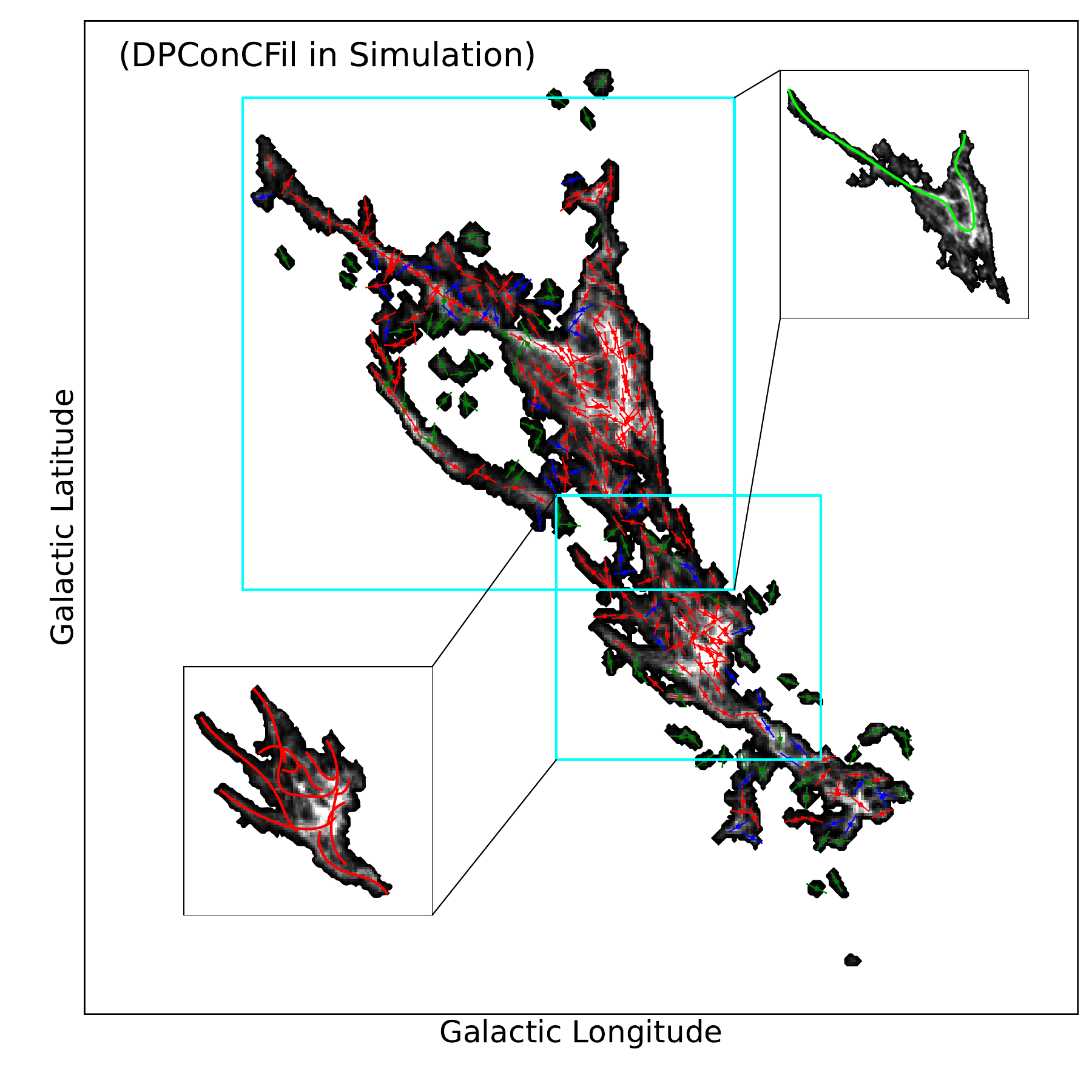}}
\end{minipage}
\caption{Filaments identified by DPConCFil from the simulated molecular clouds. The background shows the integrated intensity of all clumps detected by FacetClumps. The red, blue, and green asterisks and line segments denote the position and direction of clumps within the filaments, clumps outside the filaments that are connected to them, and clumps outside the filaments that are not connected, respectively. The zoomed-in views highlight the two largest filaments, with the lime curve in the upper right corner representing the longest intensity skeletons, and the various red curves in the lower left corner representing the intensity skeletons of different substructures.}
\label{Simulation_HD}
\end{figure*}

Figure \ref{Simulation_HD} displays the filaments identified by DPConCFil in the simulated molecular cloud. A total of 383 clumps are detected by FacetClumps, with 66.2\% located within the filaments, 12.6\% outside but connected to them, and 21.2\% outside and not connected. These molecular clouds are highly filamentary, exhibiting a higher proportion of clumps within the filaments compared to the proportion (about 50\%) from observational data. DPConCFil identifies a total of 14 filaments, with the two largest being highlighted in a zoom-in view, showcasing the longest intensity skeleton and the intensity skeletons of various substructures. 

\begin{figure*}
\centering
\vspace{0cm}
\begin{minipage}[t]{0.45\textwidth}
    \centering
    \centerline{\includegraphics[width=3.5in]{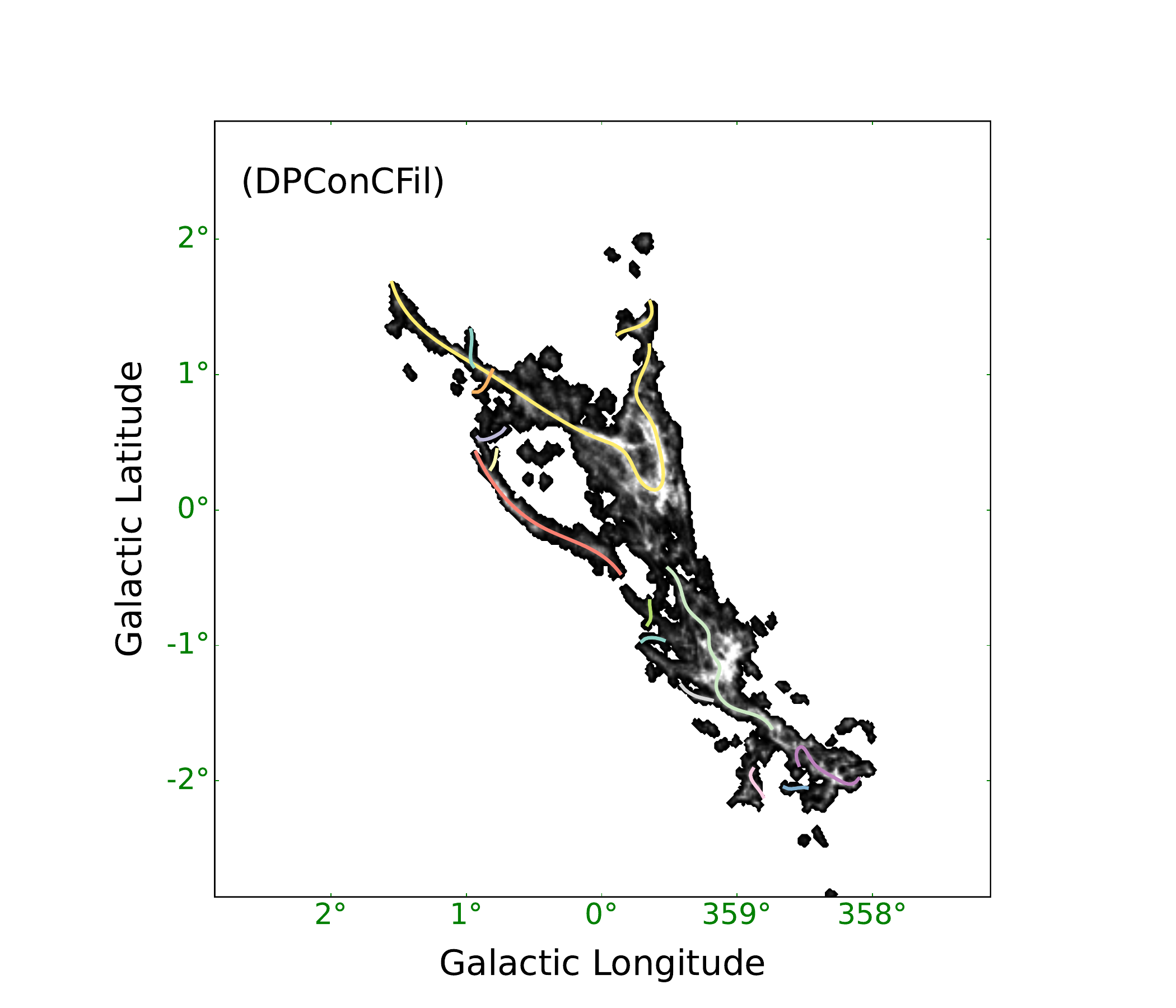}}
\end{minipage}
\begin{minipage}[t]{0.45\textwidth}
    \centering
    \centerline{\includegraphics[width=3.5in]{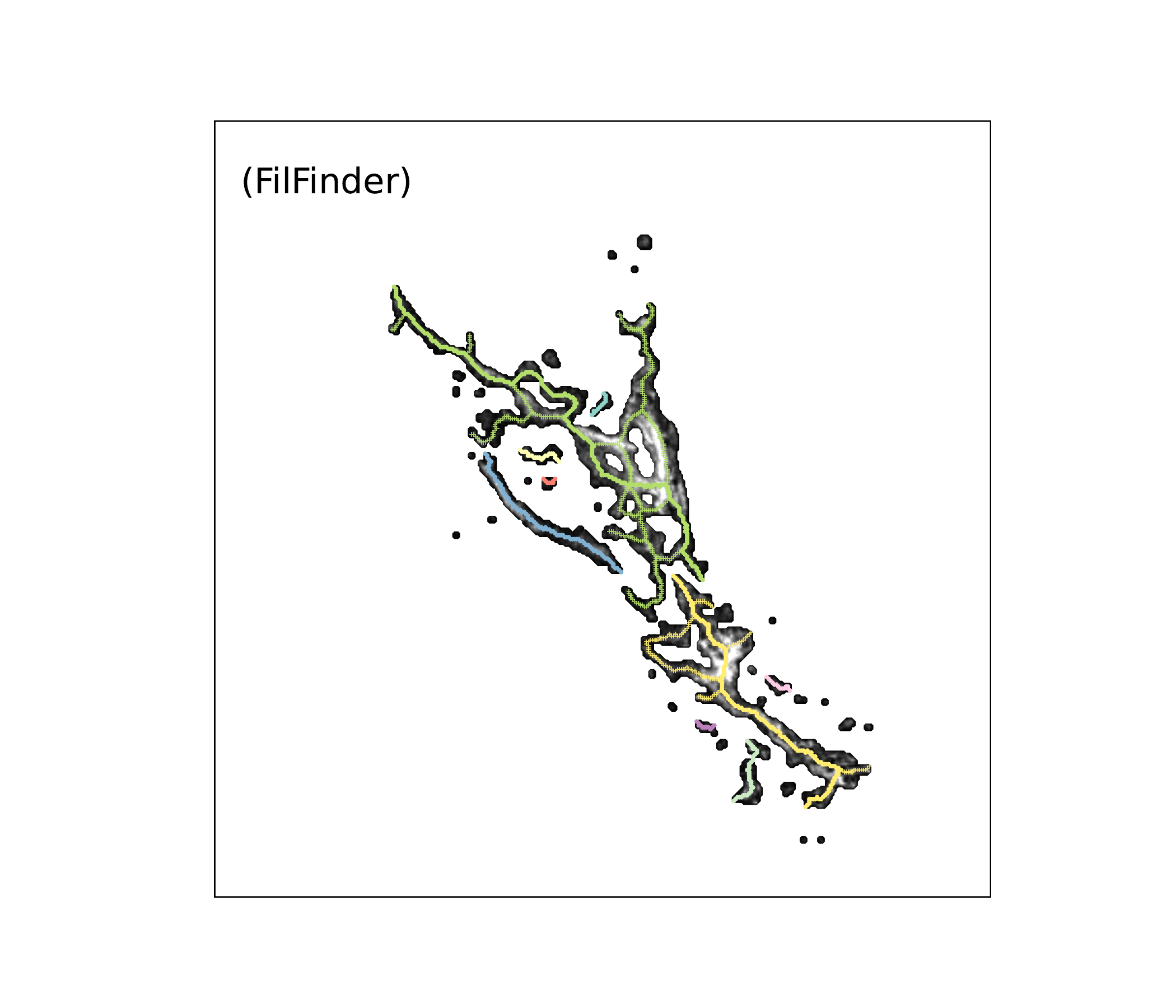}}
\end{minipage}

\begin{minipage}[t]{0.45\textwidth}
    \centering
    \centerline{\includegraphics[width=3.5in]{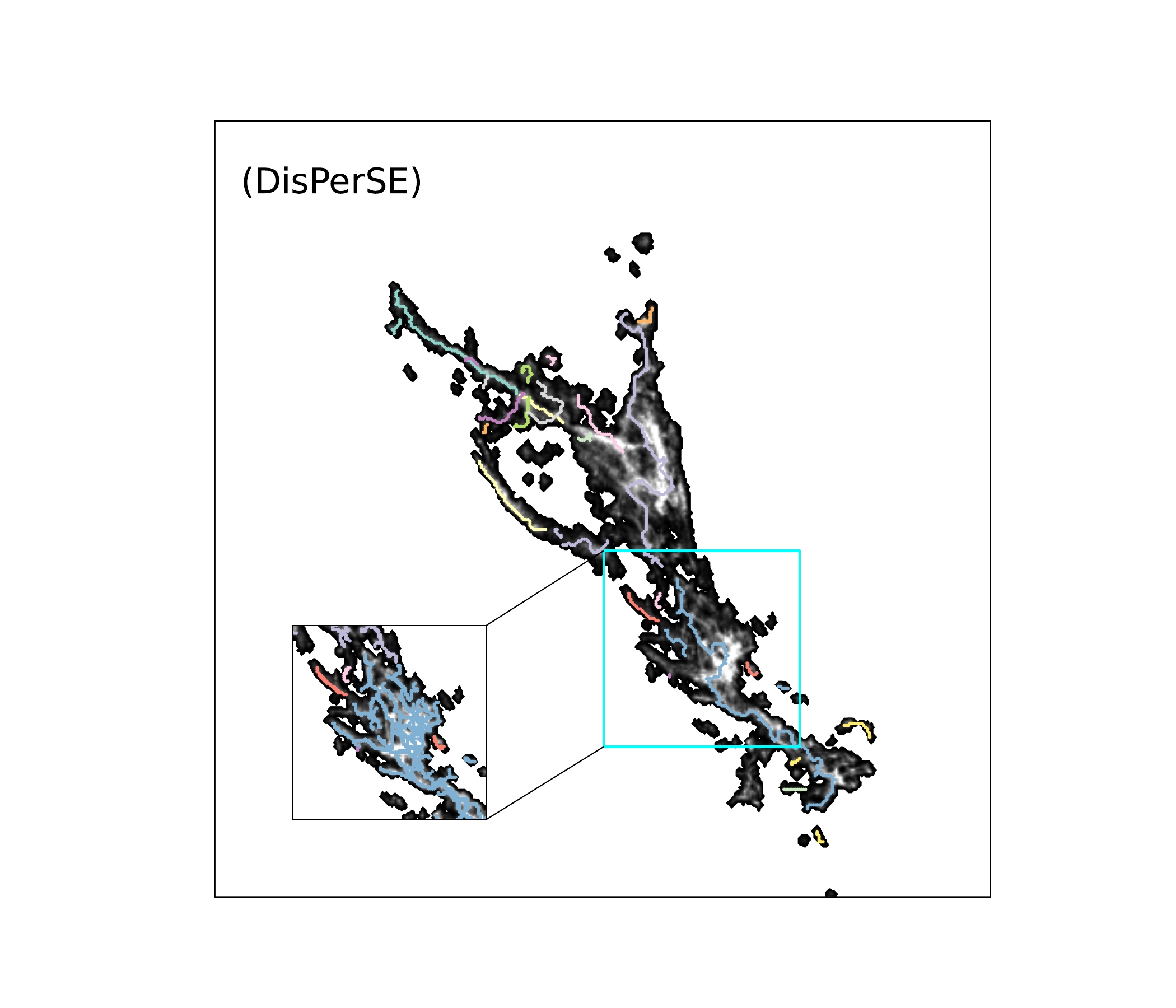}}
\end{minipage}
\begin{minipage}[t]{0.45\textwidth}
    \centering
    \centerline{\includegraphics[width=3.5in]{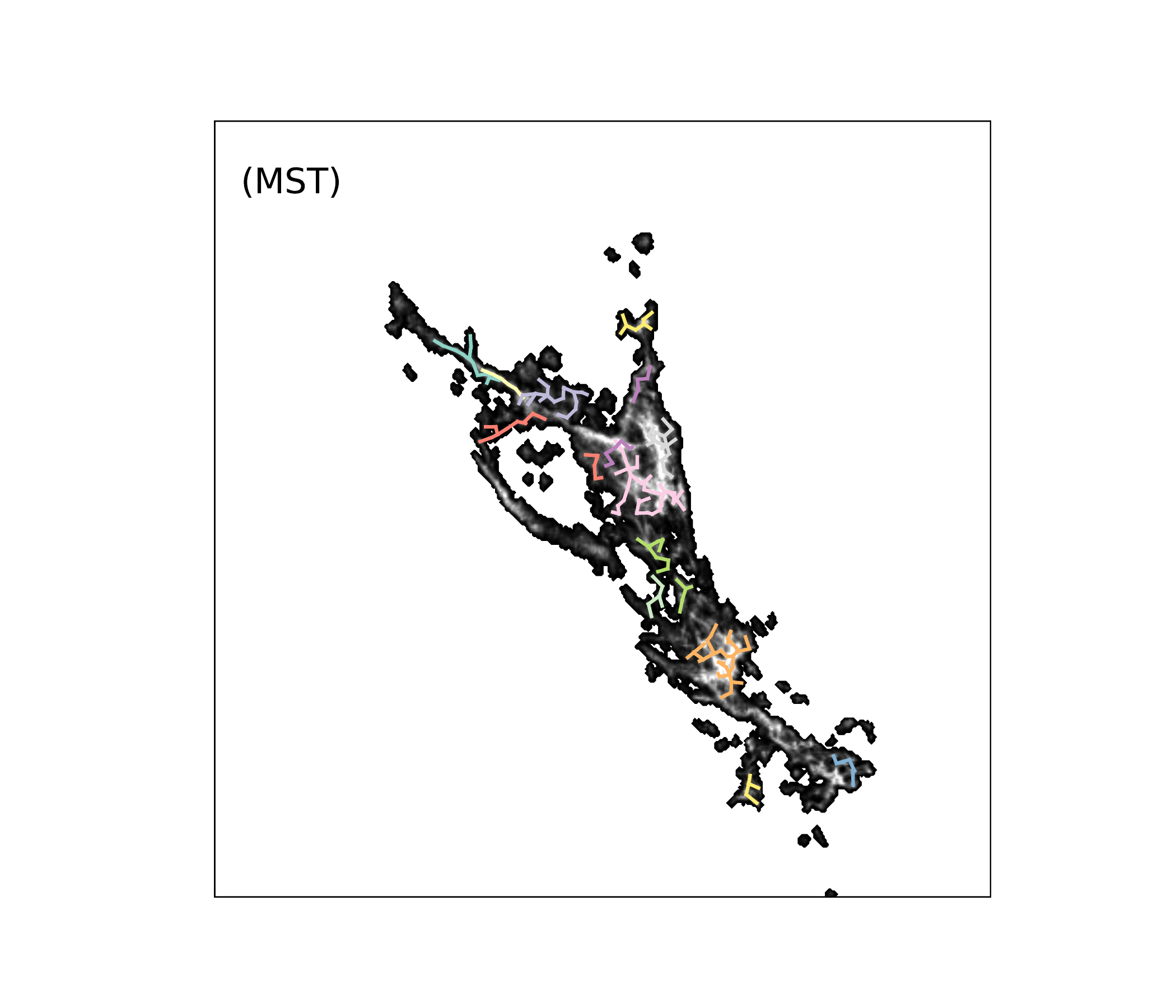}}
\end{minipage}
\caption{The structures isolated by various algorithms in the simulated molecular clouds. Different colors indicate distinct structures. DPConCFil: The background shows the integrated intensity of all clumps, and the curves denote the longest skeletons of filaments. FilFinder: the curves denote the skeletons extracted from the mask generated by FilFinder on the velocity-integrated map, similar to the left panel of Figure \ref{FilFinder_Structures}. DisPerSE: the curves denote the longest skeletons of filaments, with the zoomed-in view showcasing the skeletons within a sub-region. MST: the curves indicate the connections among clumps within the structures, which have not been filtered according to linearity checks, similar to the first panel of Figure \ref{MST_Structures}. }
\label{Algorithms_In_Simulation}
\end{figure*}

Figure \ref{Algorithms_In_Simulation} displays the structures isolated by different algorithms in the simulated molecular clouds. All algorithm parameters are consistent with those outlined in Appendix \ref{Comparison}. The code and results of the comparative experiments are available in the GitHub repository\footnote{\href{https://github.com/JiangYuTS/DPConCFil/tree/main/Comparative_Files}{https://github.com/JiangYuTS/DPConCFil/Comparative\_Files}}. FilFinder has a smaller area primarily determined by its adaptive threshold method, leading to a greater loss of flux. When encountering local regions with weak signals, DisPerSE may divide a filamentary structure into multiple filaments. In more complex regions, the filament skeletons generated by DisPerSE become excessively distorted. Resulting from the large distances between clumps, MST fails to identify several prominent elongated structures. 

These experiments support the widespread consistency observed across various simulated data \citep{Filament_Inertia_Matrix}. Furthermore, the identification and analysis methods of DPConCFil can be effectively applicable in simulated molecular clouds. 

\section{Configuration Parameters}\label{Parameters_FacetClumps}

The parameters of FacetClumps used in this work are presented in Table \ref{FacetClumps parameters}. 

\begin{table}
\centering
\caption{FacetClumps parameters.}
\label{FacetClumps parameters}
\begin{tabular}{l}
    FacetClumps.RMS=RMS(\textasciitilde0.22)\\
    FacetClumps.Threshold=5*RMS\\
    FacetClumps.SWindow=3\\
    FacetClumps.KBins=35\\
    FacetClumps.FwhmBeam=2\\
    FacetClumps.VeloRes=2\\
    FacetClumps.SRecursionLBV=[16,5]\\
\end{tabular}
\end{table}

\end{document}